\documentclass[twocolumn]{aastex631}
\usepackage{graphicx}
\usepackage{psfrag}
\usepackage{amsmath}
\usepackage{amssymb}
\usepackage{color,xcolor}
\usepackage{hyperref}
\hypersetup{
    colorlinks=true,
    citecolor=blue,
    filecolor=black,
    linkcolor=blue,
    urlcolor=magenta,
    linktocpage=true,
    breaklinks = true
}

\usepackage{enumitem}
\usepackage{subfigure,float}
\usepackage{multirow}
\usepackage{makecell}
\usepackage[normalem]{ulem}

\usepackage{array,booktabs}

\def\apjl{Astrophys. J.}

\def\aap{Astron. Astrophys.}
\def\apjs{Astrophys. J. Suppl.}

\newcommand{\beq}{\begin{equation}}
\newcommand{\eeq}{\end{equation}}

\newcommand{\ber}{\begin{eqnarray}}
\newcommand{\eer}{\end{eqnarray}}

\newcommand{\ba}{\begin{align}}
\newcommand{\ea}{\end{align}}

\def \nit {N_{\rm it}}

\def \dte {\Delta t_{\rm est}}

\def \mut {\mu_{\rm true}}
\def \dtt {\Delta t_{\rm true}}

\def \muc {\mu_{\rm try}}
\def \dtc {\Delta t_{\rm try}}

\def \dt{\Delta t}

\def \F{\mathcal{F}}

\def \p {\color{purple}}

\def \A {\mathcal{A}}

\def \frec {F_{\rm rec}}
\def \ft {F_{\rm true}}
\def \ftI {f_{1,\rm true}}
\def \ftII {f_{2,\rm true}} 
\def \frI {f_{1,\rm rec}} 
\def \frII {f_{2,\rm rec}}

\begin{document}
\title{Identifying lensed quasars and measuring their time-delays from unresolved light curves}

\author[0000-0003-0141-606X]{Satadru Bag}
\email{satadru@kasi.re.kr}
\affiliation{Korea Astronomy and Space Science Institute, Daejeon 34055, Korea}

\author[0000-0001-6815-0337]{Arman Shafieloo}
\email{shafieloo@kasi.re.kr}
\affiliation{Korea Astronomy and Space Science Institute, Daejeon 34055, Korea}
\affiliation{University of Science and Technology, Daejeon 34113, Korea}

\author[0000-0002-4359-5994]{Kai Liao}
\affiliation{School of Physics and Technology, Wuhan University, Wuhan 430072, China}

\author[0000-0002-8460-0390]{Tommaso Treu}
\affiliation{Physics and Astronomy Department, University of California, Los Angeles CA 90095, USA}

\begin{abstract}
Identifying multiply imaged quasars is challenging due to their low density in the sky and the limited angular resolution of wide field surveys.  We show that multiply imaged quasars can be identified using unresolved light curves, without assuming a light curve template or any prior information. 
After describing our method, we show using simulations that it can attain high precision and recall when we consider high-quality data with negligible noise well below the variability of the light curves.
As the noise level increases to
that of the Zwicky Transient Facility (ZTF) telescope, 
we find that precision can remain close to $100\%$ while recall drops to  $\sim 60\%$.
We also consider some examples from the Time Delay Challenge 1 (TDC1) and demonstrate that the time delays can be accurately recovered from the joint light curve data in realistic observational scenarios.
We further demonstrate our method by applying it to publicly available COSMOGRAIL data of the observed lensed quasar SDSS J1226-0006. We identify the system as a lensed quasar based on the unresolved light curve and estimate a time delay in good agreement with the one measured by COSMOGRAIL using the individual image light curves. The technique shows great potential to identify lensed quasars in wide field imaging surveys, especially the soon to be commissioned Vera Rubin Observatory.

\end{abstract}


\section{Introduction}

In the field of astronomy and astrophysics,  strong gravitational lenses have emerged as a powerful tool for determining several key factors, for example the initial mass function (IMF), the dark matter distribution and its time evolution in the lensing galaxies etc \citep{Mao:1997ek, Metcalf:2001ap,Dalal2002, Pooley_2009,Oguri2014, Jim_nez_Vicente_2019}. For more applications of strong lensing, we refer to the review by~\citet{Treu2010}. Remarkably, strongly lensed variable sources can also be a source of precision cosmology~\citep{2016A&ARv..24...11T}.
Precise time delay measurements can provide cosmological information, such as the value of Hubble constant ($H_0$) \citep{Refsdal1964_1,Refsdal1964_2,Saha,Oguri_2007, Bonvin2017,Wong:2019kwg,2020A&A...643A.165B,Birrer}. The time delay measurement of $H_0$ is independent of all other methods and can thus shed light on the ongoing tension \citep{Verde:2019ivm} between local measurements \citep{Riess:2019cxk} and the $H_0$ inferred from early universe probes like the cosmic microwave background (CMB) \citep{Planck:2018vyg}.  

So far, time delay cosmography has mainly relied on lensed quasars (QSOs). Hundreds of lensed QSOs have been detected~\footnote{https://research.ast.cam.ac.uk/lensedquasars/, https://strides.astro.ucla.edu/}, and  monitored~\footnote{http://cosmograil.org} and analyzed~\footnote{http://tdcosmo.org/}, making them the primary sources for current time delay cosmography efforts. 
However, other lensed transients~\citep{2019RPPh...82l6901O} such as supernovae~\citep{2020A&A...644A.162S}, repeating fast radio bursts~\citep{2018NatCo...9.3833L} and even gravitational waves~\citep{2017NatCo...8.1148L} will become significant contributors in the near future.  The first examples of multiply imaged supernovae have already been discovered \citep{Kelly:2014mwa,Goobar:2016uuf,2021NatAs.tmp..164R}.

Traditionally, lensed QSOs are detected using either imaging or spectroscopy based methods \citep[e.g.,][]{Huchra1985,Browne2003,Treu2018,Lemon2020}. 
Usually, measuring the time delay requires observations of light curves of the resolved images using high resolution telescopes for several years \citep[e.g.,][]{Tewes2013}. Substantial effort has gone into developing methods for extracting the time delays from the resolved image light curves \citep{Press1992,Pelt1996,2007A&A...464..471H,Kelly2009,Hirv2011,Hojjati2013,Aghamousa:2014uya}. However, detecting lensed quasars and measuring the time delays are very challenging tasks. For example, it is often difficult to distinguish binary QSOs from lensed doubly-imaged quasars only using imaging or spectroscopy data \citep{Peng_1999, Mortlock1999}. Comparing the similarity of the light curves of close images via measuring the time delay was proposed as a way to distinguish any lensed QSOs from star or QSO pairs~\citep{Pindor2005} as well. However, this approach requires resolved images, and monitoring the image light curves at high angular resolution can be very expensive in terms of telescope time. 

An alternative approach consists of using the observed unresolved (joint) light curve for detecting the lensed QSO systems and  measuring the time delay subsequently. In this approach, one does not require the systems to be resolved a priori. Therefore, this approach can take advantage of the data from ongoing time domain surveys such as ZTF \citep{ztf}, Pan-STARRS1 \citep{Pan-STARRS1}. The unresolved light curve approach will have broad application when the Vera Rubin Observatory  \citep{lsst1,lsst2} starts the Legacy Survey of Space and Time (LSST). In principle, of course, images with sufficiently large separations can be resolved with advances in image processing
techniques, e.g. the image deconvolution method developed
by~\citet{1998ApJ...494..472M}, that allows for optical monitoring of systems
in which the image separation was small compared to the seeing \citep{Millon:2020xab}. For example, for LSST, $\theta_{PSF}=0.75$ arcs, ~\citet{2010MNRAS.405.2579O} adopted 2/3$\theta_{PSF}$ as the minimum separation that the surveys can resolve. However, a method based on unresolved images can be extremely powerful and complementary. First, the unresolved method would allow the detection of images with separations too small to be identified in ground based imaging. Second, systems that are partially resolved could be difficult to identify in survey data, because deblending algorithms could partition them in a variable number of astronomical objects depending on seeing. A robust algorithm based on unresolved light curves could for example be applied on data that have been convolved to the lowest angular resolution of the time series in order to mitigate deblending issues and increase signal to noise ratio.

In the past couple of decades, several methods have been proposed to detect lensed quasars and to measure their time delays using the joint unresolved light curves \citep{Geiger1996, Shu2020,Springer:2021yhe,Springer:2021jhm, 2021arXiv211001012B}.
However, since quasar light curves are highly stochastic and display broad variety, approaches based on forward modelling the light curves under specific assumptions may be powerful but restricted in their application only to the light curves well described by the assumptions, and may thus be incomplete and biased\footnote{The challenges in model agnostic approaches to extract time delays from the unresolved lensed supernova (SN) light curves are quite different since one knows the broad shape of the SN light curves (that follow a mere rise and fall in the flux); see \citet{Bag:2020pbg,Denissenya:2021cpz}.}. 
Therefore, these attempts are likely to be less successful than estimated if the underlying assumptions are not a good description of real light curves, and result in lower precision and purity.  

With the goal of attaining a more general method, and hopefully reaching higher completeness, we develop a new technique to identify lensed quasars and measure their time delays from unresolved light curves, without assuming any template or model for the quasar light curves, and without relying on additional information.  Our method builds on that originally proposed by \citet{Geiger1996}, adding a statistical procedure to identify the true lenses and time delays by minimizing fluctuations in the reconstructed light curve\footnote{This minimization is conceptually similar to that applied by the \citet{Pelt1996} method to resolved light curves, although different in implementation.}

The paper is organized as follows. In section~\ref{sec:method} we describe the methodology that consists of reconstructing the underlying images from the observed joint light curve. We illustrate a mathematical degeneracy of the problem and then demonstrate a method to break the degeneracy partially and identify the lens systems and measure the time delay. Next, in section~\ref{sec:validation} we validate our method on simulated data with virtually noiseless data (i.e. in perfect conditions).
In section~\ref{sec:td_withnoise} we show that our method works on data with ZTF-like observational noise. In Section~\ref{sec:TDC1} we apply our technique to realistic simulations from the Time Delay Challenge \citep[][hereafter TDC]{tdc1,tdc2} 1 (Rung 0 and 1).  In section~\ref{sec:obs_J1226}, we apply our method to the light curve of the lens systems SDSS J1226-0006 obtained from the COSMOGRAIL database. In section~\ref{sec:conclusion} we summarize and conclude the work.

\section{Methodology: reconstructing the underlying light curves}
\label{sec:method}
For simplicity, let us consider a strong-lensed quasar system that is composed of two images only. Therefore, the observed flux of the lensed system would be sum of the light curves of individual images,
\beq \label{eq:lensed_flux}
F(t)=f_1(t)+f_2(t)\;.
\eeq
Intrinsic light curves of all images can be described by a common function, say $f(t)$, with different magnifications ($\mu_i$) and time delays ($t_i$),
\begin{align} \label{eq:ind_fs}
f_1(t)&=\mu_1 f(t-t_1)\;, \\ \nonumber
f_2(t)&=\mu_2 f(t-t_2)=(\mu_2/\mu_1)f_1(t-[t_2-t_1])\;,\\ 
&=\mu f_1(t-\dt)\;, 
\end{align}
where $\mu\equiv\mu_2/\mu_1~{\rm and}~\dt \equiv t_2-t_1$ so that Eq.~\eqref{eq:lensed_flux} can be recast as
\beq \label{eq:tot_flux}
F(t)=f_1(t)+\mu f_1(t-\dt)\;.
\eeq
Here $\mu$ and the $\dt$ are the magnification and time delay of the image 2 with respect to image 1. Or, in other words, they are the magnification ratio and the time delay between the two images. 

From Eq.~\eqref{eq:tot_flux} we try to reconstruct the flux of image 1,
\begin{align}
f_1(t)&=F(t)-\mu f_1(t-\dt) \;, \\ \nonumber
      &=F(t)-\mu \left[F(t-\dt)-\mu f_1(t-2\dt) \right]\;,
\end{align}
where we substitute the expression for $f_1(t-\dt)$ from the functional form of $f_1(t)$ in the first line while arriving to the second step.
We can continue this substitution to obtain an infinite series expression for $f_1(t)$ in terms of the combined flux,
\begin{align}\label{eq:f1_rec}
f_1(t)=&F(t)-\mu F(t-\dt)+\mu^2 F(t-2\dt) \nonumber \\ 
& \qquad \qquad -\mu^3F(t-3\dt)+\mu^4F(t-4\dt)-....\;, \nonumber \\ 
      =& \sum_{n=0}^{\infty} \left( -\mu \right)^n F(t-n\cdot \dt)\;.       
\end{align}
 Up to this point, we do not impose any restriction other than assuming that the lensed system has two images. If $\mu<1$, the infinite sum converges and, in principle, we can determine the light curve of image 1.
 
 Without any loss of generality, let us refer to the image with larger magnification ($\mu_i$) as image 1 and the other as image 2 (this can be easily generalised to more than 2-image systems by sorting the $\mu_i$'s). This ensures $\mu \equiv \mu_2/\mu_1 <1$ which in turn guarantees that the sum in Eq.~\eqref{eq:f1_rec} converges. Now a positive (negative) $\dt$ means image 2 (the image with smaller magnification) arrives later (earlier) in time with respect to image 1. 
 
 Therefore, given any choices of $\mu$, $\dt$, we can reconstruct the light curve of image 1 (which has the larger magnification) using Eq.~\eqref{eq:f1_rec} and then we can calculate $f_2(t)$ using Eq.~\eqref{eq:ind_fs}. Let us call these reconstructed underlying light curves $\frI(t),\frII(t)$.  In this approach it is ensured that $ \frec(t) \equiv \frI(t)+\frII(t) = F(t)$ (the observed light curve of the lensed system). The key point is that for any choice of $(\mu,~\dt)$ one can obtain a unique solution for $f_1(t)$ that satisfies Eq.~\eqref{eq:tot_flux} exactly\footnote{Although we test doubly imaged lensed systems in this article for simplicity but this approach can be generalised to more than 2-image systems. }.

 We use cubic interpolation to obtain the flux in between two observed points as required in the sum in Eq.~\eqref{eq:f1_rec}. However, the higher order terms in the summation also require the flux outside the observed range. Since we cannot predict the quasar light curve beyond the observation range (owing to the fact that we know little about the time variability of quasar light curves in general), we assume that $F(t)$ remains flat outside the observation range for simplicity. Below we briefly discuss that this assumption has a negligible effect on the reconstruction at one boundary \cite[as demonstrated by][]{Geiger1996}.

 \subsection{Mathematical degeneracies}
 \label{sec:rec}

 \begin{figure}
\centering
\includegraphics[width=\linewidth]{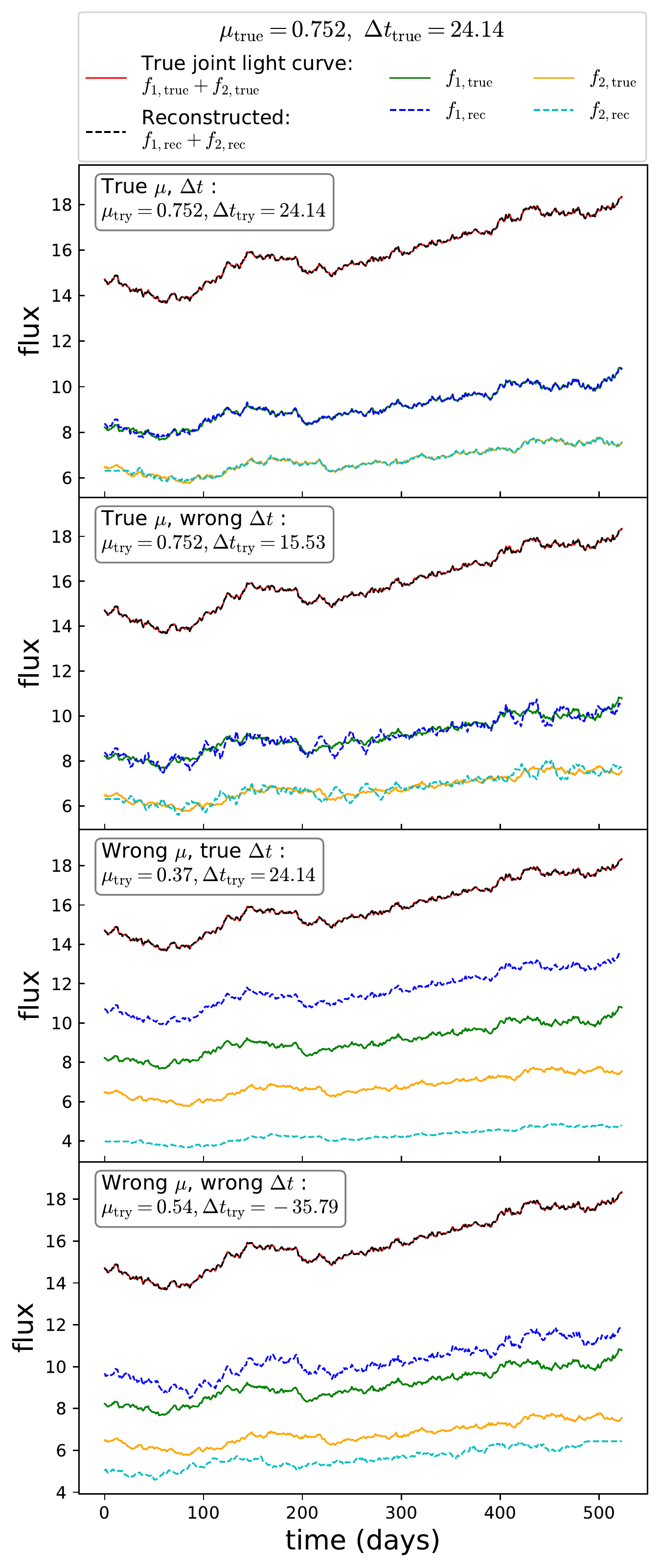}
\caption{Light curves of a typical doubly imaged quasar (in arbitrary flux units) with negligible observational noise (system 6: $\mut=0.752$, $\dtt=24.14$ days, no. of data points=$524$). We compare the reconstructed light curves (of the 2 individual images and the joint one) for 4 choices of $\lbrace \muc, \dtc \rbrace$ shown in the 4 panels. 
Only for the choice $\muc=\mut, ~\dtc=\dtt$ (top panel), the reconstructed light curves match the true ones, i.e. $\frI \approx \ftI$ and $\frII \approx \ftII$.}
\label{fig:noerror_system6}
\end{figure}

In this section we demonstrate that the  reconstruction technique described above yields mathematically degenerate lensed solutions from any joint light curve. First let us consider high quality data with negligible observational noise, well below the variability of the light curves, so that we have $F_{\rm true}(t)=f_{1, \rm true}(t)+\mu f_{1, \rm true}(t-\dt)$. Then we reconstruct the light curve of the first image from $F_{\rm true}(t)$ for different choices of the trial $\lbrace \muc, \dtc \rbrace$ using Eq.~\eqref{eq:f1_rec}. 

In order to illustrate the mathematical degeneracy we consider a simulated lensed system composed of two images with the true time delay $\dtt=24.14$  days and the magnification ratio $\mut=0.752$. 
This system is labeled as system 6 in the simulated set that we use for training and validation purpose in the next section. The details of the simulation are given in Section~\ref{sec:validation}. 
We compare the reconstructed light curves of the underlying images and the combined one ($\frI,~\frII$ and $\frec$) with the corresponding true light curves ($\ftI, ~\ftII$ and $\ft$) in Figure~\ref{fig:noerror_system6} for 4 choices of $\lbrace \muc, \dtc \rbrace$. In all the 4 panels, the reconstructed joint light curve matches accurately with the truth, i.e. $\frec=\ft$ (notice that the red and the black dashed curves coincide in all the panels).
In the top panel, we choose the values of trial magnification ratio and time delay same as the true values, $\muc=\mut, ~\dtc=\dtt$, finding that the reconstructed image light curves match the corresponding true curves, i.e. $\frI \approx \ftI$ and $\frII \approx \ftII$. In the other panels, $\frI$ and $\frII$ are different from $\ftI$ and $\frII$ respectively, although $\frec(t)=\ft(t)$ is maintained in all 4 panels. Note that flux is shown in arbitrary units in figures throughout the article since only the relative time variability of the light curve matters.
 
We estimate the precision of the reconstruction of the combined flux by calculating
 \begin{equation}
     E_F=\frac{1}{N_D}\sum_{i}^{N_D} ~\bigg|\frac{\ft(t_i)-\frec(t_i)}{\ft(t_i)}\bigg|\;,
 \end{equation}
where $N_D$ is the number of data points in a light curve (number of observation epochs).
 For any choice of $\lbrace \muc, \dtc \rbrace$ we find that $E_F \sim \mathcal{O} (10^{-15})$ which is basically a manifestation by the numerical round-off error. This clearly tells us that for any choice of $\lbrace \muc, \dtc \rbrace$ we always get back the combined flux ($\frec(t)=\ft(t)$) exactly.
 Using the truths, $\lbrace \muc, \dtc \rbrace \to \lbrace \mut, \dtt \rbrace$, we can reconstruct the underlying images, $\frI \to \ftI$, $\frII \to \ftII$.  If the trial $\lbrace \muc, \dtc \rbrace$ are different from the true values, the joint light curve is still reconstructed exactly but the light curves of individual images are wrong.

We do not discuss the cases in the presence of significant amount of noise/uncertainty in the data yet because even with negligible noise in the data (in perfect conditions) we get degenerate solutions. Later in the article, we include ZTF-like noise in the data while we try to estimate the time delay using a version of this approach, modified to break the degeneracy.

As we assume the observed joint light curve to be flat outside the observed time range, some error is introduced in the reconstruction of first image ($\frI$) near one of the boundaries even for true choices: $\muc=\mut$ and $\dtc=\dtt$. 
E.g. a positive $\dtc$ introduces maximum error in $\frI$ near the left edge, i.e. for $t\leq t_0+\dtc$ where $t_0$ is starting time, and the reconstruction gradually becomes more accurate for larger $t$ as more terms from the observed range start to contribute. The same thing happens but from the opposite boundary when $\dtc$ is negative. This conclusion is demonstrated in Figures~3 and~4 of \citet{Geiger1996}.

\citet{Geiger1996} correctly emphasised that one gets infinite solutions due to a mathematical degeneracy and  concluded that it is impossible to determine the correct time delay without assuming any additional information such as the intrinsic variability of the quasar light curves. 
However, as we detail in the next subsection,  there is a way to break the degeneracy ( partially), identify the lensed systems, and measure the time delay without requiring any additional information.

\subsection{Identifying the true solution}
\label{sec:dist_true}

\begin{figure*}
\centering
\includegraphics[width=\linewidth]{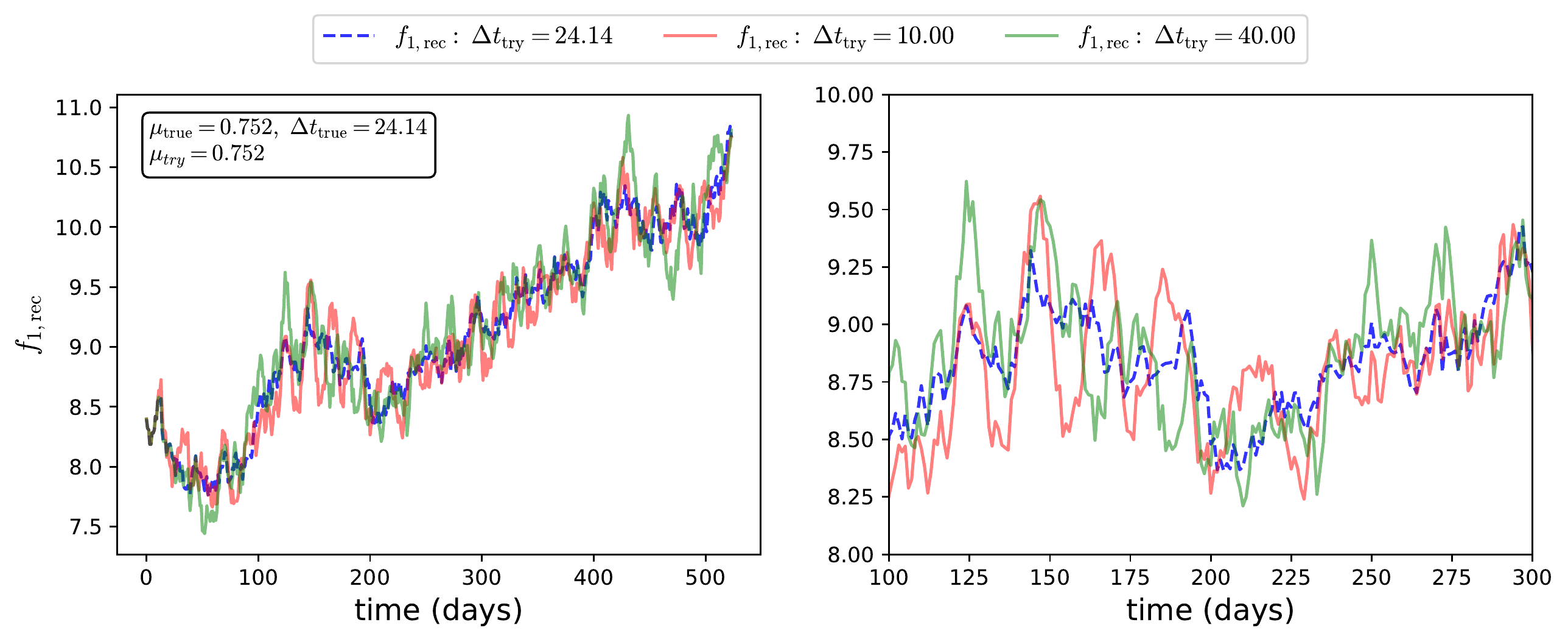}
\caption{The reconstructed light curves of the first image ($\frI$, in arbitrary unit) corresponding to different $\dtc$ are compared for system 6 (considering data with negligible noise). The trial magnification is fixed to the true value, $\muc=\mut=0.752$. The right panel zooms in to a shorter interval to show that for any $\dtc \neq \dtt$ more fluctuations are found in the reconstructed light curve. }
\label{fig:show_fluc_sys6_mut_NE}
\end{figure*}

\begin{figure*}
\centering
\includegraphics[width=\linewidth]{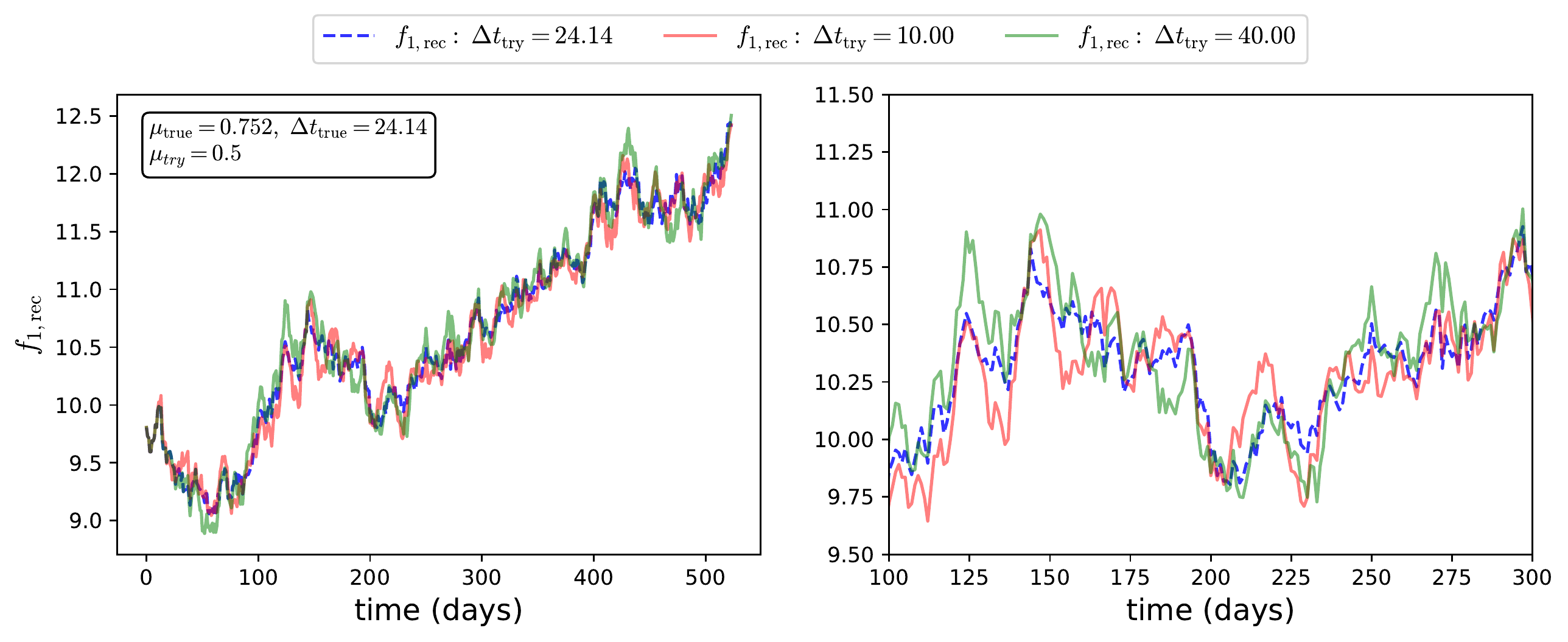}
\caption{Same as Figure~\ref{fig:show_fluc_sys6_mut_NE} but with $\muc=0.5$ which is different from the $\mut$. Again we find that the light curves $\frI$'s corresponding to $\dtc \neq \dtt$ have more fluctuations than the true one.}
\label{fig:show_fluc_sys6_mu3_NE}
\end{figure*}

In this section we show that the true time delay can be identified by breaking the degeneracy with  $\dtc$ via minimization of fluctuations in the reconstructed light curve.

To illustrate this approach, we first consider the cases with negligible noise. Having demonstrated that the approach works on perfect data we then apply it to cases with substantial observational noise. 

For this illustration, we focus on the same example as in previous section (lensed quasar system 6) taken from the training set. We will not use the values of $\mut$ and $\dtt$ anywhere in the analysis anymore; however, the true values will help us train our algorithm. Since the reconstructed light curve of the second image, i.e. $\frII$, is just a scaled (by $\muc$) and shifted (by $\dtc$) version of the light curve of the first image ($\frI$), from now onward we only focus on $\frI$. Remember that we refer to the brighter (fainter) image as first (second) image.

In Figure~\ref{fig:show_fluc_sys6_mut_NE} we compare the reconstructed light curves of the first/brighter image ($\frI$) corresponding to different choices of $\dtc$ while keeping $\muc=\mut$ fixed. The left panel shows the light curves for the full period of observation, whereas the right panel zooms into the time interval between ($100, 300$) days. The dashed blue curve in both panels represents $\frI$ corresponding to $\dtc=\dtt$ ($\muc$  is already fixed to $\mut$) that matches very well the true light curve of the first image $\ftI$  (Figure~\ref{fig:noerror_system6}).
Crucially, for values of $\dtc$ that are different than $\dtt$ we get more overall fluctuations in the respective $\frI$ curves around their mean values (notice that the red and green curves in the right panel of Figure~\ref{fig:show_fluc_sys6_mut_NE} show more fluctuations than the dashed blue curve).
 
Figure~\ref{fig:show_fluc_sys6_mu3_NE} shows the same set of curves as in Figure~\ref{fig:show_fluc_sys6_mut_NE} but for another choice of $\muc=0.5$ (which is different from $\mut$). Naturally, all the reconstructed light curves now lie above the true light curve since $\muc$ (which measures the contribution of the second image) is smaller than $\mut$. However, we observe the same feature that $\frI$'s for $\dtc \neq \dtt$ exhibit more fluctuations than the $\frI$ corresponding to $\dtc=\dtt$. Also, we observe that the $\frI$'s are simply somewhat scaled according to different choices of $\muc$.

\begin{figure*}
\centering
\includegraphics[width=\textwidth]{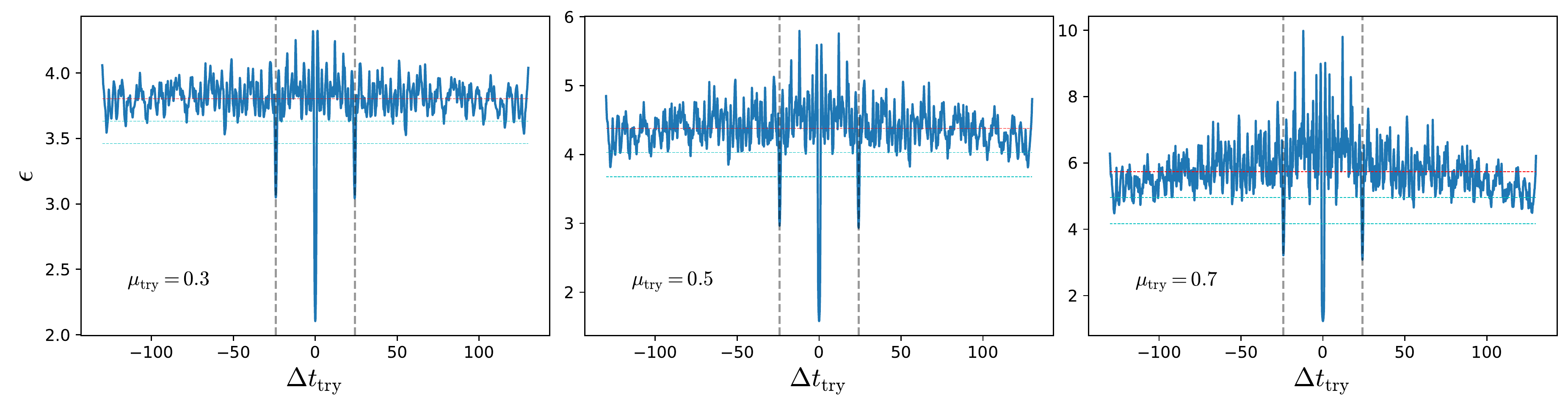}
\caption{Amount of fluctuation in $\frI$ (estimated via the $\epsilon$ statistic) as a function of $\dtc$ for case study system 6 (considering negligible noise in the data).
The three panels correspond to three fixed values of the magnification ratio, $\muc=0.3, 0.5, 0.7$ from left to right. 
The horizontal red and cyan lines in all the panels show the mean and the $1\sigma$, $2\sigma$ values of the respective $\epsilon (\dtc)$ curve.
The dashed vertical lines mark $\dtc = \pm \dtt$ around where $\epsilon (\dtc)$ in each panel shows a pair of prominent secondary minima. The minimum at $\dtc=0$ is a generic  feature corresponding to the `unlensed' solution for which the reconstructed light curve is a scaled version of the observed light curve.}
\label{fig:nonoise_sys6_eps}
\end{figure*}

\begin{figure}
\centering
\includegraphics[width=\linewidth]{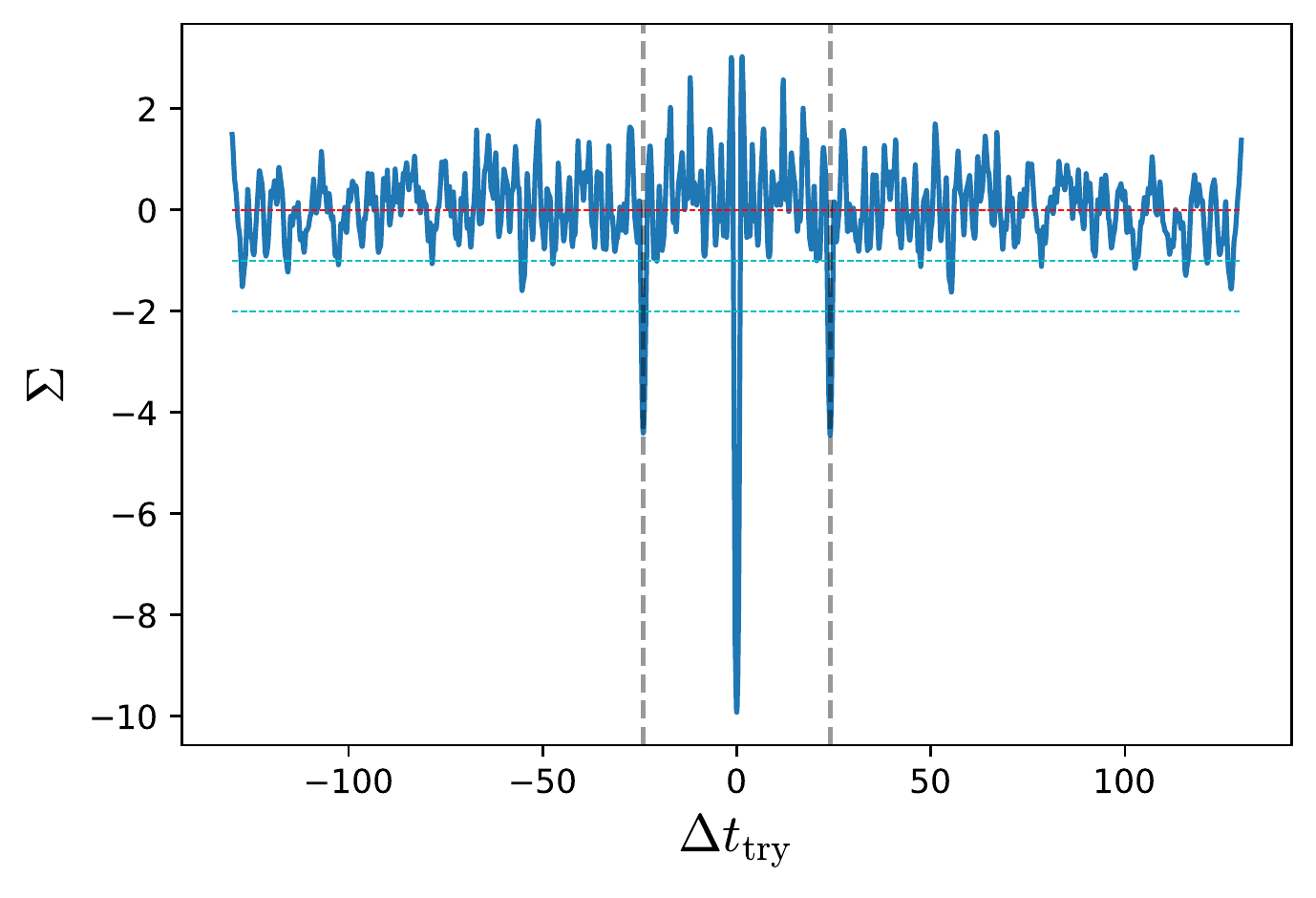}
\caption{Dimensionless fluctuation estimator $\Sigma$  as a function of $\dtc$ with fixed $\muc=0.3$ for system 6 (considering negligible noise in the data). The horizontal red and cyan lines show the mean and the $1\sigma$, $2\sigma$ values, respectively.
The dashed vertical lines mark the true time delay, $\dtc = \pm \dtt$.}
\label{fig:nonoise_sys6_mu0.3_sig}
\end{figure} 
 
For $\dtc=\dtt$, one gets minimal fluctuation in the reconstruction of individual image light curves, e.g. $\frI$. For other trial time delays, $\dtc \neq \dtt$,  $\frI$'s tend to exhibit more fluctuations. The only exception, as we find later, is $\dtc=0$ for which the reconstructed light curve is just a scaled version of the observed light curve and thus produces a generic minimum in the fluctuations which can be ignored. The effect of choosing different $\muc$'s appears to be negligible in this regard. We find that all the systems show the above characteristic.   This is not unexpected: light curves that are reconstructed using the wrong trial time delay will mix fluxes at different intrinsic times, which are expected to differ more than fluxes at the same intrinsic time. Conceptually, this is the same principle adopted by \citet{Pelt1996} to identify the time delays from spatially resolved light curves.

In conclusion, we can distinguish the true underlying light curve among all the reconstructions by measuring the amount of fluctuations in the reconstructed light curves, e.g. in $\frI$'s, corresponding to different $\dtc$ (while keeping $\muc$ fixed). We also tested the effect of trying different choices of $\muc$.
For a fixed $\muc$, we estimate the {\em amount of fluctuation} in $\frI$ corresponding to different $\dtc$ using the expression
 \begin{equation}\label{eq:fluction_raw}
     \epsilon(\dtc)=\sum_{i}^{N_D} \left(\frI (t_i) - \frI (t_{i+1})\right)^2\;,
 \end{equation}
 where $N_D$ is the number of data points (observation epochs). Note that the quantity $\epsilon$ has the dimension of flux squared. For completeness, we experimented with other metrics (alternative to \eqref{eq:fluction_raw}) for quantifying the fluctuation, e.g. using the aggregated deviation in $\frI$ with respect to its smoothed version. Although different metrics produce consistent results, we do not find any that performs better than the  simple formalism given in \eqref{eq:fluction_raw}, especially in the presence of substantial amount of noise in the data.

As an illustration we again consider the case of the lensed system 6. We compute $\frI$ for all $\dtc \in (-130.0,130.0)$ days with a resolution of $0.1$ days with a fixed $\muc$. For each $\dtc$, we separately calculate the corresponding value of $\epsilon$ using Eq.~\eqref{eq:fluction_raw}. The three panels of Figure~\ref{fig:nonoise_sys6_eps} show $\epsilon$ as a function of $\dtc$ for system 6 for three fixed values of trial magnification ratio, $\muc=0.3,~0.5,~0.7$ from left to right.
 The gray dashed vertical lines in each panel mark $\dtc=\pm \dtt$. The red and two cyan lines represent the mean and the $1\sigma$, $2\sigma$ values of entire $\epsilon$ curve in each panel.
 We observe the followings.
 \begin{itemize}
     \item The $\epsilon(\dtc)$ curve looks quite symmetric around $\dtc=0$ and the minima appear at similar $|\dtc|$ values in the positive and the negative domains. 
     
     \item At $\dtc=0$, which gives essentially 1-image solution (let us call it `unlensed'), we get a global minimum. This global minimum, implying least fluctuations present in the unlensed solution, always appears at $\dtc=0$ for all the systems (even in the presence of substantial amount of noise in the joint light curve data). 
     
     \item At $\dtc \approx \pm \dtt$ (marked by the vertical dashed lines), we notice a pair of prominent secondary minima. One can determine if the system is lensed by identifying this pair of prominent secondary minima in addition to the global minimum.
     
     \item The three observations above are valid for all reasonable choices of $\muc$. Moreover, the positions of the minima (including the pair of secondary minima) do not change significantly with $\muc$.

 \end{itemize}

Since the target secondary minimum in the $\epsilon(\dtc)$ curve appears both at $\pm \dtt$, we cannot determine if the time delay is positive or negative from this approach. This is expected since from unresolved data we cannot establish whether the brighter or fainter image is the leading one\footnote{In our convention, the brighter image is tagged as `first".}. Furthermore, the positions of the minima in the $\epsilon(\dtc)$ curve are quite insensitive to the choice of $\muc$, as evident from the three panels of Figure~\ref{fig:nonoise_sys6_eps}. Thus, our method can estimate the time delay and confirm the lensing nature of an unresolved source, but cannot be used to estimate the magnification ratio.
Since different choices of $\muc$ do not affect the final results, we fix $\muc=0.3$ throughout the article\footnote{One can also combine the fluctuation curves corresponding to different $\muc$ to increase the signal in our analysis that helps us identify the lensed systems, but for simplicity we use just $\muc=0.3$ in this article.}.
 
We conclude this section by introducing a dimensionless version of the fluctuation statistics: 
  \begin{equation}\label{eq:fluc_sig}
     \Sigma(\dtc)= \frac{\epsilon(\dtc) - \langle \epsilon(\dtc) \rangle}{\sigma_{\epsilon}}\;,
 \end{equation}
 where we subtract the mean from $\epsilon(\dtc)$ and then scale it with the standard deviation of the whole $\epsilon(\dtc)$ curve, $\sigma_{\epsilon}$. Therefore, $\Sigma$ essentially measures the fluctuations in units of the standard deviation in the $\epsilon(\dtc)$ curve. As an illustration, we re-plot the left panel of Figure~\ref{fig:nonoise_sys6_eps} in terms of $\Sigma$ as a function of the trial time delay in Figure~\ref{fig:nonoise_sys6_mu0.3_sig}, for system 6 with fixed $\muc=0.3$.

 \section{Validating the method on simulations}
 \label{sec:validation}
In this section and in the following one we train and validate the method, proposed in the previous section, on a number of simulated lensed systems. 
The simulations are taken from the Time Delay Challenge 1 (TDC1). However, for simplicity, we ignore the microlensing effect and only consider doubly-imaged systems. Simulating the observation of a doubly-imaged but unresolved quasar therefore involves three conceptual steps \citep[see][for details]{tdc1,tdc2}: 1) The quasar's intrinsic light curve in a given band is generated at the accretion disk of the black hole in an active galactic nucleus (AGN) and modelled by a Damped Random Walk (DRW) process. 2) The foreground lens galaxy causes multiple imaging, leading to two light curves that are offset from the intrinsic light curve (and each other) in both amplitude (due to magnification), and time. 3) Since we assume the images are unresolved, we combine the individual image light curves. In this section we consider high quality light curves with negligible noise  and one day cadence. The next section deals with data with realistic noise, at the level one can expect for, e.g., ZTF.
 
 \subsection{Testing the algorithm on a training set}

\begin{figure*}
\centering
\subfigure[Lensed systems]{
\includegraphics[width=0.485\linewidth]{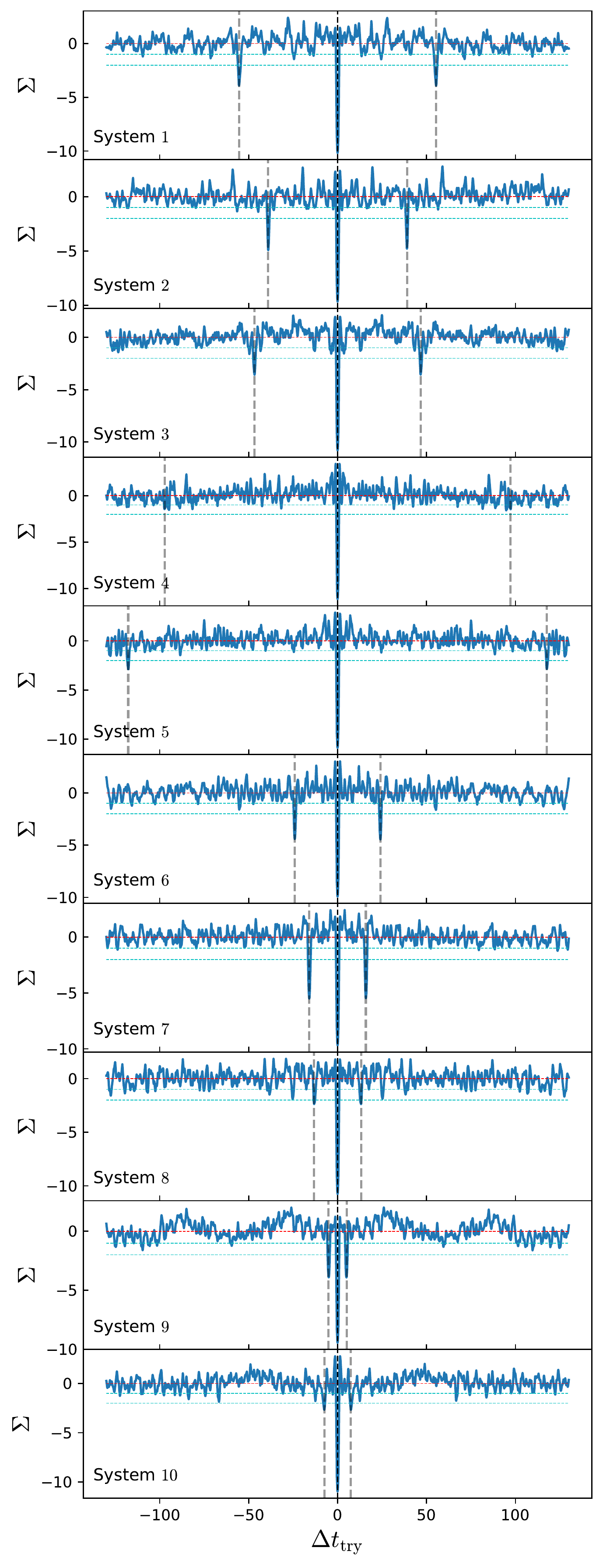}\label{fig:nna}}
\subfigure[unlensed systems]{
\includegraphics[width=0.485\linewidth]{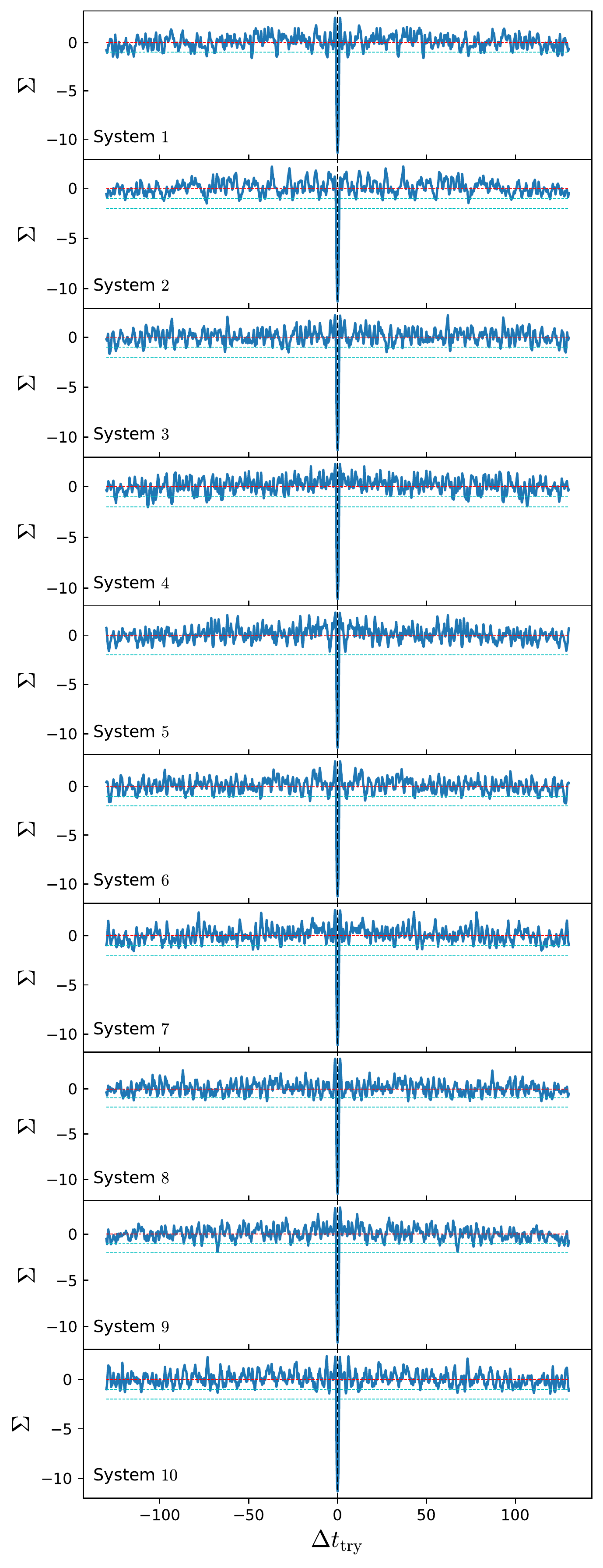}}
\caption{Fluctuation statistics for the training set with negligible noise: lensed (left panel), unlensed (right panel). Apart from the global minima at $\dtc=0$, we find pairs of prominent secondary minima (more than $2\sigma$ deep) near $\dtc=\pm \dtt$ only for the lensed systems. The trial magnification ratio is kept fixed at $\muc=0.3$.}
\label{fig:nonoise_set2_sig_comb}
\end{figure*}

First we analyse  a `training' set comprising of 10 lensed and 10 unlensed (with just 1-image, hence no-lensing) systems with known time delays (and magnification ratios) that have been used for optimizing the algorithm. The set includes a diversity of intrinsic light curves.
The normalised fluctuation curves, $\Sigma$ as function of trial time delay ($\dtc$) are shown in the left panel of Figure~\ref{fig:nonoise_set2_sig_comb} for all the lensed system for $\muc=0.3$. The right panel shows the same $\Sigma(\dtc)$ curves for the 10 unlensed systems (again with $\muc=0.3$).  

\textbf{Lensed set:}
For each of the lensed systems we notice that: (i) $\Sigma(\dtc)$ is approximately symmetric around $\dtc=0$, (ii) a global minimum is found at $\dtc=0$, (iii) a pair of secondary minima is found at $\dtc \approx \pm \dtt$. For $9$ out of $10$ lensed systems, both the minima at $\dtc \approx \pm \dtt$ have $\Sigma<-2.0$. The only exception is system 4 for which the minima near $\dtc=\pm \dtt$ are somewhat shallower $\Sigma \sim -1.5$. System 4 has by far the small magnification ratio, $\mut=0.14$ -- the other 9 $\mut$'s in the training set are distributed between $0.263$ (system 8) and $0.92$ (system 5). Thus, for system 4, the contribution from the second/fainter image is only $12\%$ of the joint light curve making it very difficult to identify the system as lensed. 

\textbf{Unlensed set:}
As expected, for the unlensed systems we find global minima at $\dtc=0$. No significant additional minima are found (i.e. deeper than $\Sigma=-2$). Therefore, we correctly identify all 10 systems as unlensed.

\textbf{Selection criteria:}
In view of the results for the training lensed and unlensed set of systems, we lay down the following {\em conservative} selection criteria to detect the lensed systems. 

\begin{enumerate}[label=\textbf{C.\arabic*},ref=C.\arabic*]
\item Apart from the global minimum, the fluctuation estimator $\Sigma$ should exhibit a pair of secondary minima at similar absolute values of the time delay with negative and positive signs, i.e. at $\dtc \approx \pm \dte$. \label{cri:c1}

\item  Both these minima should have depth $\Sigma(-\dte)<-2.0$. No other minima (except that at $\dtc=0$ of course) should have $\Sigma \leq -2$. \label{cri:c2}
\end{enumerate}

If both the above criteria are met, we identify the system as lensed with the estimated time delay $\dte$; otherwise it is identified as an unlensed system.

Following the above {\it conservative} selection criteria we identify $9$ out of $10$ lensed systems and none of the $10$ unlensed systems as lensed. Thus we have $100\%$ precision\footnote{Precision is defined as the fraction of true positive cases among all the positive outcomes (includes the false positive outcomes). Recall is the fraction of positive cases in a sample recovered correctly.} and a recall of $90\%$ for the training set considering high quality data with negligible noise. 

The estimated time delays for the $9$ lensed systems are presented in Table~\ref{tab:knownSet2_lensed} and compared with the corresponding true time delays. The time delay estimations are accurate and consistent with the respective truths within the sampling resolution of the trial time delay ($0.1$ days). We empirically take the time delay sampling resolution as the uncertainty in the estimate in this case.

We note that the uncertainty estimate is substantially more complex for data with realistic noise, as we describe in the next section.
A proper error estimation for this type of statistics requires analysing a large number of simulations which is beyond the scope of the present paper and will be carried out in detail in the future.

\begin{table}
\centering
 \begin{tabular}{||c|c|c||} 
 \hline
 System No. & True time delay & Estimated time delay  \\[-0.25ex]
  &  $\dtt$ in days &  $\dte$ in days
 \\[1.0ex]
 \hline\hline
 $1 $ & $ 55.37 $  & $ 55.4 \pm 0.1$    \\ 
 \hline
 $2 $ & $ 39.1 $  & $  39.0 \pm 0.1$    \\
 \hline
 $ 3$ & $ 46.7 $  & $ 46.7 \pm 0.1$    \\
 \hline
  $4 $ & $ 97.19 $  & $ -- $    \\
 \hline
  $ 5$ & $ 117.7 $  & $ 117.7 \pm 0.1$    \\
 \hline
  $6 $ & $ 24.14 $  & $ 24.1 \pm 0.1$    \\
 \hline
  $7 $ & $ 15.9 $  & $ 15.9 \pm 0.1$    \\
 \hline
  $8 $ & $ 13.27 $  & $ 13.2 \pm 0.1$    \\
 \hline
  $9 $ & $ 5.13 $  & $ 5.0 \pm 0.1$    \\
 \hline
  $ 10$ & $ 7.42 $  & $ 7.5 \pm 0.1$    \\
 \hline
 \hline
\end{tabular}
\caption{ Training set with negligible noise.  We successfully identify $9$ out of 10 systems as lensed, and we estimate the time delay with high accuracy (within the adopted uncertainty corresponding to the sampling of the time delay trial). For system 4 - the one with the lowest magnification ratio - we found minima in the fluctuation parameter ($\Sigma$) at the true time delay, but they were not deep enough to pass our stringent selection criteria.}
\label{tab:knownSet2_lensed}
\end{table}

\subsection{Testing the algorithm on a blind set}
\begin{figure*}
\centering
\subfigure[Lensed systems identified]{
\includegraphics[width=0.485\linewidth]{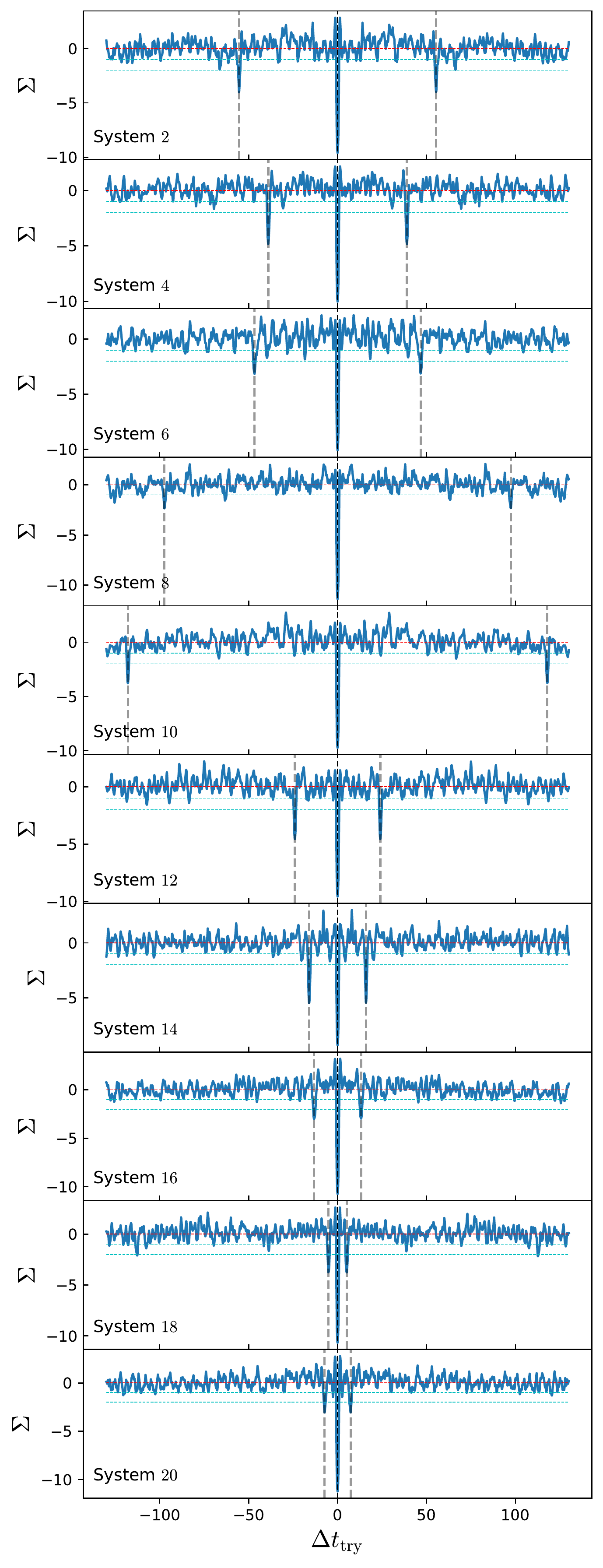}}
\subfigure[Unlensed systems identified]{
\includegraphics[width=0.485\linewidth]{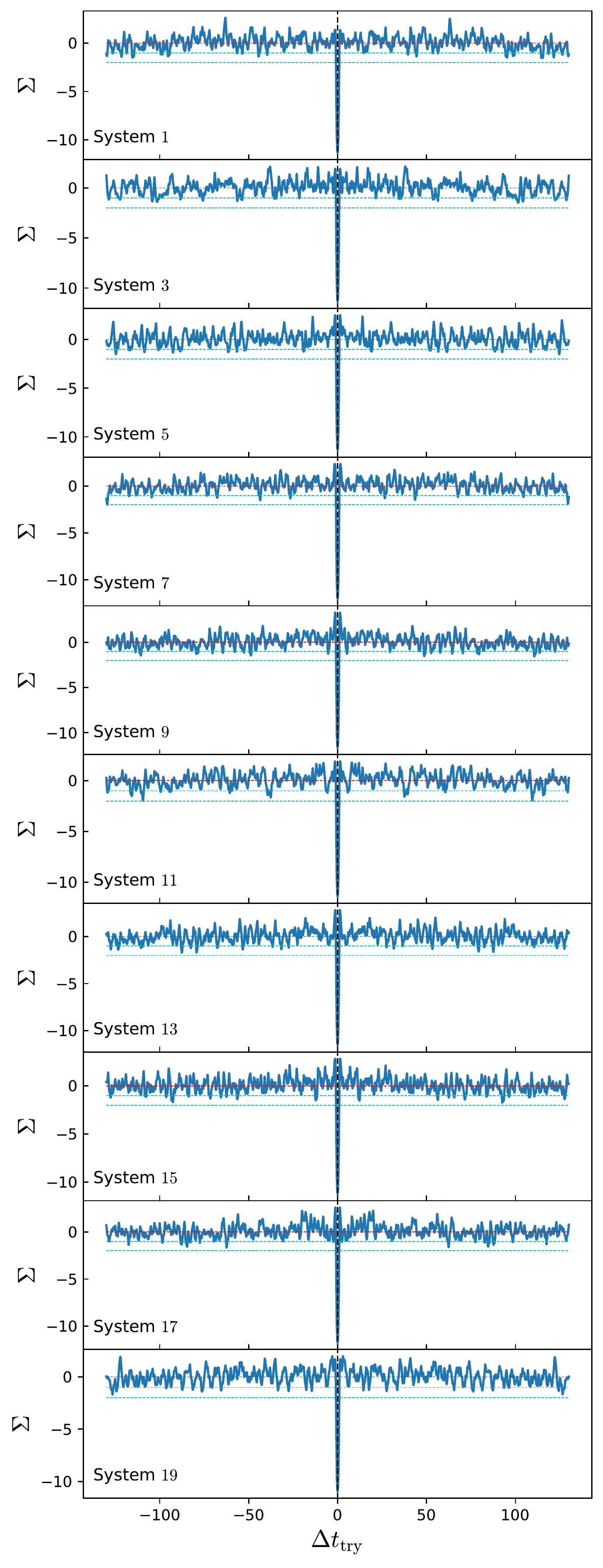}}
\caption{Fluctuation statistics for the  blind set (negligible noise): lensed (left panel), unlensed (right panel). We identify $10$ systems as lensed by detecting a pair of prominent secondary minima (more than $2\sigma$ deep) near $\dtc=\pm \dte$ (ignoring the global minima at $\dtc=0$). Note that the trial magnification ratio is kept fixed at $\muc=0.3$.}
\label{fig:nonoise_blind_set4_sig_comb}
\end{figure*}

In order to test the method and prevent experimenter bias, we carried out the following blind test. One coauthor simulated a set of 20 lensed and unlensed systems without revealing anything but the joint light curves to the rest of the team. The true time delays (of the lensed systems only) are disclosed only after the results have been frozen.

By applying the conservative selection criteria introduced above we find $10$ lensed systems and $10$ unlensed systems. The $\Sigma(\dtc)$ curves for the systems we identify as lensed are shown in the left panel of Figure~\ref{fig:nonoise_blind_set4_sig_comb}. The right panel shows the same for the systems we identify as unlensed. The estimated time delays of the $10$ lensed systems are given in Table~\ref{tab:blindSet4_nonoise}. Building on our findings for the training set we report the sampling resolution ($0.1$ days) as the uncertainty in the time delay estimation for the blind case too.   

After unblinding, we discover that there are 10 lensed and 10 unlensed systems in the blind set. We identify all of the lensed cases correctly and accurately estimate their time delays. 
Therefore, combining the training and blind sets, we have correctly classified $39$ out of $40$ systems in total ($19$ out of $20$ lensed cases and all the $20$ the unlesed cases), corresponding to $100\%$ precision and  $95\%$ recall.
This exercise establishes that in perfect conditions (when observational noise is negligible and cadence is much higher than the time delay)  one can identify the lensed cases with extreme precision and measure the time delays very accurately. 

\begin{table}
\centering
 \begin{tabular}{||c|c|c||} 
 \hline
 System No. & True time delay & Estimated time delay  \\[-0.25ex]
  &  $\dtt$ in days &  $\dte$ in days
 \\[1.0ex]
 \hline\hline
  $ 1$ & Unlensed  & $ -- $  \\ 
 \hline
  $ 2$ & $ 55.4 \pm 0.1$  & $ 55.37 $  \\ 
 \hline
  $ 3$ & Unlensed  & $ -- $  \\ 
 \hline
  $ 4$ & $ 39.0 \pm 0.1$  & $ 39.1  $  \\ 
 \hline
  $ 5$ & Unlensed  & $ -- $  \\ 
 \hline
  $ 6$ & $ 46.7 \pm 0.1$  & $ 46.7 $  \\ 
 \hline
  $ 7$ & Unlensed   & $ -- $ \\ 
 \hline
  $ 8$ & $ 97.4 \pm 0.1$  & $ 97.19 $  \\ 
 \hline
  $9$ & Unlensed   & $ -- $ \\ 
 \hline
  $ 10$ & $ 117.9 \pm 0.1$ & $ 117.7 $   \\ 
 \hline
  $ 11$ & Unlensed  & $ -- $  \\ 
 \hline
  $12 $ & $ 24.0 \pm 0.1$   & $ 24.14 $ \\ 
 \hline
  $ 13$ & Unlensed  & $ -- $  \\ 
 \hline
  $14 $ & $ 16.0 \pm 0.1$ & $ 15.9 $   \\ 
 \hline
  $15 $ & Unlensed  & $ -- $ \\ 
 \hline
  $ 16$ & $ 13.2 \pm 0.1$  & $ 13.27 $  \\ 
 \hline
  $ 17$ & Unlensed  & $ -- $  \\ 
 \hline
  $18 $ & $ 5.1  \pm 0.1$  & $ 5.13 $  \\ 
 \hline
   $ 19$ & Unlensed  & $ -- $ \\ 
 \hline
  $20 $ & $ 7.3 \pm 0.1$  & $ 7.42 $  \\ 
 \hline
 \hline
 \hline
\end{tabular}
\caption{Blind set (negligible noise). For each system we report the identification and for the identified lenses the estimated time delay. After unblinding we find that all systems are correctly identified and the time delays estimated within the uncertainty.}
\label{tab:blindSet4_nonoise}
\end{table}

 \section{Dealing with noisy data}
 \label{sec:td_withnoise}
 
 In this section we introduce realistic noise in the joint light curve data, $F_{\rm obs}(t)$. In the  Time Delay Challenge 1 (TDC1) simulations, in order to add photometric noise expected for LSST, an rms photometric uncertainty was drawn first from a
Gaussian of mean $0.053$ and width $0.016$ nanomaggie,
and then a noise value was drawn from a
Gaussian of width equal to the above rms.
In order to validate our method on data quality of smaller aperture telescopes,
such as the ongoing ZTF survey,  
we keep the noise level in our simulations three times larger than that of LSST. In other words, we draw rms photometric uncertainty from a normal distribution with mean $0.159$ and width of $0.048$ nanomaggie\footnote{Most of the combined images that we simulate in this work have brightness ranging from $8.62$ to $22.45$ nanomaggies ($20.16-19.12$ magnitudes).}. 

We notice that applying the procedure described in the previous section directly to the noisy data leads to significant spurious features and mis-identification. However, we find that the performance of the algorithm is dramatically improved by smoothing the observed light curve prior to the application of our method.
 
In practice, we follow the iterative smoothing algorithm \citep{Shafieloo:2005nd, Shafieloo:2007cs, Shafieloo:2009hi, Aghamousa:2014uya}, summarized in Appendix~\ref{app:smoothing}. The choice of smoothing scale is important. Instead of choosing just a single smoothing scale, we choose three smoothing scales ($\delta=3.0,~4.0,~5.0$ days\footnote{The light curves are observed in an interval of one day. The typical features in the intrinsic light curves are of the scale $\mathcal{O}(10)$ days. So choosing $\delta=3.0,~4.0,~5.0$ does not destroy the intrinsic features (in the light curves) that the method uses as signal in our analysis. }) and combine the fluctuation curves corresponding to each $\delta$ while keeping $\nit=10$ (the results are relatively insensitive to the choice of $\nit$)\footnote{In this way we are adding signals coming from multiple smoothing scales.}. 

After smoothing, we identify the lensed systems and then estimate the corresponding time delays by tracking the minima of the final normalised fluctuation curve $\Sigma(\dtc)$. 
 
 \begin{figure}
\centering
\includegraphics[width=0.485\textwidth]{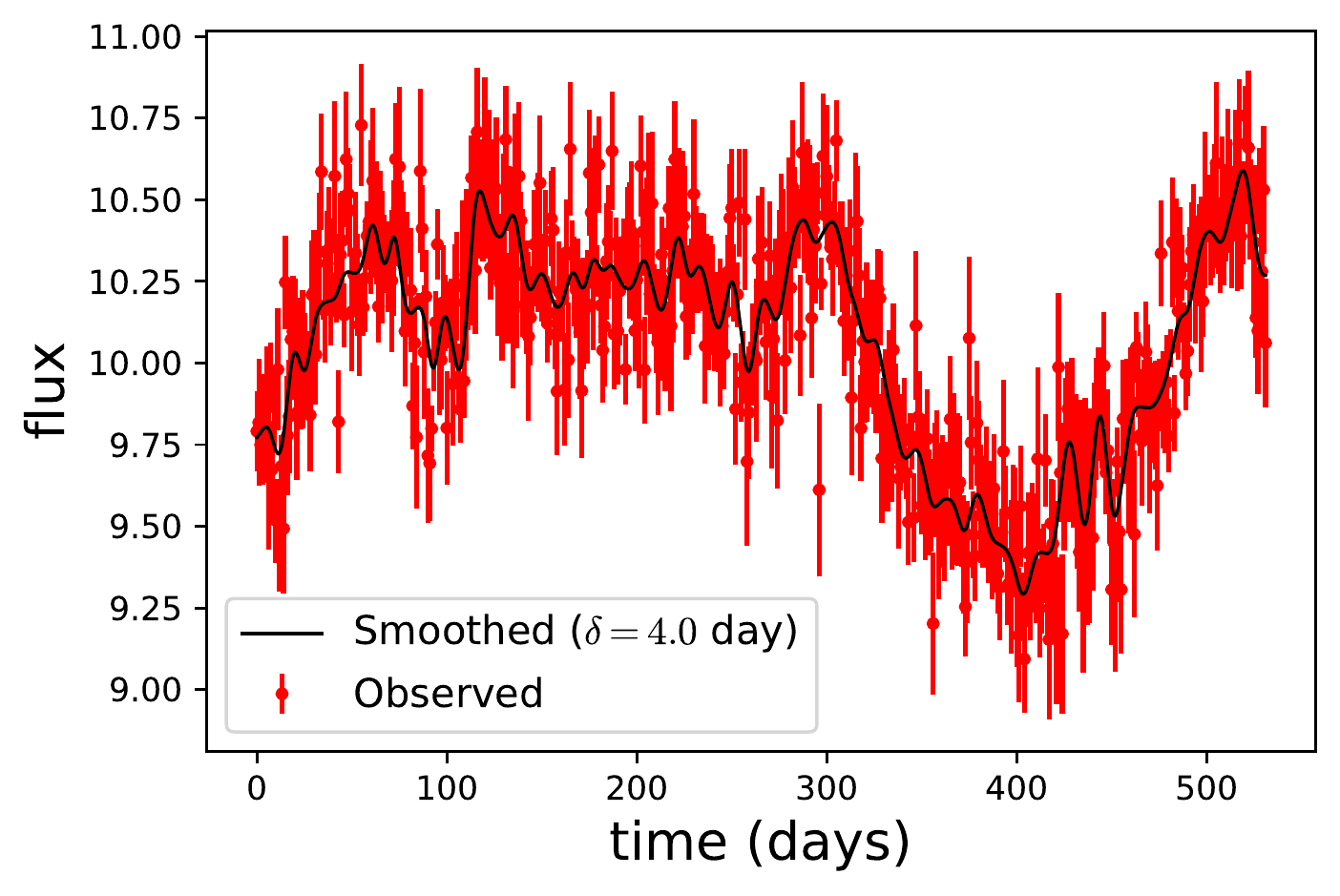}
\caption{Joint light curve (flux in arbitrary unit) in the presence of ZTF-like noise for the example system 7. 
The solid black curve represents the smoothed flux corresponding to the smoothing scale $\delta=4.0$ days, following the iterative smoothing algorithm described in Appendix~\ref{app:smoothing}.}
\label{fig:withnoise_sys7_smooth}
\end{figure}
  
 \subsection{Data with ZTF-like noise: training set}
 \begin{figure}
\centering
\includegraphics[width=\linewidth]{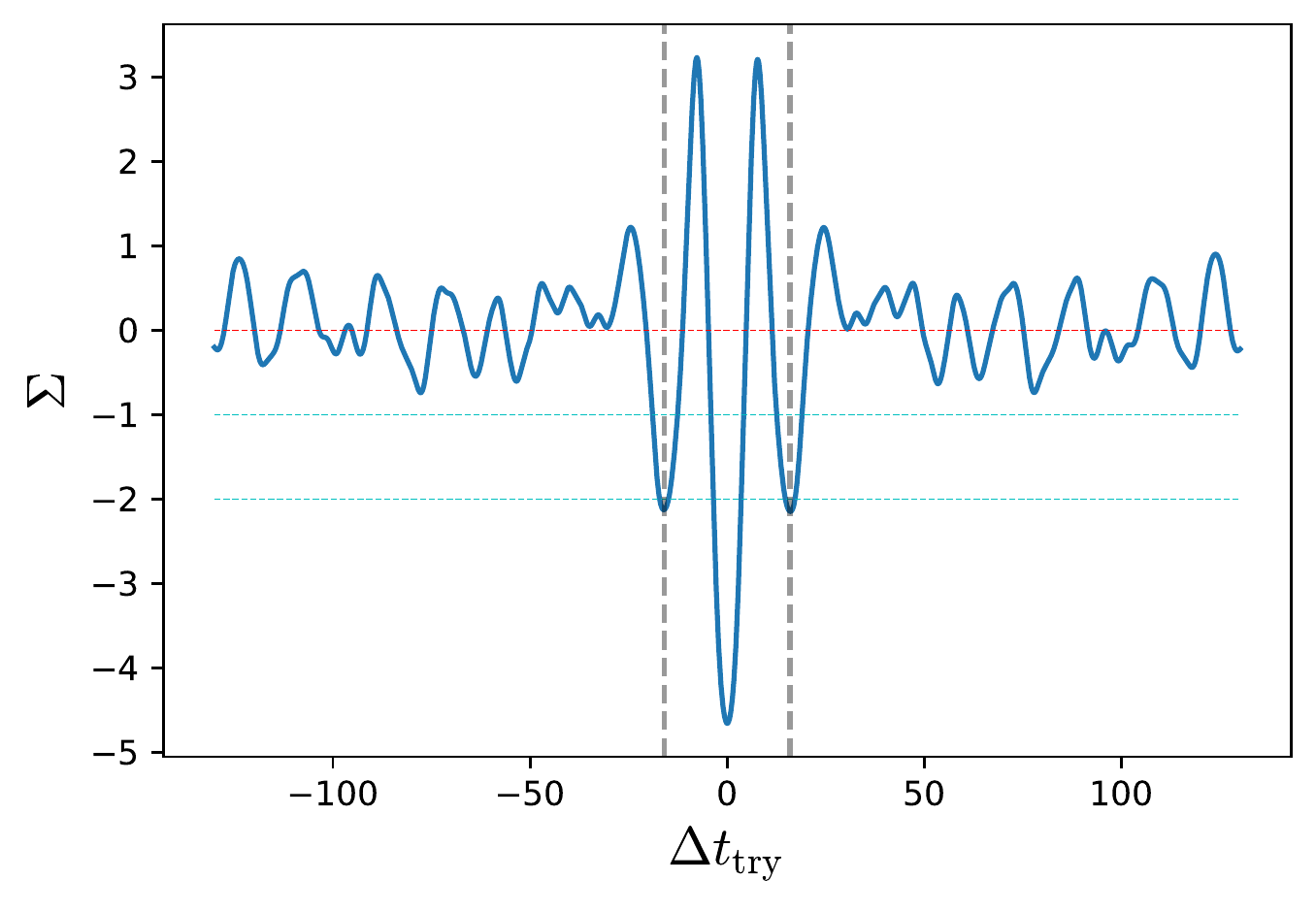}
\caption{Fluctuation estimator $\Sigma$ as a function of $\dtc$ for the example system 7. 
We clearly find a pair of prominent secondary minima at $\dtc \approx \pm \dtt=15.9$ days, shown by the dashed vertical lines. We fix $\muc=0.3$ as in previous examples.}
\label{fig:withnoise_sys7_sig}
\end{figure}

 \begin{figure*}
\centering
\subfigure[Lensed systems]{
\includegraphics[width=0.485\textwidth]{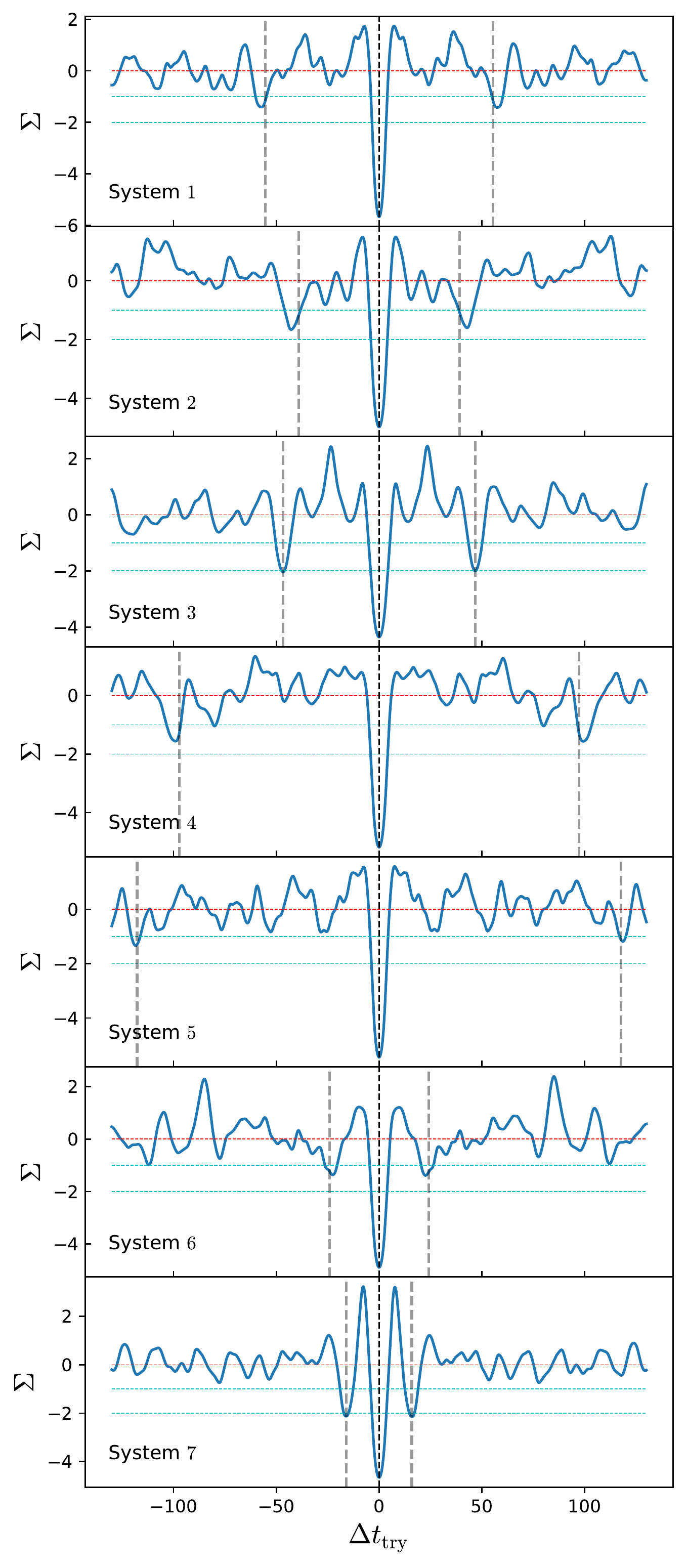}\label{fig:wna}}
\subfigure[Unlensed systems]{
\includegraphics[width=0.485\textwidth]{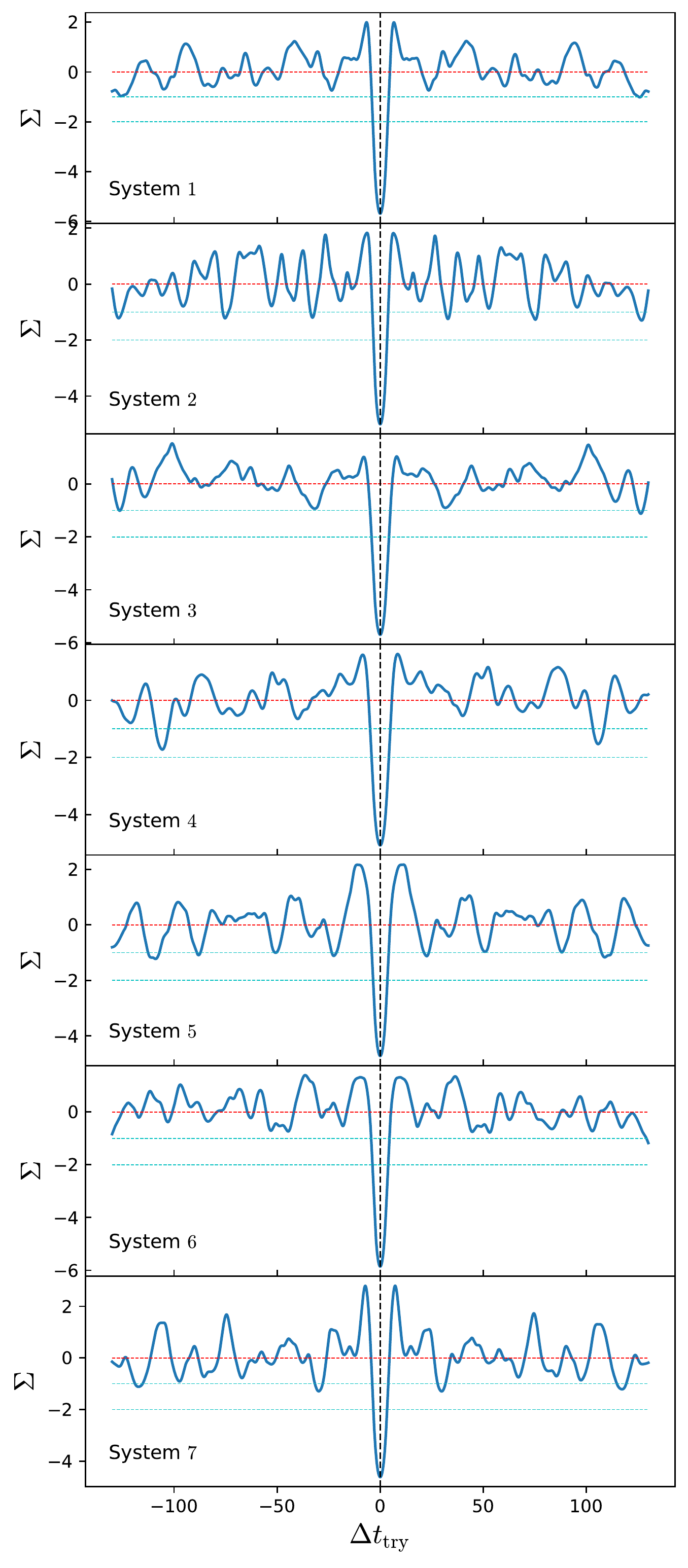}}
\caption{Fluctuation results for some of the systems from the training set considering ZTF-like noise in the data: lensed (left panel), unlensed (right panel). Apart from the global minima at $\dtc=0$, we find a pair of prominent minima (more than $2\sigma$ deep) at $\dtc \approx \pm \dtt$ for two lensed systems (system 3 and 7). However, with the relaxed criteria we identify four other systems (1, 2, 5, 6) as highly probably lensed and one system (4) as probably lensed. Furthermore, using the conservative criteria \ref{cri:c2} all unlensed systems are correctly identified as unlensed, while the relaxed criteria gives one false probable case (system 4). The trial magnification ratio is kept fixed at $\muc=0.3$.}
\label{fig:withnoise_set2_sig_comb}
\end{figure*}
 
We consider the same training lensed and unlensed sets (each has 10 systems) discussed in the noiseless case, adding ZTF-like noise. Let us take the example of system 7 from the lensed set. The smoothed light curve corresponding to the smoothing scales $\delta=4.0$ days is shown in Figure~\ref{fig:withnoise_sys7_smooth} as a black curve. Smoothed light curves corresponding to $\delta=3.0,~5.0$ days are quite similar to that for $\delta=4.0$ days, while smoothed light curve corresponding to lower $\delta$ exhibits more fluctuations.
 Using the smoothed light curves we measure the fluctuation in the respective reconstructions ($\frI$) separately using Eq.~\eqref{eq:fluction_raw}. Then, we calculate the fluctuation estimator $\Sigma(\dtc)$, from the sum of the $\epsilon(\dtc)$ curves using Eq.~\eqref{eq:fluc_sig}. Hence, we combine the information from the three smoothing scales.
 
The fluctuation estimator $\Sigma$ for system 7 is shown as a function of $\dtc$ with fixed $\muc=0.3$ in Figure~\ref{fig:withnoise_sys7_sig}.
Due to our smoothing procedure, the $\Sigma(\dtc)$ curve has fewer minima than in the negligible noise case (compare the plot for system 7 in the left panel of Figure~\ref{fig:nonoise_set2_sig_comb} with Figure~\ref{fig:withnoise_sys7_sig}). Again, we notice the global minima at $\dtc=0$ which corresponds to unlensed solution and hence can be ignored. In addition, we find a pair of secondary minima at $\dtc=-16.1$ and $16.0$ with depths $\Sigma=-2.13$ and $-2.15$ respectively. Since the minima satisfy the conservative selection criteria \ref{cri:c2}, the system is identified as lensed with estimated time delay $\dte=16.05$ days which matches very well the true time delay, $\dtt=15.9$ days.

We then analyse all the lensed and unlensed light curves from the training set with ZTF-like noise.  We identify 2 out of 10 lensed systems correctly (system 3 and 7) which have secondary minima deeper than $\Sigma=-2.0$, thus satisfying the conservative selection criteria \ref{cri:c2}. Five other systems also have prominent pairs of secondary minima in their $\Sigma(\dtc)$ curves but the minima are not deep enough to satisfy the conservative selection criteria. The left panels of Figure~\ref{fig:withnoise_set2_sig_comb} show  $\Sigma(\dtc)$ for these seven systems exhibiting a prominent pair of secondary minima near the true time delay.

Furthermore, we identify all the unlensed systems correctly using the conservative criteria \ref{cri:c2}. The right panels of Figure~\ref{fig:withnoise_set2_sig_comb} show $\Sigma(\dtc)$ for some unlensed systems. No significant pairs  of  secondary  minima is found.

\subsubsection{Relaxed selection criteria}
\label{sec:relaxed_criteria}
From the left panel of Figure~\ref{fig:withnoise_set2_sig_comb} it is evident that for five lensed systems (1, 2, 4, 5, 6) the secondary minima near the true time delay are much deeper than the other minima (false minima) but not deeper than the conservative selection threshold $\Sigma =-2.0$.
Also, from the training set analyses we observe that for unlensed cases there are many false shallow minima with comparable depths in the final $\Sigma (\dtc)$ curves. Based on these findings we formulate a set of relaxed selection criteria aimed at increasing the recall, possibly in return for a decrease in precision. These criteria are illustrative. In practice, the optimal strategy will depend on the data quality and whether recall or precision is the main goal of a search. We leave the tailoring of the criteria to specific needs for future work. Our illustrative relaxed criteria are:

\begin{enumerate}
\item We identify the pair of secondary minima in $\Sigma(\dtc)$ curve if the deepest minimum (ignoring the global minimum at $\dtc=0$) in the positive and negative domains of $\dtc$ occur at similar absolute values, i.e. at $\dtc \approx \pm \dte$. If such a pair cannot be found, we identify the system as {\em confirmed unlensed}.  

\item If both the secondary minima are deeper than $2\sigma$, i.e. both have $\Sigma<-2.0$, we detect the system as {\em confirmed lensed}. This is the conservative selection criterion introduced previously.

\item One the other hand, if either of the secondary minima is shallower than $1\sigma$, i.e. either has $\Sigma > -1.0$, we identify the system as {\em confirmed unlensed}.

\item If both the minima are deeper than $1\sigma$, we measure the difference in depths between the secondary minimum and the next deepest minimum (`third minimum') in both positive and negative $\dtc$ domains separately. If the secondary minima are at least $50\%$ deeper than the third minimum in their respective domains, we consider the system as a probable lensed case. However, depending on the depth of the third (false) minima we classify the following cases:

\begin{itemize}
    \item If there is no minimum other than the pair of secondary minima deeper than $\Sigma=-1.0$, we identify the system as {\em highly probable lensed case}.
    
    \item If the third minimum on either of positive or negative $\dtc$ domains is also deeper than $\Sigma=-1.0$, we identify the system as {\em probable lensed case}.
\end{itemize}

\item If either of the secondary minima is not sufficiently deeper than the third minima, we identify the system as {\em probable unlensed case}.
\end{enumerate}

Using these relaxed criteria we identify the two systems -- 3 and 7 -- from the lensed training set as confirmed lensed cases. Four other systems (1, 2, 5, 6) are identified as `highly probable lensed' whereas system 4 is identified as probably lensed. The remaining three systems are identified as unlensed. Thus we identify all the seven systems from the left panel of Figure~\ref{fig:withnoise_set2_sig_comb} as lensed with varying degrees of certainty (confirmed, highly probable and probable cases).
The estimated time delays for these systems are given in Table~\ref{tab:withnoise_knownSet2_lensed}. We can see that the estimated time delays are accurate even in the presence of significant uncertainty in the data. Note that the relaxed criteria incorrectly identify only one system among the ten unlensed systems in the training set. Thus, these relaxed criteria produce $87.5\%$ precision with a recall of $70\%$ when we consider ZTF-like noise in the data for the training sets.

\subsubsection{Error estimation}

The uncertainty in our time delay estimations depends on several factors, such as the intrinsic time variability of the light curves, the noise in data, cadence, number of observation epochs  etc. 
Thus, a proper statistical determination of the uncertainty requires a large number of simulations spanning a range of conditions. Clearly, that is beyond the scope of the present paper, aimed at introducing the method as a proof of concept. However, we found that the estimated time delays are consistent with the truth within $5\%$ for all the cases we study in this article considering reasonable amount of noise in the data (apart from these simulations, that also include the simulations from TDC1, one observed system from the COSMOGRAIL database as described below).  Therefore, heuristically $5\%$ seems to be a reasonable initial assessment of the uncertainty of the estimated time delay for the time being. In the future, when applying the method to search for new lenses and measure their time delay, we plan to do a proper error analysis based on large number of simulations, mimicking the actual observing conditions.

\subsubsection{False negatives}

\begin{figure}
\centering
\includegraphics[width=0.485\textwidth]{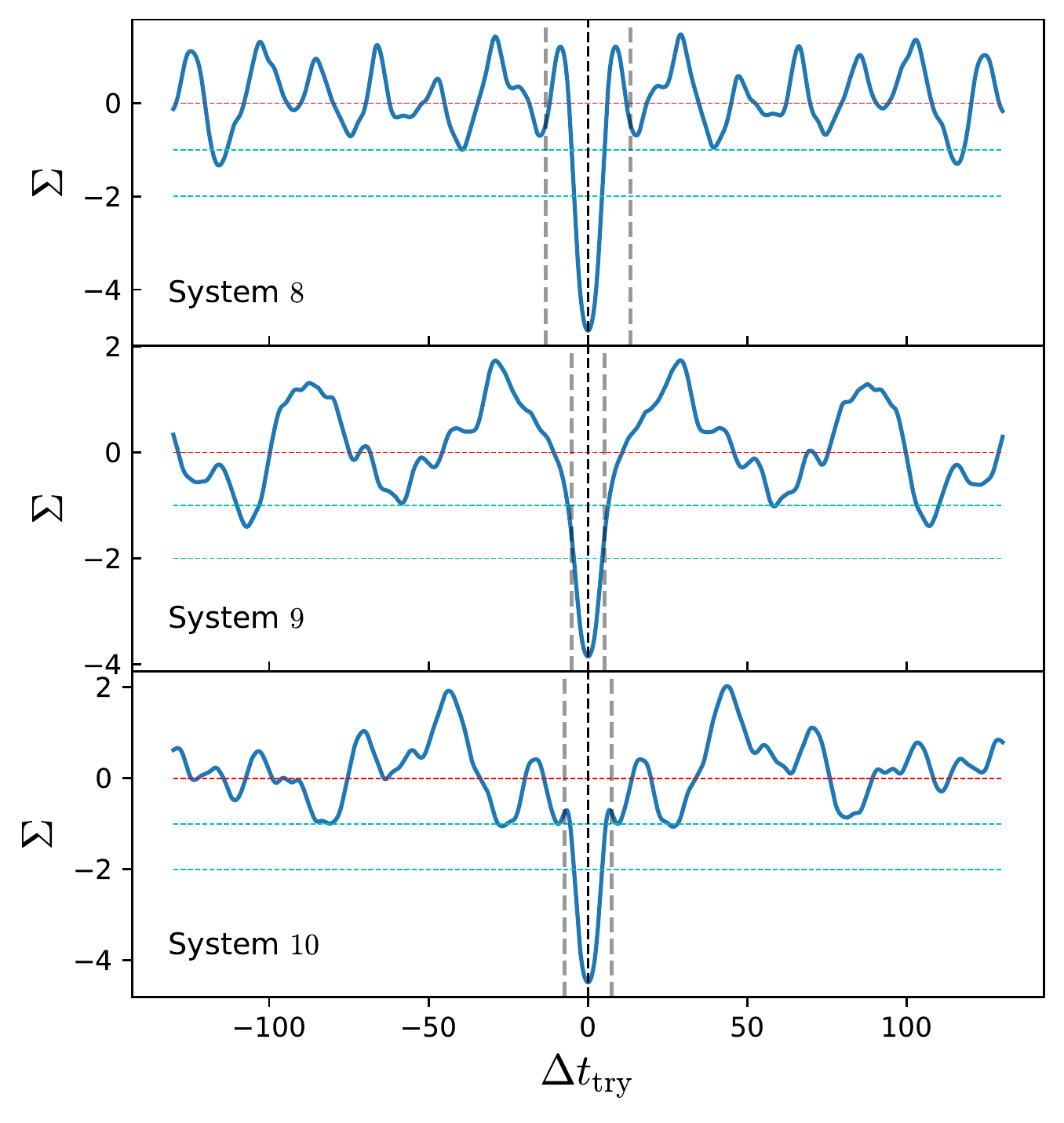}
\caption{Fluctuation statistics of the false positives from the training set. 
Note that for system 8 and 10 we still have a pair of secondary minima at $\dtc \approx \pm \dtt$ (marked by the dashed vertical lines). However, the minima are not deep enough for identification of the systems as lensed. Note that all cases have small $\dtt$ compared to the width of the central minimum.}
\label{fig:withnoise_kset_others}
\end{figure}

Next we look deeper into the three lensed systems in the training set (system 8, 9 and 10) that we missed, i.e. the false negative cases. 
Figure~\ref{fig:withnoise_kset_others} shows the $\Sigma (\dtc)$ curves for these systems.
Two systems show a pair of secondary minima at $\dtc \approx \pm \dtt$ (system 8  and 10). However, these pair of minima are either not dominant, or not deep enough. For system 9, the actual time delays are blurred within the central minimum. These three systems (system 8, 9 and 10) have relatively smaller time delays, $\dtt=13.27, 5.13$ and $7.42$ days respectively, and are likely lost due to the effect of smoothing.

\begin{figure*}
\centering
\includegraphics[width=\textwidth]{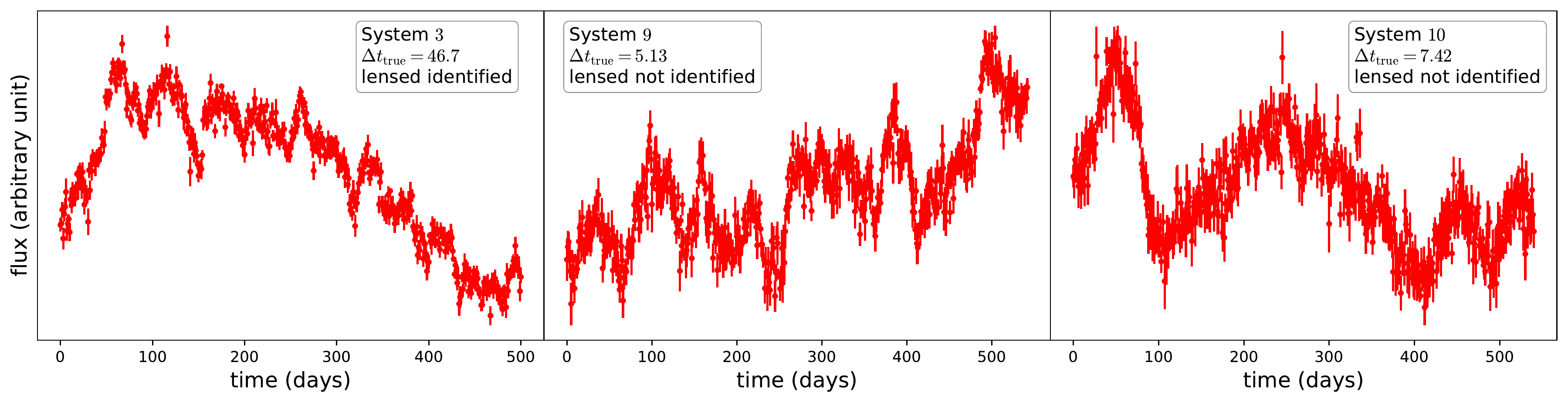}
\caption{Comparison of the light curves of 3 lensed systems from the training set. The system in the left panel is identified as lensed, while the ones in the middle and right panels are false negatives, likely due to a combination of noise and short intrinsic time delay.
}
\label{fig:withnoise_Set2_fluxes}
\end{figure*}

To gain further insight, in Figure~\ref{fig:withnoise_Set2_fluxes} we compare the observed light curves of three lensed systems. The system (3) on the left is identified as lensed, but the middle and right ones are false negatives. Comparing the light curves,  the false negative cases are likely due to a combination of (i) noise suppressing the intrinsic fluctuation in the light curves; (ii) short time delays that are confused with the global minimum at $\dtc=0$ due to the noisy data and smoothing process.

\begin{table}
\centering
 \begin{tabular}{||c|c|c||} 
 \hline
 System No. & True time delay & Estimated time delay  \\[-0.25ex]
  &  $\dtt$ in days &  $\dte$ in days
 \\[1.0ex]
 \hline\hline
 $ 3$ & $ 46.7 $  & $ 46.70 \pm 2.33 $    \\
 \hline
  $7 $ & $ 15.9 $  & $ 16.05 \pm 0.80 $    \\
 \hline
 
 $ 1$ (*) & $55.37 $  & $ 57.35 \pm 2.87 $    \\
 \hline
 
 $2$ (*) & $ 39.1 $  & $  42.80 \pm 2.14$    \\
 \hline
 
 $ 5$ (*) & $117.7 $  & $118.45 \pm 5.92  $    \\
 \hline

  $6 $ (*) & $ 24.14 $  & $ 22.60 \pm 1.13$    \\
 \hline
 
  $4 $ (**) & $97.19 $  & $ 99.15 \pm 4.96 $    \\
 \hline
 \hline
\end{tabular}
\caption{ Training set with  ZTF-like noise.
The estimated time delays are compared with the truths for the seven systems correctly identified as lensed. We identify the top two systems as confirmed lensed. The next four systems, marked with a star, are identified as highly probable lenses. System 4 is identified as probably lensed, i.e. with lesser certainty (marked with double stars). For all the seven systems, the estimated time delays are accurate within a few percent.}
\label{tab:withnoise_knownSet2_lensed}
\end{table}

\subsection{Data with ZTF-like noise: blind set}
Next we analyse a blind set consisting of $20$ light curves with ZTF-like noise level as in the training set. 
The conservative selection criteria identify only one confirmed lensed case with a pair of secondary minima deeper than $\Sigma=-2.0$. However, the relaxed criteria detect two additional highly probable lensed systems and two additional probable lensed cases. 
The estimated time delays for these five systems are presented in~Table \ref{tab:blindSet4_withnoise}. Furthermore, we identify all the 10 unlensed systems correctly, so we do not have any false positive. In conclusion, combining the training and blind sets, the method results in a precision of $92.3\%$ with $60\%$ recall.

\begin{table}
\centering
 \begin{tabular}{||c|c|c||} 
 \hline
  System No. & True time delay & Estimated time delay  \\[-0.25ex]
  &  $\dtt$ in days &  $\dte$ in days
 \\[1.0ex]
 \hline\hline
  $ 11$ & $ 24.45 \pm 1.22$  & $24.14 $ \\ 
 \hline
   $ 7$ (*) & $ 96.55 \pm 4.83 $  & $97.19 $  \\ 
 \hline
  $ 9$ (*) & $ 122.10 \pm 6.11$  & $117.7 $ \\ 
 \hline
   $ 3$ (**) & $ 40.10 \pm 2.01$  & $39.1 $  \\ 
 \hline
  $ 13$ (**) & $ 15.80 \pm 0.79 $  & $ 15.9$  \\ 
 \hline
 \hline
\end{tabular}
\caption{ Blind set (considering ZTF-like noise in the light curve data):
The estimated time delays for the five systems those we identify as lensed in the blind set. 
We identify the only system at the top as confirmed lensed. Two systems (7,9), marked with a star, are identified as highly probable lensed cases. Another two systems (3 and 13) are identified as probably lensed, i.e. with lesser certainty (marked with double stars).}
\label{tab:blindSet4_withnoise}
\end{table}

\section{Applying the method to Time Delay Challenge 1 (TDC1) simulations}
\label{sec:TDC1}
The simulated data used so far to establish the method are very well sampled with 1 day cadence. In Appendix~\ref{app:cadence} we test the method on simulated data with cadence of 3 days. We found that even with the poorer cadence one can still identify the lensed systems, however the signal in the $\Sigma(\dtc)$ curves is reduced, as expected. Nevertheless, in realistic scenarios, quasar light curves are observed seasonally in multiple years offering multiple patches in the data. These patches can be used separately in order to boost the signal in our analysis. 

To test the performance on realistic multi-year data, we analyse a number of light curves for double systems taken from the Time Delay Challenge 1 (TDC1)  simulations \citep{tdc1,tdc2}. In this section, we show results for a subset of systems selected to have relatively low noise levels compared to their variability from Rung 0 and Rung 1 of TDC1 as examples. Both rungs have good cadence (3 days on the average with a dispersion of 1 day) and around $400$ observation epochs. The systems in Rung 0 are observed for 5 years, while those in Rung 1 are observed for 10 years.

We stress that this paper is meant only to introduce and illustrate the technique. Therefore, for computational reasons we restrict ourselves to a subset of light curves taken from Rung 0 and 1 of TDC1.
We leave for future work a systematic exploration of all the systems in TDC1.

\subsection{TDC1, Rung 0}

\begin{figure*}[hbt]
\centering
\includegraphics[width=0.9\textwidth]{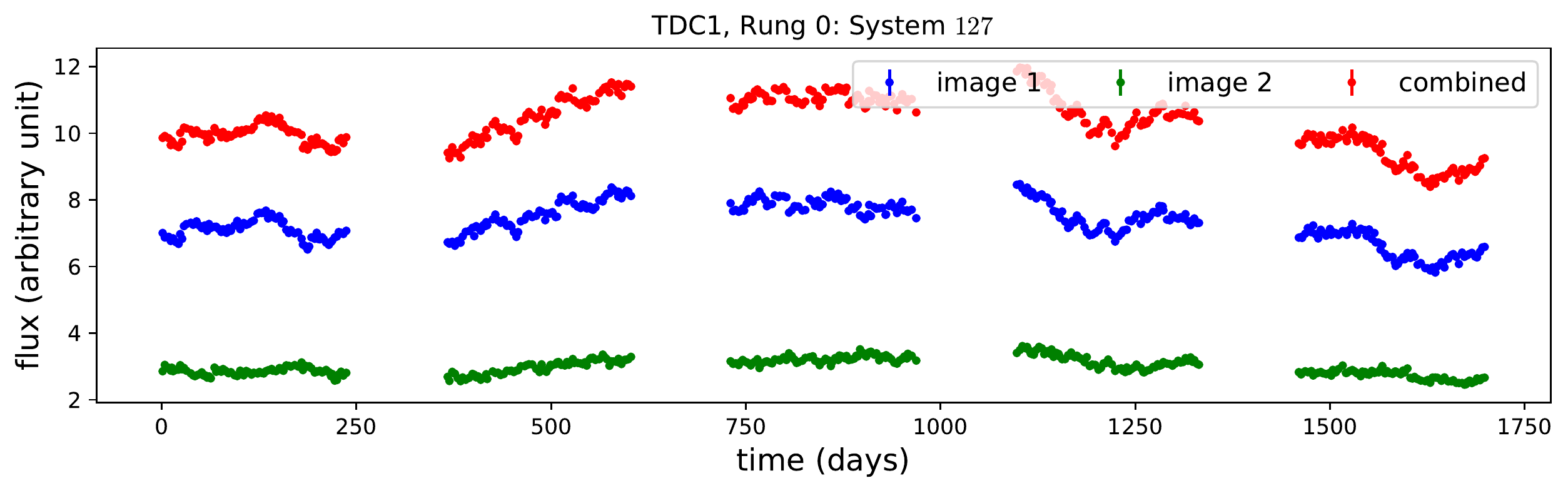}
\caption{Simulated light curve of a doubly imaged lensed quasar (system 127) taken from TDC1, Rung 0. The light curves of the first and second images are shown by the blue and green lines respectively. The system has a time delay of $38.33$ days. We add these two fluxes to construct the equivalent unresolved light curve (red line), and use it for our analysis.}
\label{fig:TDC1_r0_sys127_flux}
\end{figure*}

In Rung 0 the light curves are sampled in roughly $400$ observation epochs over a period of 5 campaign years.
Let us first consider the example of system $127$. The simulated light curves of the two images are shown by blue and green curves in Figure~\ref{fig:TDC1_r0_sys127_flux}. We add the two fluxes to construct the equivalent unresolved light curve, shown in red in the figure, and use it to test our method.

 \begin{figure*}
\centering
\includegraphics[width=0.8\textwidth]{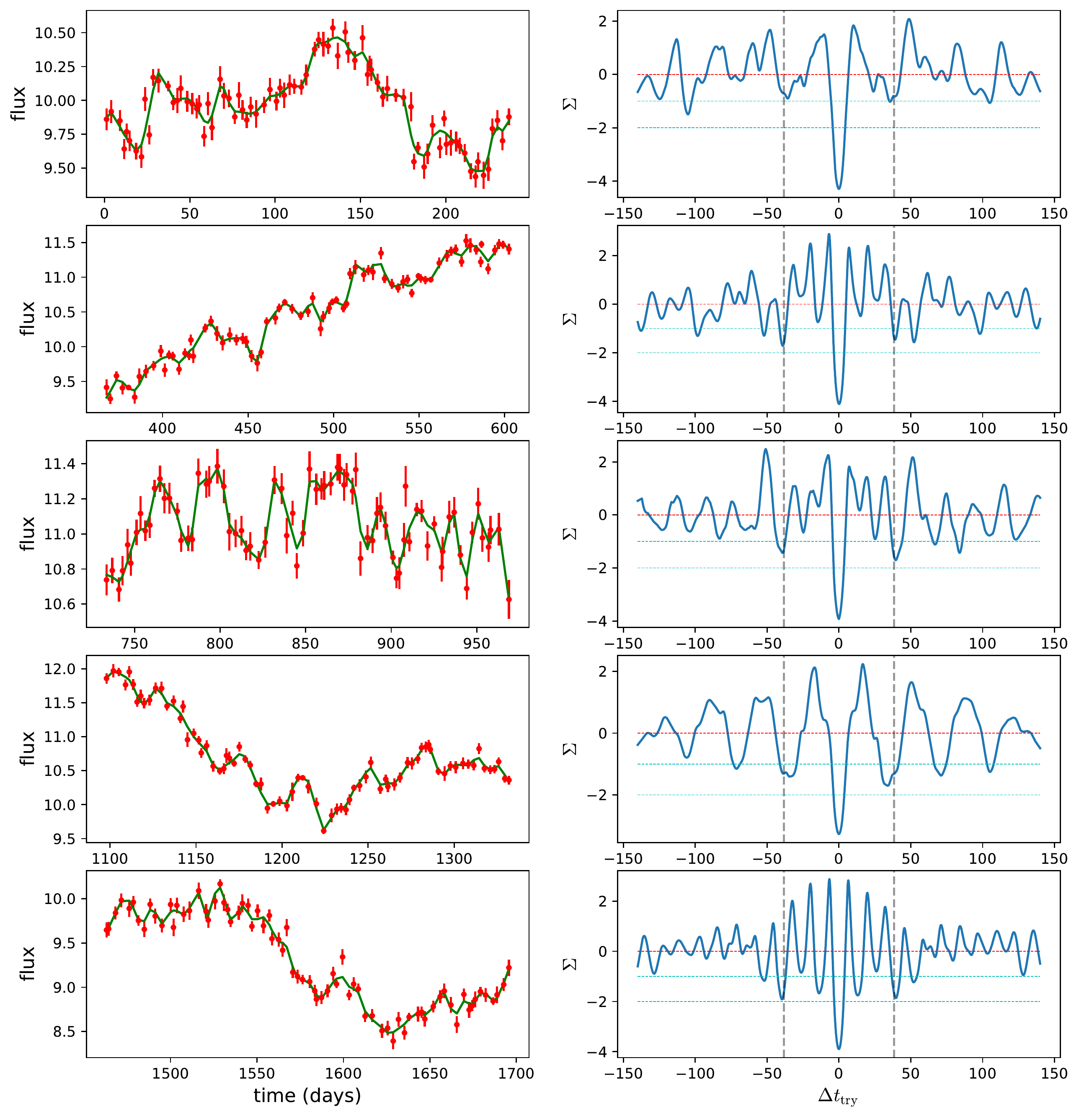}
\caption{The left panels show the light curve of system 127 from TDC1 (Rung 0) in the five observing seasons separately. The green curves represent the smoothed fluxes with a smoothing scale of $\delta=4$ days and $\nit=10$. The right panels show the fluctuation estimator $\Sigma$ as a function of $\dtc$ for each of the seasons using the corresponding smoothed light curve. Note that all the five patches yield a strong pair of minima in the fluctuations near the correct time delay, $\dtc = \pm 38.33$ days, shown by the dashed vertical lines. They also show some false minima at incorrect time delays.}
\label{fig:TDC1r0_sys127_patches}
\end{figure*}

The joint light curve has five patches (corresponding to the five observing seasons) which are shown separately in the left panels of Figure~\ref{fig:TDC1r0_sys127_patches}. Each patch has observation time range of roughly $240$ days with approximately $80$ observation epochs.  The green curves in the left panels represent the smoothed light curves for each patch with smoothing scale $\delta=4.0$ and $\nit=10$ (the average cadence is 3 days approximately). The right panels show the fluctuation estimator $\Sigma$ as a function of $\dtc$ for each of the seasons using the corresponding smoothed light curve. We note that for each season we obtain a pair of secondary minima near the true time delay, i.e. at $\dtc \approx \pm \dtt=38.33$ days, which is shown by the vertical dashed lines. However, for all the seasons we also get some `false minima' at incorrect time delays. 

\begin{figure}
\centering
\includegraphics[width=\linewidth]{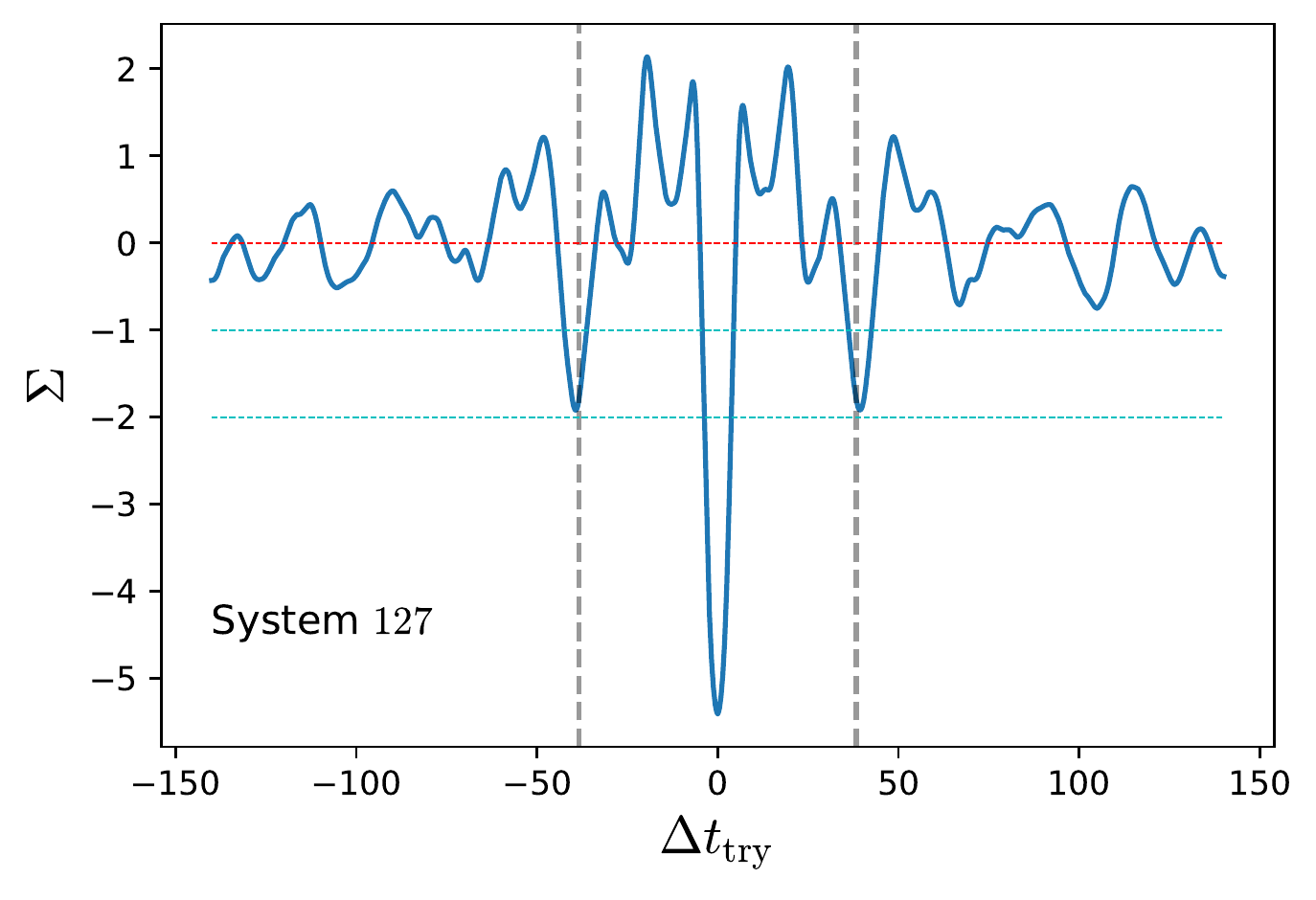}
\caption{Fluctuation statistics for system 127 from TDC1, Rung 0, based on all the 5 seasons of data and multiple smoothing scales ($\delta=3.0,~4.0,~5.0$). Two prominent minima are evident at $\dtc=-39.3,~39.4$ days with depth $\Sigma \approx -1.9$. This pair of prominent secondary minima not only identifies the system as lensed but also estimates the true time delay (vertical dashed lines) within 3\%.}
\label{fig:TDC1r0_sys127_comb_lensed}
\end{figure}

For some observing seasons, e.g. the 3rd and the 5th, the pair of secondary minima near the true time delay is more prominent (with respect to the false minima) as compared to that for other patches. To take advantage of all the available information, we sum the fluctuation measurements, $\epsilon(\dtc)$ calculated using Eq.~\eqref{eq:fluction_raw}, coming from all the seasons. Furthermore, we use multiple smoothing scales, $\delta=3.0,4.0,5.0$ days, instead of using a single smoothing scale and combine the $\epsilon(\dtc)$ curves for each smoothing scale too. Finally, we calculate the $\Sigma(\dtc)$ curve from this combined $\epsilon(\dtc)$ curve using Eq.~\eqref{eq:fluc_sig}. Figure~\ref{fig:TDC1r0_sys127_comb_lensed} shows the combined fluctuation estimator $\Sigma$ as a function of the trial time delay. The dashed lines again mark the true time delay, $\dtc=\pm \dtt$. We find the pair of secondary minima, occurring at $\dtc=-39.3,39.4$, are now very prominent both having the depth $\Sigma \approx -1.9$. Importantly, the false minima have been suppressed by combining information from different patches and various smoothing scales. Thus, we can identify the object as a highly probable lensed system. The estimated time delay $\dte=39.35$ is within 3\% of the true time delay.

\begin{figure}
\centering
\includegraphics[width=\linewidth]{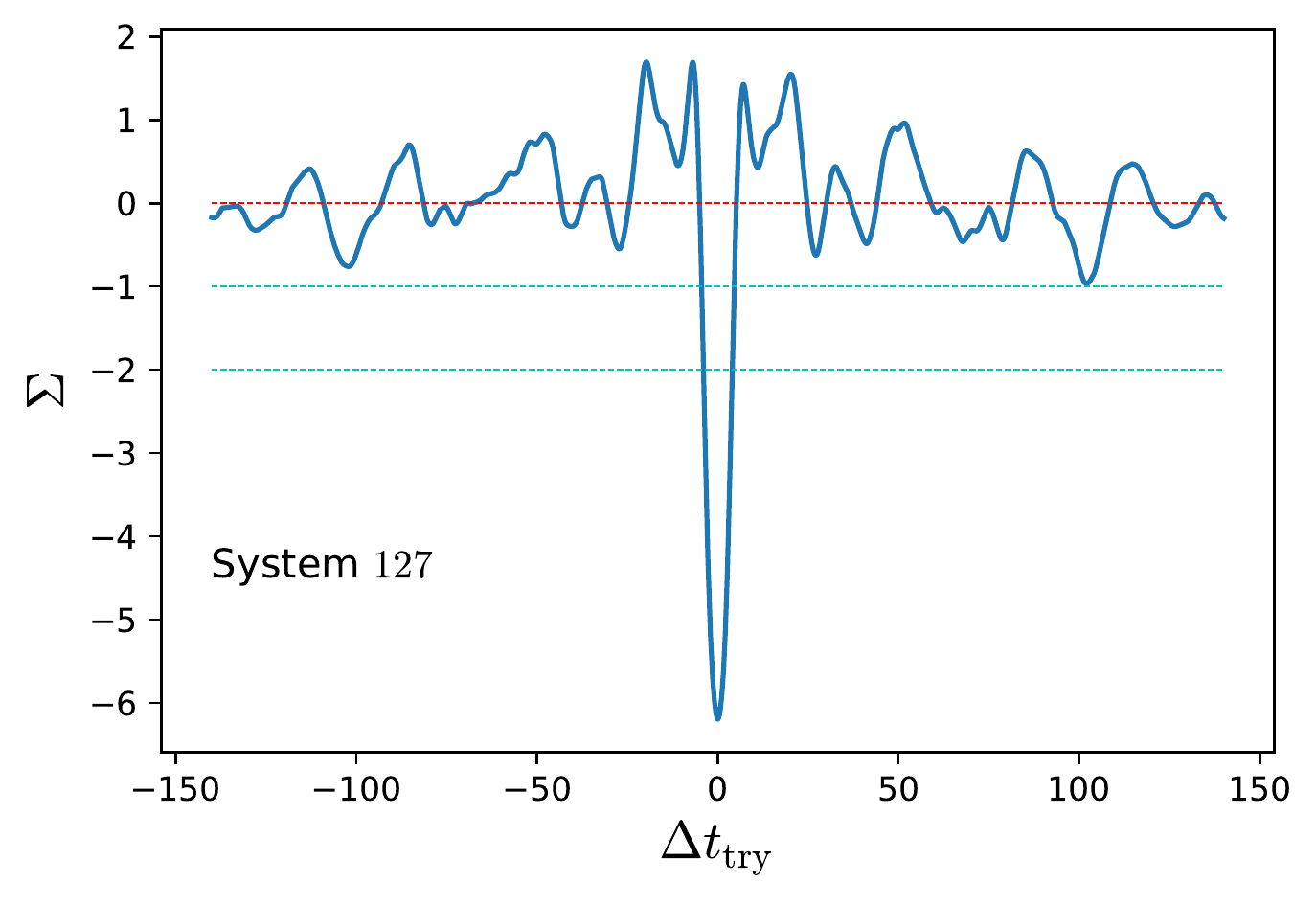}
\caption{Same as~Figure \ref{fig:TDC1r0_sys127_comb_lensed} for the light curve of only the brightest image (blue light curve in Figure~\ref{fig:TDC1_r0_sys127_flux}), mimicking a true negative case. No prominent pair of minima is found, identifying the light curve correctly as unlensed.}
\label{fig:TDC1r0_sys127_comb_unlensed}
\end{figure}

To illustrate the behaviour with unlensed sources, we also analyse the light curve of the first/brightest image by itself (shown in blue in Figure~\ref{fig:TDC1_r0_sys127_flux}). Again, we consider the data patches separately and follow the same procedure as described above for the joint light curve. The combined $\Sigma (\dtc)$ curve for this unlensed case is shown in Figure~\ref{fig:TDC1r0_sys127_comb_unlensed}.
The foremost pair of minima in the $\Sigma (\dtc)$ curve occurs at $\dtc=-102.2,102.1$ and has depths $\Sigma=-0.76,~-0.96$ which are sufficiently shallow so that one can readily identify this as an unlensed case.

\begin{figure*}
\centering
\subfigure[using the joint light curve (lensed)]{\includegraphics[width=0.485\textwidth]{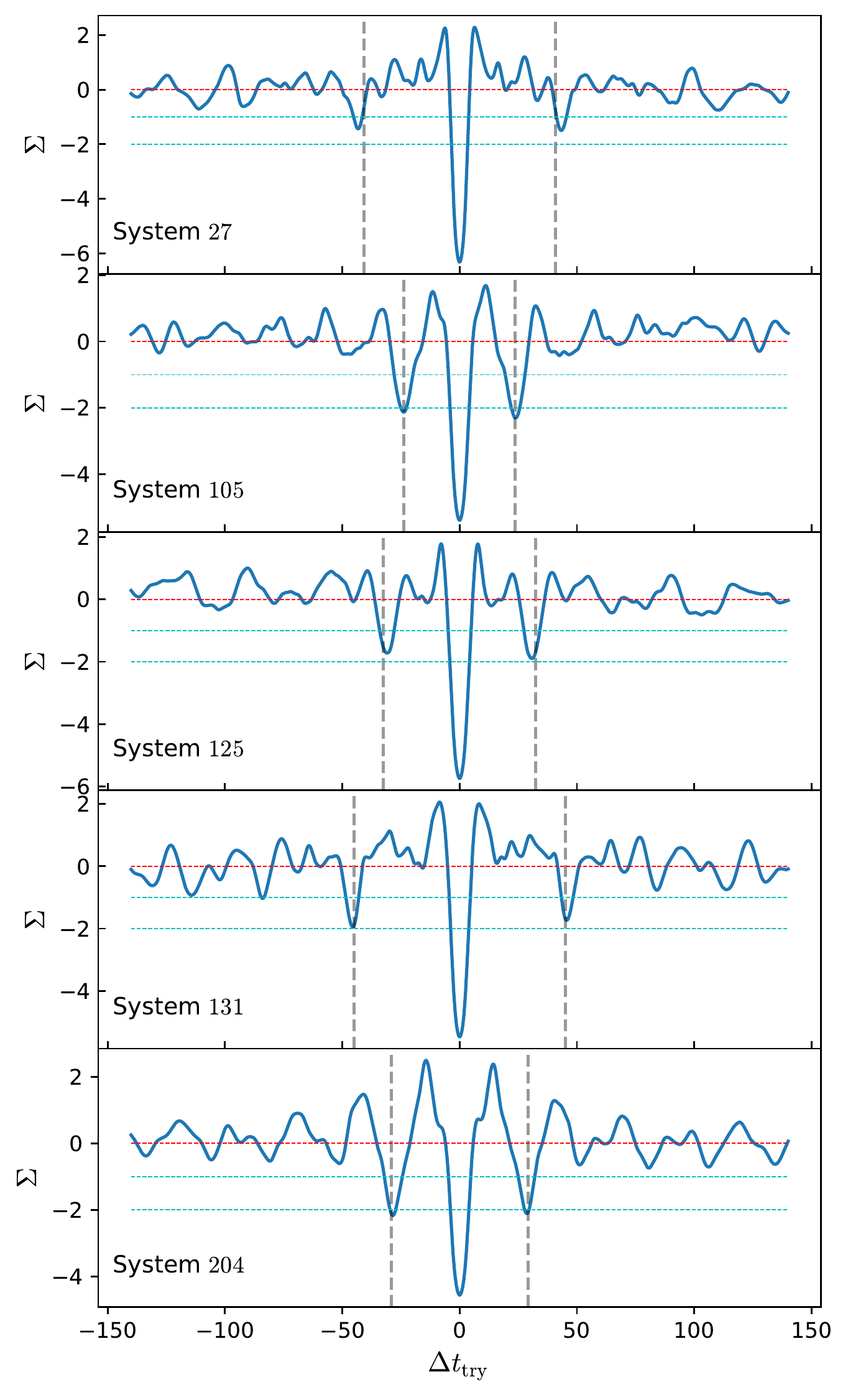}}
\subfigure[using the brightest image light curve only (no lensing)]{\includegraphics[width=0.485\textwidth]{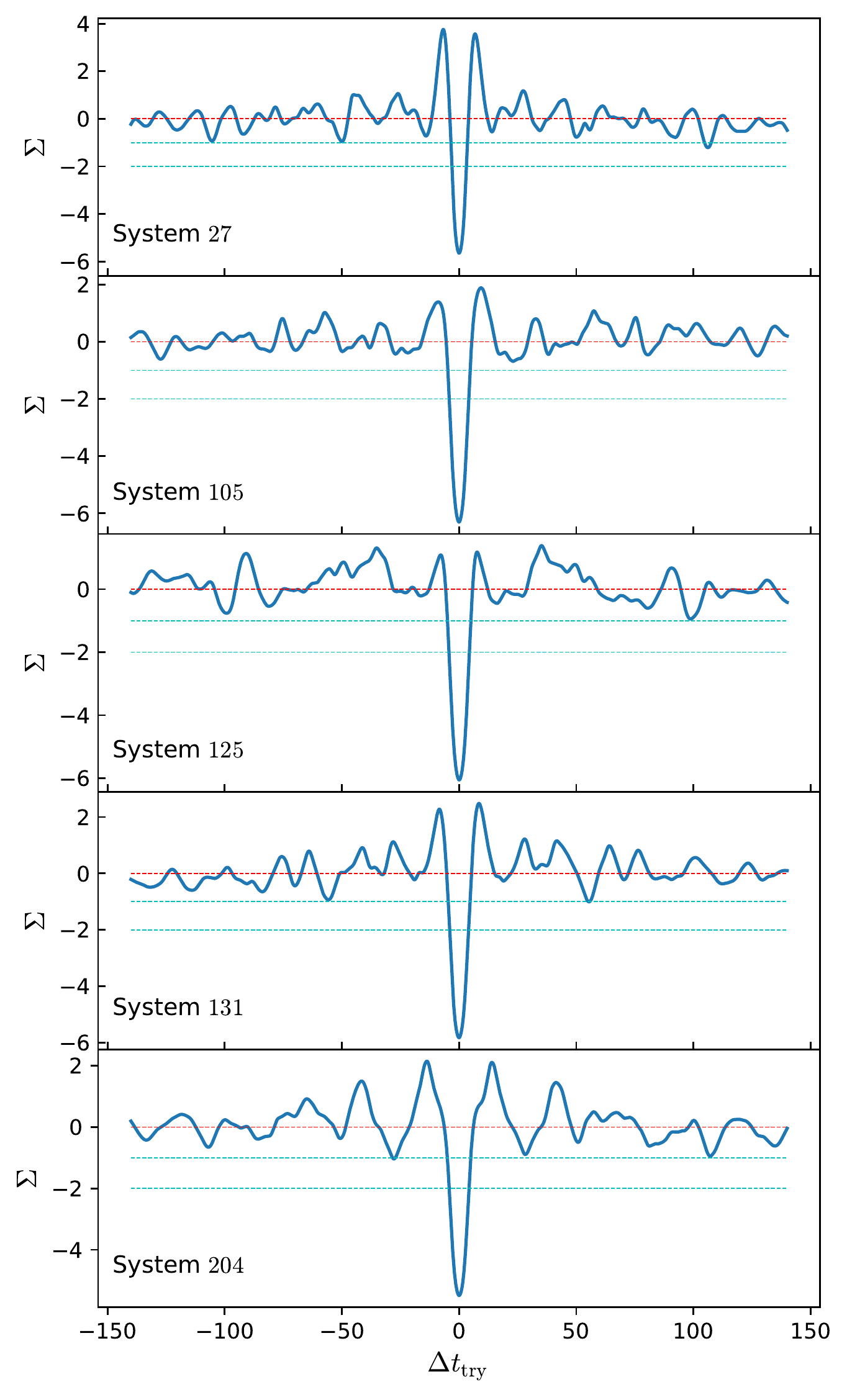}}
\caption{Fluctuation estimator as a function of time delay for some example systems taken from TDC1, Rung 0, based on the combination of the individual fluctuations from the 5 seasonal patches. The panels on the left show the results for the actual lenses, while the panels on the right show the results for true negatives constructed by analysing only the brighter image for each system.}
\label{fig:TDC1r0_extra_systems}
\end{figure*}

\begin{table*}
\renewcommand{\arraystretch}{1.4}
\centering
 \begin{tabular}{ccc} 
 \hline
  System No. & True time delay & Estimated time delay  \\[-0.25ex]
  &  $\dtt$ in days &  $\dte$ in days
 \\[1.0ex]
 \hline
 \multicolumn{3}{c}{TDC1 (Rung 0): observed in $\sim 400$ epochs over a period of $5$ years }\\
 \hline
  $27 $ & $ -40.79 $  & $43.15 \pm 2.16$    \\
 
 $105$  & $ 23.73 $  & $ 23.80 \pm 1.19$    \\

  $125 $  & $ 32.47 $  & $30.85  \pm 1.54$    \\
  $ 127$ & $ 38.33 $  & $  39.35 \pm 1.97$    \\
 
  $ 131$ & $ -45.03 $  & $ 45.55 \pm 2.28$    \\
 
  $ 204$ & $ 29.1 $  & $ 28.75 \pm 1.44$    \\
 \hline

 \multicolumn{3}{c}{TDC1 (Rung 1): observed in $\sim 400$ epochs over a period of $10$ years }\\
 \hline
   $5 $ & $ 32.47 $  & $ 33.00 \pm 1.65$    \\
   $ 102$ & $  -13.04$  & $ 13.50 \pm 0.68$    \\
    $202 $ & $ 50.81 $  & $ 49.35 \pm 2.47$    \\
    $ 208$ & $ 39.93 $  & $ 39.80 \pm 1.99$    \\
    $246 $ & $ 31.38 $  & $ 32.25 \pm 1.61$    \\
    $254 $ & $ -44.97 $  & $44.10  \pm 2.21$    \\
    $ 358$ & $ 47.26 $  & $ 46.10 \pm 2.31$    \\
 \hline
 
\end{tabular}
\caption{ Time delays estimated from the joint unresolved light curves compared with the true time delays for doubly imaged systems taken from the TDC1 simulations. The top and bottom parts of the table show the results from Rung 0 and Rung 1 respectively. All the time delays are estimated within 3\%. Note that this approach yields the absolute value of the time delay, not the sign. The quoted errors are $5\%$ of the estimated time delays.}
\label{tab:tdc1}
\end{table*}

We studied five additional randomly chosen systems (with relatively low noise) from TDC1 (Rung 0) using the same algorithm. The left panels in Figure~\ref{fig:TDC1r0_extra_systems} show the combined $\Sigma(\dtc)$ curves for these five systems. The right panels show the corresponding results using the brightest image only, used as true negatives (unlensed). For each of the lensed cases the $\Sigma(\dtc)$ curve exhibits a pair of prominent minima that stand out from the other minima, near the true time delay, shown by the dashed vertical lines. Hence, we correctly identify these systems as lensed cases. For the systems 105 and 204, the pair of secondary minima have depths more than $\Sigma=-2.0$. None of the unlensed cases shows a prominent pair of secondary minima in the $\Sigma(\dtc)$ curves. Therefore, the method correctly identifies all the unlensed light curves. 

The estimated time delays for all the six lensed systems from the TDC1 (Rung 0) are compared to the corresponding true time delays in the top part of Table~\ref{tab:tdc1}, with excellent agreement. For systems 105 and 204 the error is much smaller than a day since these light curves have features that stand out particularly well against the noise.

\subsection{TDC1, Rung 1}
\label{ref:tdc1_r1}
Next, we analyse a random subset of light curves from Rung 1 which has cadence of 3 days on the average with 1 day dispersion and $400$ observation epochs. The light curves in Rung 1 are sampled over a period of 10 years. Thus each of the 10 patches in the data is roughly $\sim 120$ days long. Thus Rung 1 is not suitable for assessing systems with time delays longer than $\dtt \gtrsim 100$ days. We use the trial time delay $\dtc \in \lbrace -80.0,80.0 \rbrace$ days with a spacing of $0.1$ days for Rung 1.

\begin{figure*}
\centering
\subfigure[using the joint light curve (lensed)]{\includegraphics[width=0.485\textwidth]{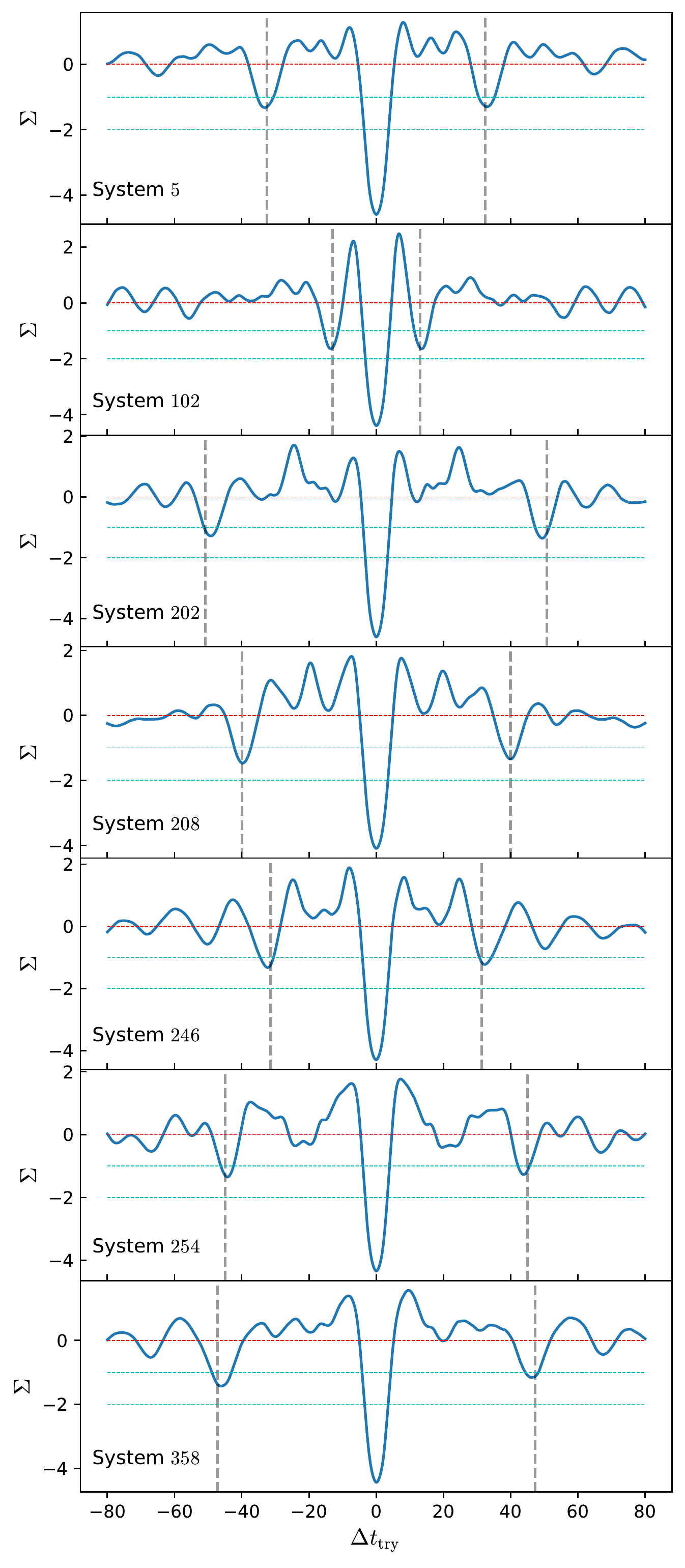}}
\subfigure[using the brightest image light curve only (no lensing)]{\includegraphics[width=0.485\textwidth]{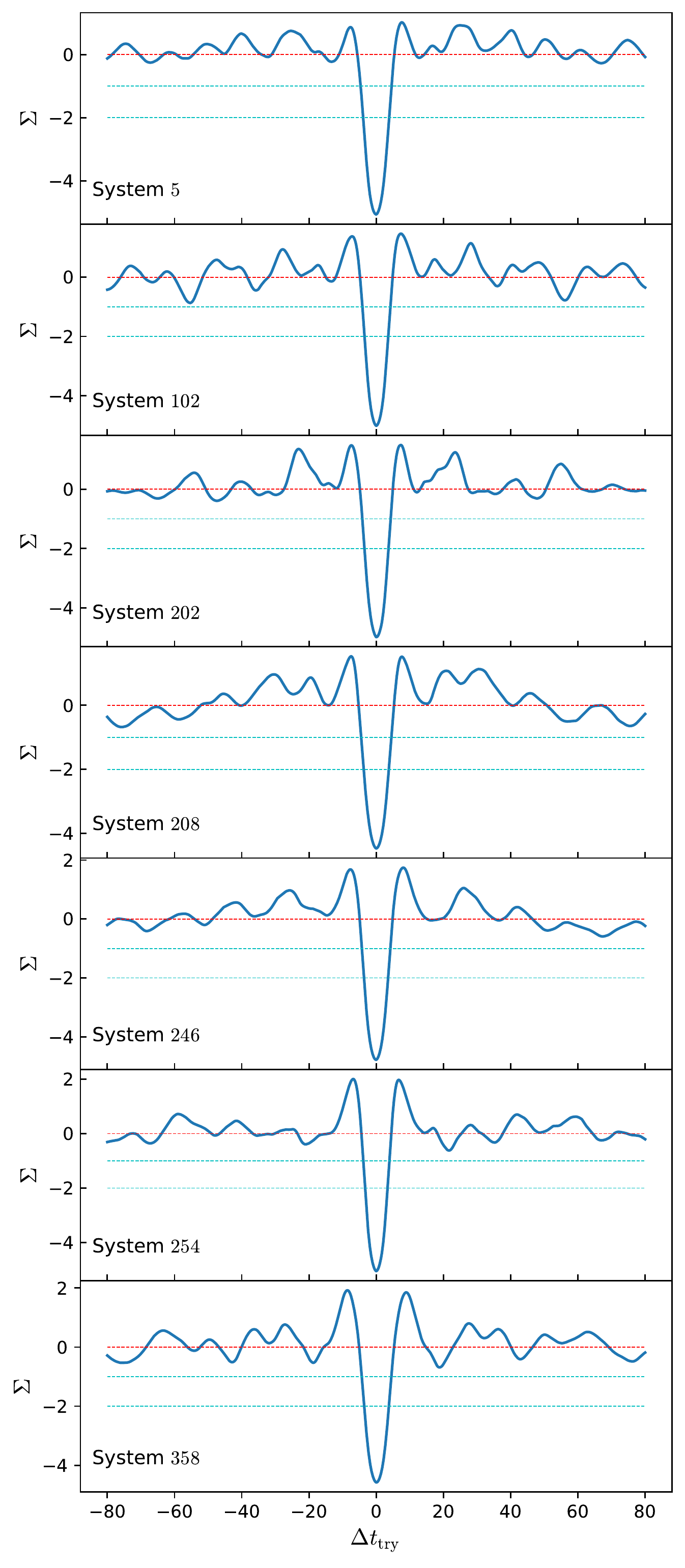}}
\caption{Fluctuation estimator as a function time delay for seven systems from TDC1, Rung 1, based on the combination of the individual fluctuations from the 10 seasonal patches. The panels on the left show the results for the actual lenses, while the panels on the right show the results for true negatives constructed by analysing only the brighter image for each system.
}
\label{fig:TDC1r1_systems}
\end{figure*}

Following the same strategy as for Rung 0, expanded to 10 seasonal patches in each light curve, we compute the fluctuation estimator $\Sigma$ as a function of trial time delay for some systems from Rung 1 as examples. The left panels in Figure~\ref{fig:TDC1r1_systems} show $\Sigma (\dtc)$ considering the joint light curves (lensed cases) for seven systems. For all the systems we find the pair of secondary minima in the $\Sigma (\dtc)$ curves near the true time delay (marked by the dashed vertical lines), i.e. $\dtc \approx \pm \dtt$. Therefore, one can straightforwardly identify the systems as lensed.
However, we see that the secondary minima are somewhat shallower than in Rung 0. This is likely to be due to the shorter overlap between delayed light curves, resulting in weaker signal.

The right panels of Figure~\ref{fig:TDC1r1_systems} show $\Sigma (\dtc)$ for true negatives built from the same systems, using only the light curve of the brighter image. Since we do not find a pair of prominent minima, we correctly identify them as unlensed cases. 

The estimated time delay for the seven lensed systems from TDC1, Rung 1 are compared with the corresponding truths in the bottom part of Table~\ref{tab:tdc1}. We find that our estimates match the truth to within 3\%, better than our fiducial 5\% error.

\section{Application to the COSMOGRAIL light curve of lensed quasar SDSS J1226-0006}
\label{sec:obs_J1226}

\begin{figure*}
\centering
\includegraphics[width=0.9\textwidth]{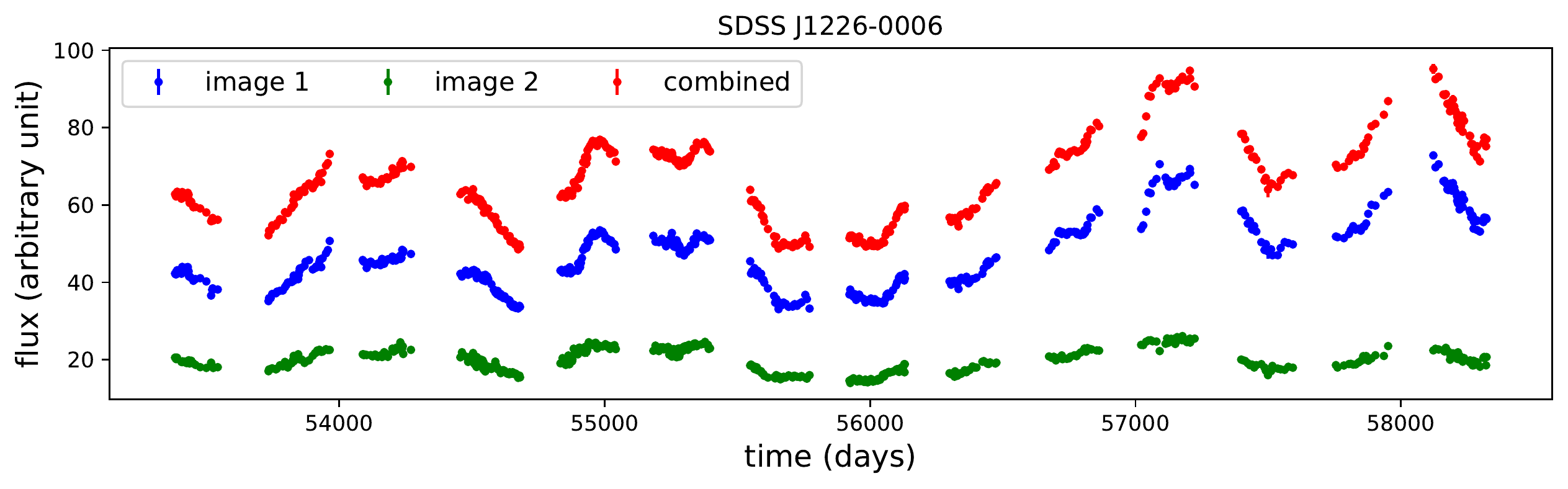}
\caption{Light curves of the lensed quasar SDSS J1226-0006 observed with the Euler telescope by the COSMOGRAIL collaboration. 
The light curves of the first and second images are shown in blue and in green respectively. We add the two to construct an unresolved light curve, shown in red.}
\label{fig:J1226_flux}
\end{figure*}

\begin{figure*}
\centering
\includegraphics[width=0.8\textwidth]{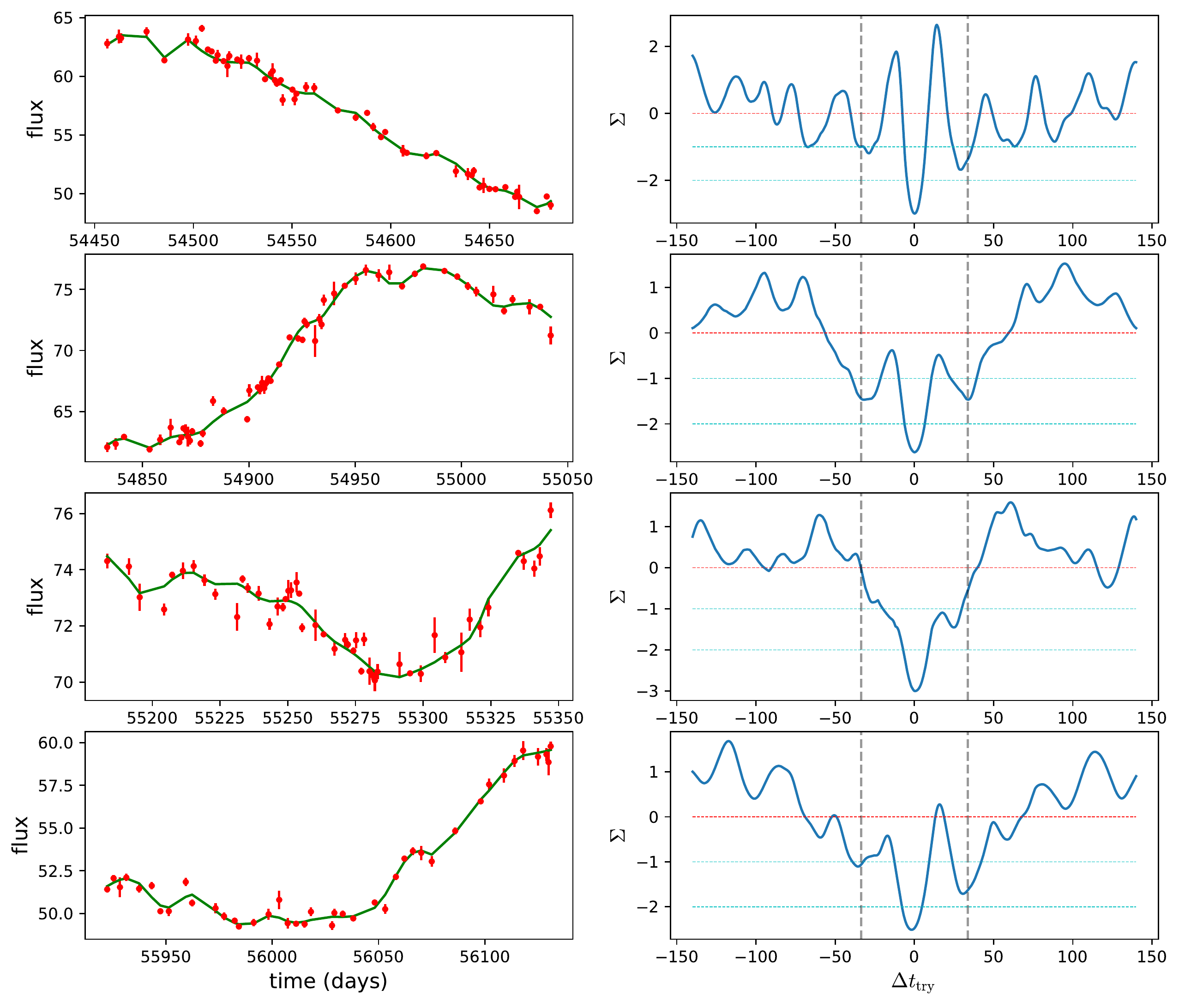}
\caption{The left panels show the light curves of SDSS J1226-0006 for four seasonal patches that meet the quality criteria for our analysis. The green curves represent the smoothed fluxes with a smoothing scale of $\delta=8$ days and $\nit=10$. The right panels show the fluctuation estimator $\Sigma$ as a function of $\dtc$ for each of these four patches. Note that three patches, except the third one from the top, exhibit a strong pair of minima near the correct time delay, $\dtc = \pm 33.7$ days, shown by the dashed vertical lines.}
\label{fig:J1226_patches}
\end{figure*}

In order to test the performance of our method on real data, we apply it to the publicly available light curves of the doubly imaged lensed quasar SDSS J1226-0006, obtained by the COSMOGRAIL collaboration using the 1.2m Euler Telescope \citep{Millon:2020xab}. The time delay estimated by the COSMOGRAIL team is $\dtt=33.7 \pm 2.7$ for this system using the observed image light curves \citep{Millon:2020xab}. We chose SDSS J1226-0006 since the data have low noise level compared to the light curve variability, sufficiently long patches after discarding large gaps, time delay significantly longer than our smoothing scales, and no evidence for strong microlensing. This system is thus a good match to the simulated light curves used in previous sections, providing a good comparison. Analysis of more COSMOGRAIL systems is left for future work.

The observed light curves of the two images are shown in Figure~\ref{fig:J1226_flux} in blue and green. 
The sum of these light curves, representing an unresolved joint light curve, is shown in red. We consider this joint light curve as our data and test if we can identify the system as lensed and if we can estimate the time delay using our method.

This system has been observed for 14 years, resulting in 14 seasonal patches. However, we cannot use all the patches since many of them have large time separations (gaps) between two consecutive observations within the patches.  Hence we select only patches with (i) a maximum gap of $16$ days (separation between any two successive observations), (ii) duration longer than $160$ days. We find four such patches which are shown in the left panels of Figure~\ref{fig:J1226_patches}. The patches have average cadence of $3.94, ~3.79, ~3.34$ and $4.97$ days  and include $57,~55,~49$ and $42$ data points respectively.

Although the average cadence in the patches is 3-4 days, there are occasionally significant gaps ($\sim 10$ days) between consecutive observations. Therefore we need to use smoothing scales that are larger than the average cadence, but not so large as to wipe out all the important features of the light curve. We choose two smoothing scales $\delta=8.0$ and $9.0$ days. The result obtained using a single smoothing scale $\delta=8.0$ days is already quite good, but the use of two smoothing scales boosts the signal. The green curves in the left panels of Figure~\ref{fig:J1226_patches} represent the smoothed fluxes, corresponding to the smoothing scale $\delta=8$ and the number of iteration $\nit=10$, in each data patch. The right panels show the fluctuation estimator $\Sigma$ as a function of $\dtc$ for these smoothed light curves in the four patches. We can see that, except for the third patch from the top, we get a strong pair of minima  near the correct time delay $\dtc \approx \pm 33.7$ day, shown by the dashed vertical lines. However, there are a number of other pairs of `false' minima. We expect the false minima to decrease in depth when we combine the seasons.  

\begin{figure}
\centering
\includegraphics[width=0.485\textwidth]{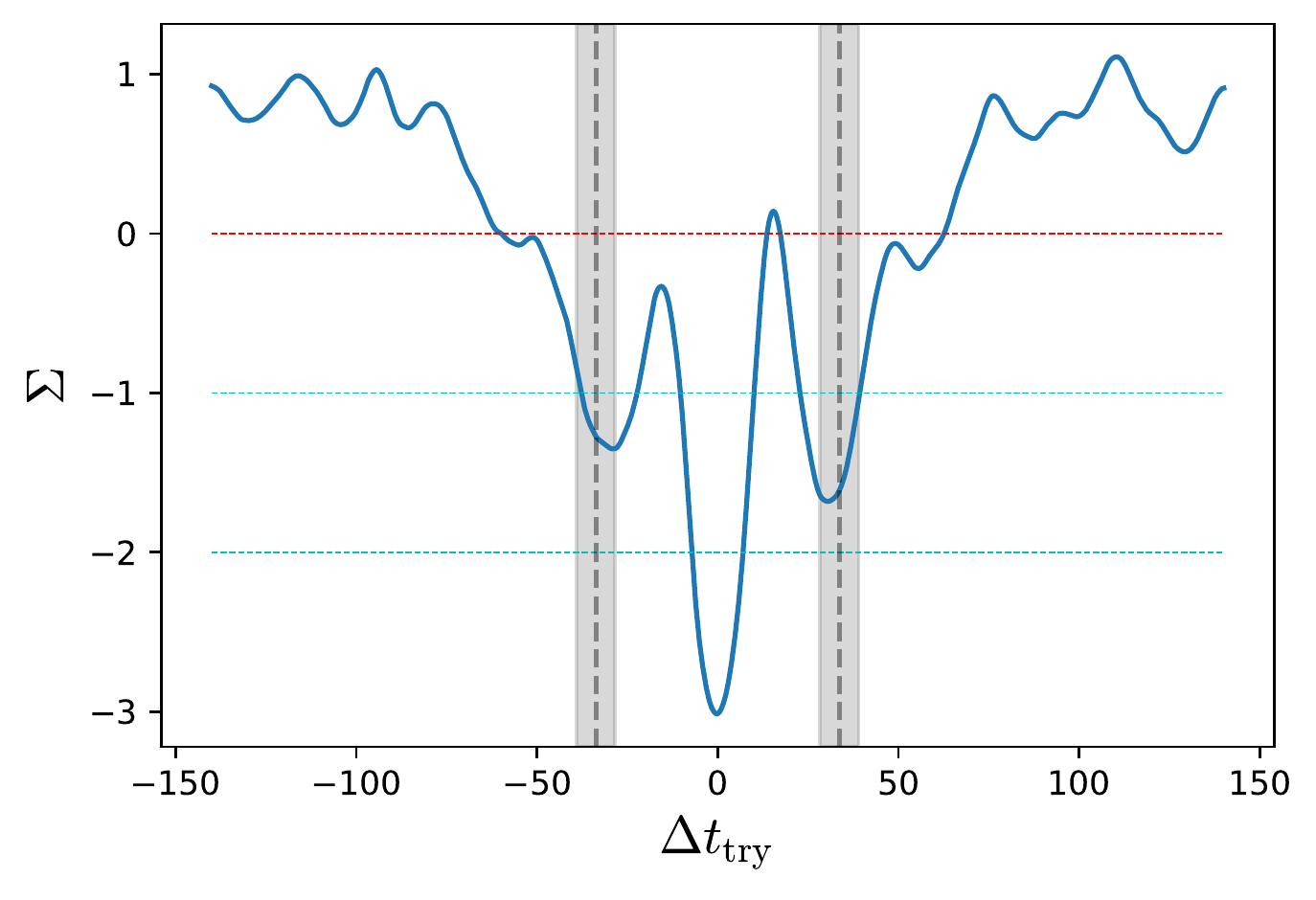}
\caption{Combined fluctuations corresponding to the four data patches and two smoothing scales $\delta=8.0,~9.0$. $\Sigma(\dtc)$ clearly shows two prominent minima at $\dtc=-28.7,30.5$ days with depths of $\Sigma=-1.35,~-1.68$ respectively. The vertical dashed lines and the shaded gray region show the COSMOGRAIL estimated time delay and the $2\sigma$ uncertainty around it.}
\label{fig:J1226_comb_lensed}
\end{figure}

The combined fluctuation estimator $\Sigma (\dtc)$ from the four seasonal patches  is shown in Figure~\ref{fig:J1226_comb_lensed}. 
We find a pair of minima at $\dtc=-28.7,~30.5$ days with depth $\Sigma=-1.35,~-1.68$ respectively. Therefore, we detect the system as `highly probable lensed' case, according to our relaxed criteria (described in section \ref{sec:relaxed_criteria}).
Our estimation of the time delay ($\dte =29.60 \pm 1.48$ day) is consistent with the COSMOGRAIL estimation, ($\dtt=33.7 \pm 2.7$) within the reported $1\sigma$ uncertainty. Note that both of the secondary minima in Figure~\ref{fig:J1226_comb_lensed} are wide enough to accommodate the time delay estimated by COSMOGRAIL. 

\begin{figure}
\centering
\includegraphics[width=0.485\textwidth]{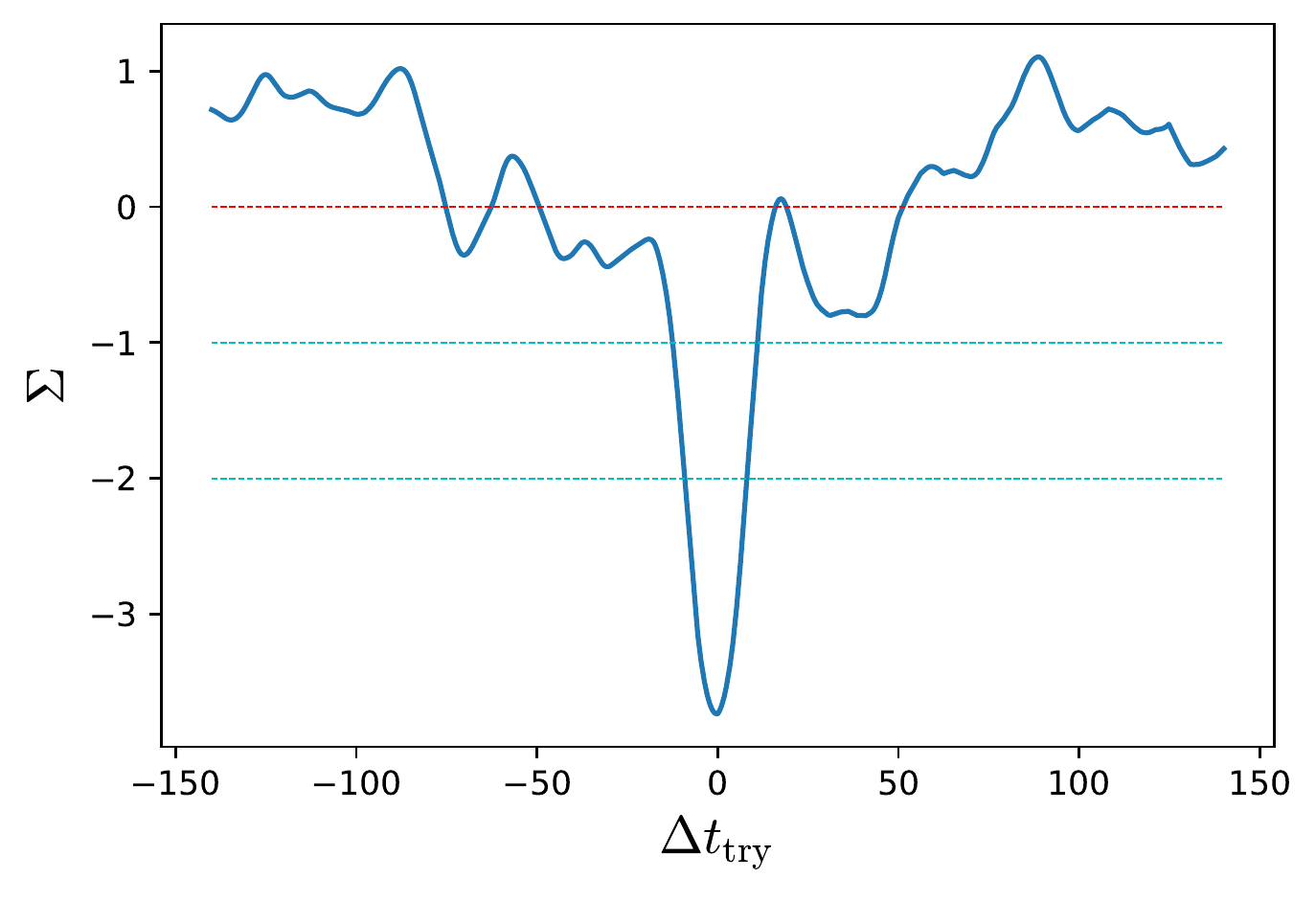}
\caption{Same as Figure~\ref{fig:J1226_comb_lensed} for the light curve of the brightest image only. There is no prominent pair of minima, correctly identifying the system as unlensed.}
\label{fig:J1226_comb_unlensed}
\end{figure}

For completeness, we analyse the observed light curve of the brightest image, which is of course an unlensed case. We consider the same data patches and follow the same procedure as described above for the lensed case. The combined $\Sigma (\dtc)$ curve is shown in~Figure~\ref{fig:J1226_comb_unlensed}. The absence of a pair of prominent minima at similar values of $|\dtc|$ correctly identifies this light curve as a true negative.

In summary, we correctly identify the system SDSS J1226-0006 as a lens only using the unresolved light and we estimate the time delay within the COSMOGRAIL uncertainties. Furthermore, our method correctly identifies the light curve of the brightest image only as a true negative.

\section{Summary and discussion}
\label{sec:conclusion}

We present a novel technique to detect lensed quasar systems and measure the time delays using only unresolved joint light curve data, without any need for assuming a model/template or additional information. Our method is general and can be applied to survey data with insufficient angular resolution to resolve the lensed quasars, thus opening up the opportunity to identify and measure time delays in a cost effective manner.

Our method builds on that proposed by \cite{Geiger1996},  partially breaking the degeneracy in the reconstructed solutions, by looking for minima in the residual fluctuations in the reconstructed light curve as a function of trial time delay. A global minimum is always found for $\dtc=0$, while doubly imaged quasars are identified by a pair of symmetric minima located at approximately $\pm$ the true time delays. The location of the pair of minima provides an estimate of the time delay.

We conduct several tests of our technique. First, we use simulations based on Damped Random Walk for the quasar light curves, with and without noise. We use training sets to define selection criteria and blind sets to test the performance of the method. Second, we use simulated light curves from the Time Delay Challenge. Third, we apply the method to the light curve of the lensed quasar SDSS J1226-0006 observed by the COSMOGRAIL collaboration using the 1.2m Euler Telescope \citep{Millon:2020xab}.

Our main results can be summarized as follows:

\begin{enumerate}
    \item For light curves with negligible noise, we find $95\%$ recall and $100\%$ precision, based on conservative criteria. The true time delays are recovered within the sampling resolution of the trial time delay 0.1 days.
    \item For light curves with ZTF-like noise, we find that smoothing the light curves using an iterative smoothing algorithm prior to applying our method greatly enhances its performance. After smoothing, we find $15\%$ recall and $100\%$ precision, based on conservative criteria. We then introduce a set of relaxed criteria, that yields precision of $92.3\%$ with a higher recall of $60\%$.  The true time delays are recovered within 3\%.
    \item For realistic LSST multi-year light curves taken from TDC1, we find that combining the fluctuation statistics from multiple years greatly improves the signal in our fluctuation analysis. We consider a number of doubly imaged systems from Rung 0 and 1 as examples and demonstrate that the method can find the systems as lensed and recover the time delays within $3\%$ of the respective truths. A follow up work will analyse all the systems from different rungs of TDC1 and study the precision and recall for TDC1 compilation comprehensively.

    \item For the COSMOGRAIL light curve of SDSS J1226-0006 we find that our method correctly identifies it as a lens from the joint unresolved light curve and estimates the time delay within the COSMOGRAIL uncertainty. We use the light curve of just one of the images to simulate a false negative, and show that it is correctly labeled by our method.
\end{enumerate}

The main strength of our method in comparison to those proposed in the literature \citep{Geiger1996, Shu2020,Springer:2021yhe,Springer:2021jhm, 2021arXiv211001012B} is that it does not require any additional information, neither any model/template for the quasar light curves, nor any spectroscopic information\footnote{ In a separate article we aim to provide the mathematical proof as to how our data driven method detects lenses and measures the time delays by minimizing the fluctuation in reconstructed image light curves. This will further allow us to compare our approach with that of the other proposed methods.}. Our method is thus well suited for detecting lensed quasar systems only using the light curves from ongoing time domain wide field surveys like Pan-STARRS1, ZTF etc and the future surveys like LSST by Vera C. Rubin Observatory, as well as from existing databases of quasar light curves. The generality of the method suggests that it should be more complete and unbiased than alternatives based on stronger assumptions. For example, our method can identify the lenses and measure the corresponding time delays independent of the power spectrum of the intrinsic quasar light curves, as demonstrated in Appendix \ref{app:PS_demo}.  

In future work, we plan to test this method on large number of simulations in a variety of scenarios, with the goal of obtaining a proper understanding of the uncertainties, and determining the precision and recall as a function of conditions. From this exercise one can also understand which observation strategy is favoured in this approach, e.g. better cadence vs longer observation time vs better noise control etc, or tailor the results to existing and planned surveys. 
Furthermore, we plan to extend the algorithm to quad systems. Finally, we plan to apply our algorithm to existing datasets and carry out a search for lensed quasars.  

\section*{Acknowledgement}
The Seondeok high performance computing cluster at KASI has been used in this work.
S.B. and A.S. thank Wuhyun Sohn, Eric V. Linder and Alex G. Kim for many useful discussions. A.S. would like to acknowledge the support by National Research Foundation of Korea NRF-2021M3F7A1082053 and the support
of the Korea Institute for Advanced Study (KIAS) grant
funded by the government of Korea.
K.L. was supported by the National Natural Science Foundation of China
(NSFC) No. 11973034. A.S. and K.L. also acknowledge the support and hospitality received from Beijing Normal University.
T.T. acknowledges support by
the National Science Foundation through grant NSF-AST-1906976 "Collaborative Research: Toward a 1\% measurement of the Hubble Constant with gravitational time delays", and by the Packard Foundation through a Packard Research Fellowship.


\appendix

\section{Iterative smoothing with exponential kernel}
\label{app:smoothing}
We smooth the observed light curve $F_{\rm obs}(t)$ iteratively with a Gaussian kernel following \citet{Shafieloo:2005nd, Shafieloo:2007cs, Shafieloo:2009hi, Aghamousa:2014uya}. 
The smoothed flux in the $n$th step is obtained from the previous step as
\begin{equation}
 F_{n}(t)=F_{n-1}(t)+ \frac{1}{N(t)} \sum^{N_D}_i\frac{\left(F_{\rm obs}(t_i)-F_{n-1}(t_i)\right)}{{\sigma^2_{\rm obs}}(t_i)}\times \exp{\left[-\frac{(t -t_i)^2}{2 \delta^2} \right]}
\end{equation}
where the normalisation term $N(t)$ is given by.
\begin{equation}
 N(t)=\sum^{N_D}_i \left(\frac{1}{{\sigma^2_{\rm obs}}(t_i)}\right) \times \exp{\left[-\frac{(t -t_i)^2}{2 \delta^2} \right]}
\end{equation}
Here $F_{n-1}(t)$ is the smoothed flux obtained in the previous step, i.e. at the $(n-1)$ step.
We start with an initial guess which can be a constant number for simplicity, $F_0(t)=$constant, and continue iterating for $\nit$ times. After a sufficient number of iterations, the smoothed flux becomes independent of the initial guess. The smoothing method has two parameters: the smoothing scale $\delta$ and the number of iteration $\nit$.

\section{Light curves with degraded sampling cadence}\label{app:cadence}

The simulated data that we use for validation in Sections~\ref{sec:validation} and~\ref{sec:td_withnoise} are sampled with daily cadence. We tested the effect of sampling the same light curves with 3-day cadence.

In general, $\Sigma(\dtc)$ become smoother with decreasing cadence, since small timescale features are erased.
For the 3-day cadence light curves with negligible noise we identify 6 out of 10 systems as lensed, compared with 9 for the daily cadence sampling.

For the light curves with ZTF-like noise, we find that the target pair of secondary minima at $\dtc \approx \pm \dtt$ is slightly shallower in the 3-day cadence case. However, we can still identify a few with the relaxed selection criteria. Fortunately, in reality, quasars are typically observed over many years leading to multiple patches in the data. We can independently use those patches since typically the patches are longer than the maximum of $\dtc$. 
As in the case of TDC1 data shown in the main text, it is likely that combining multiple years of observations will improve precision and recall of the estimator. A systematic investigation of the dependency of precision and recall on sampling is left for future work.

\section{Examples of flat and blue power spectra}
\label{app:PS_demo}

\begin{figure*}
\centering
\includegraphics[width=0.485\linewidth]{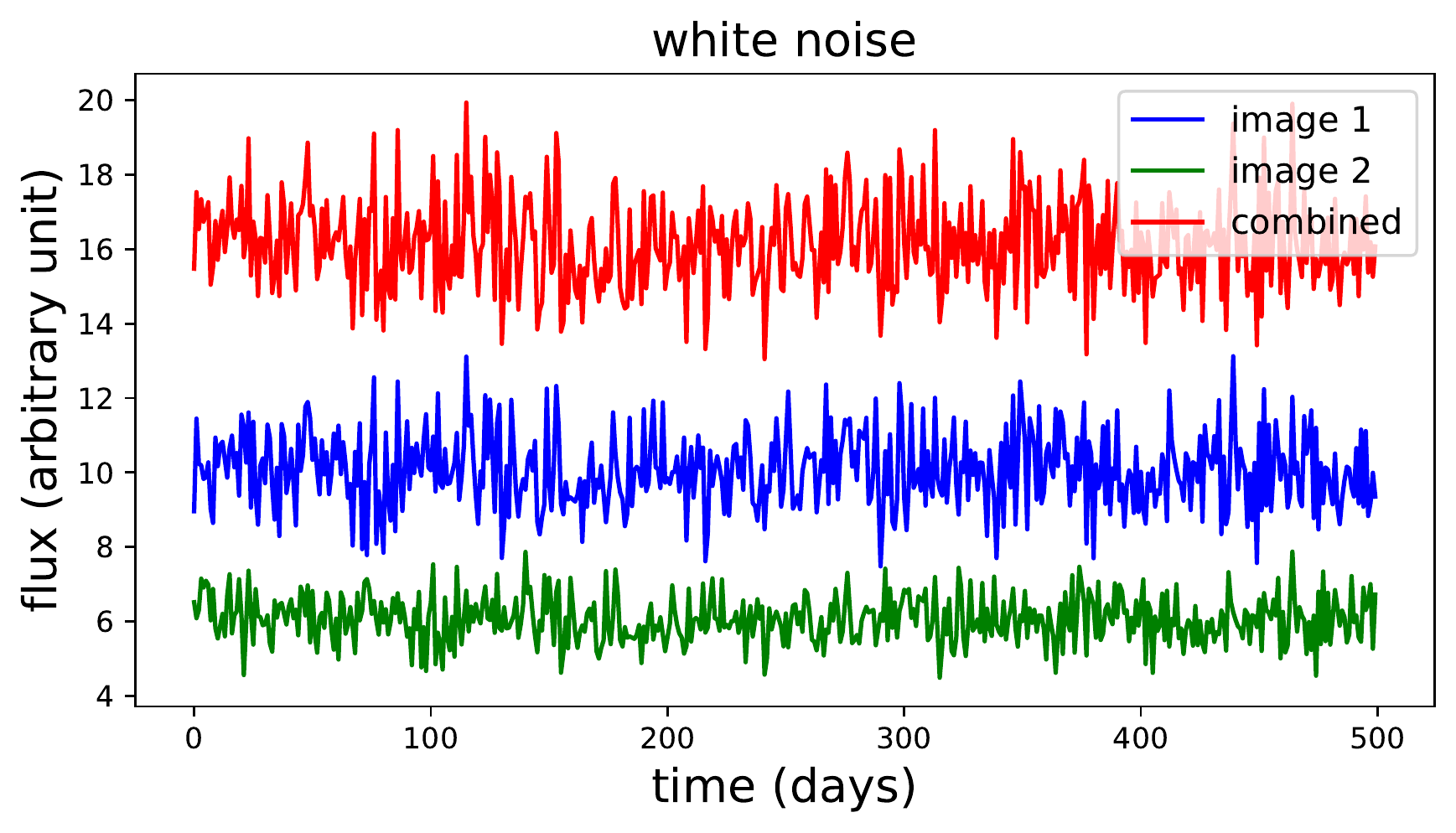}
\includegraphics[width=0.485\linewidth]{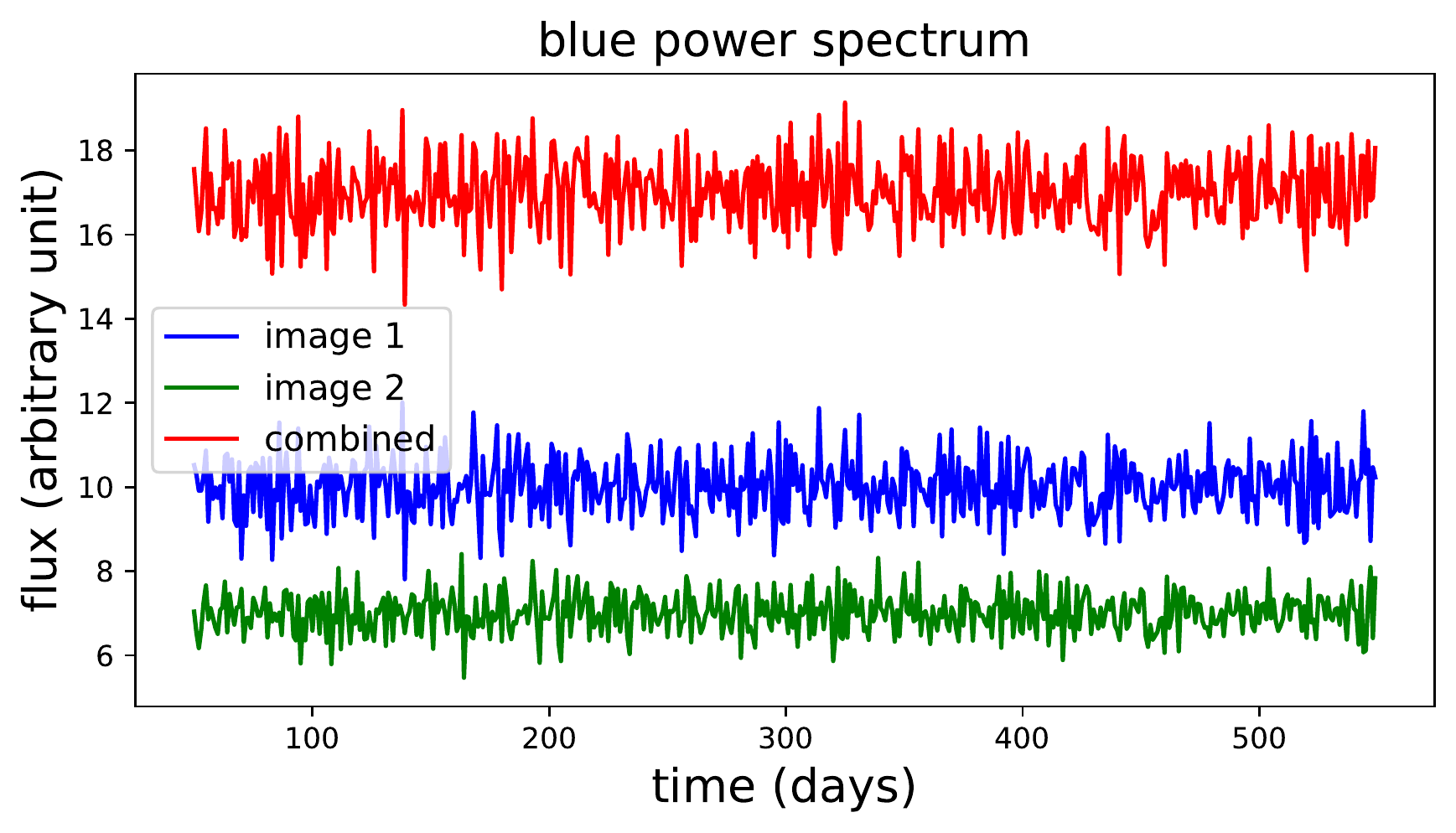}
\subfigure[white noise (flat power spectrum, $P(\omega)=$ constant)]{
\includegraphics[width=0.485\linewidth]{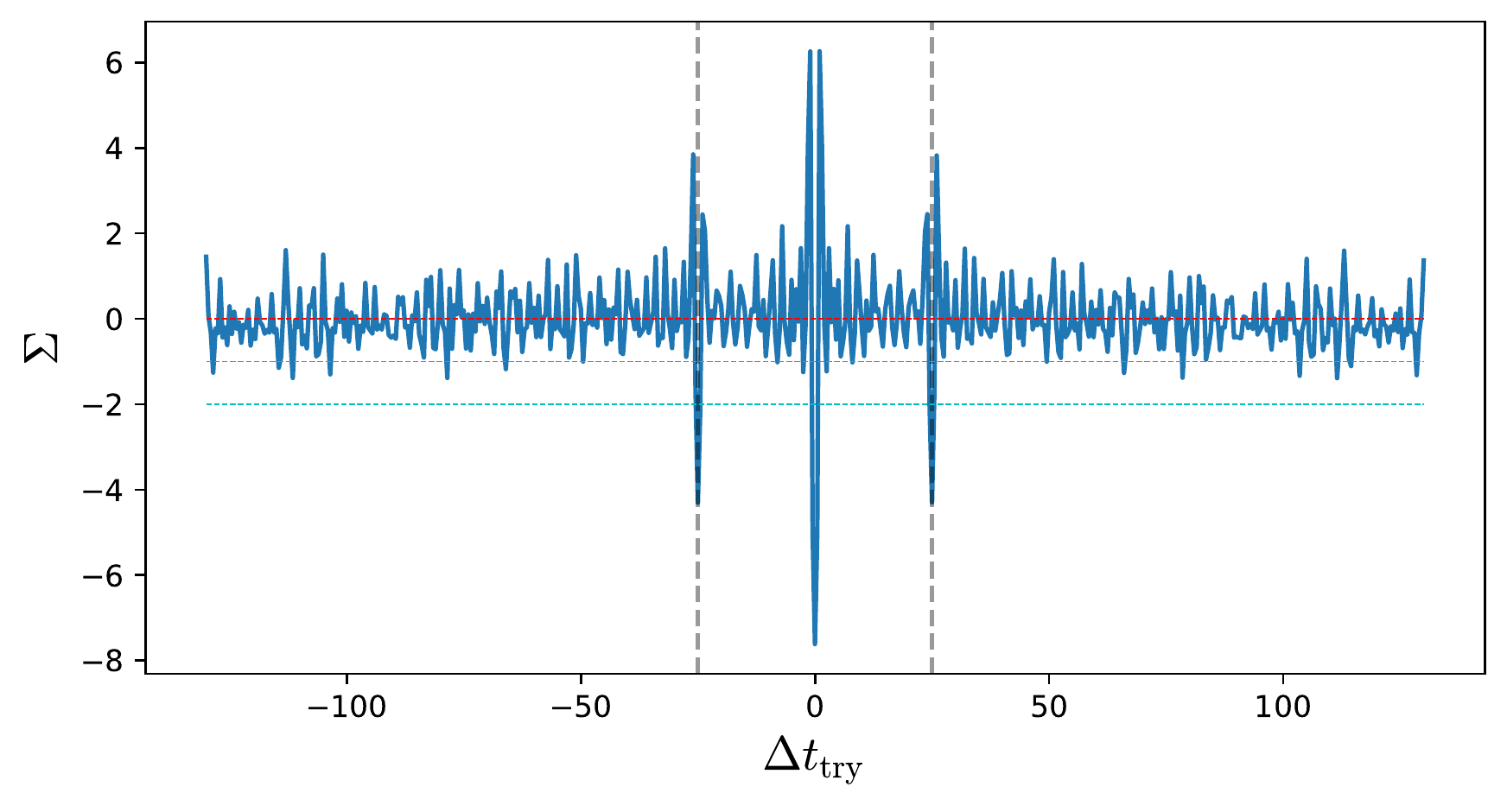}}
\subfigure[blue power spectrum ($P(\omega) \propto |\omega|^2$)]{
\includegraphics[width=0.485\linewidth]{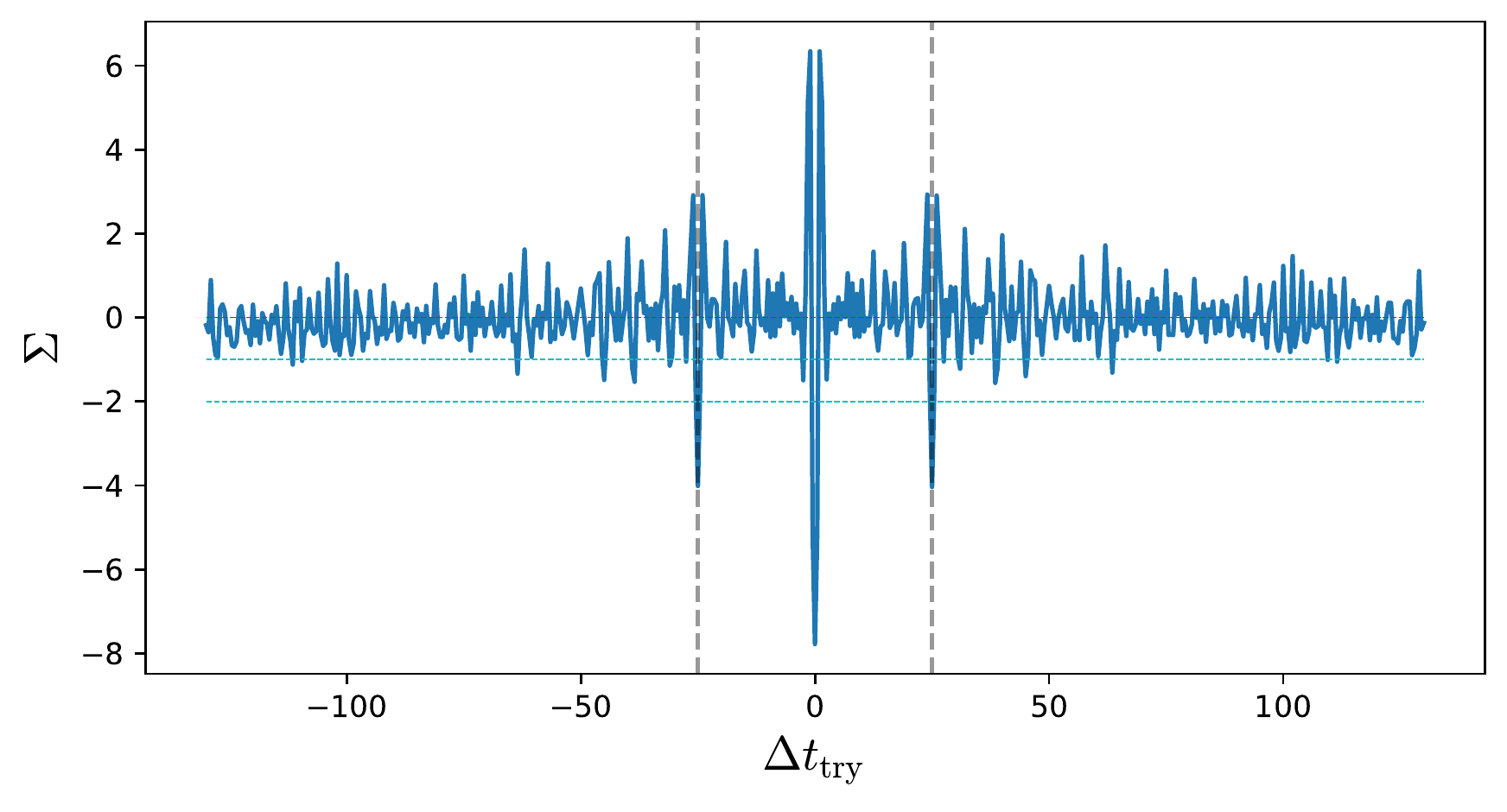}}
\caption{The figure demonstrates that our method performs well even for intrinsic light curves being generated from white noise (left panels) and blue power spectrum (right panels). For simplicity, we assume negligible observational noise and one day of cadence for simplicity. The vertical dashed lines represent $\dtc=\pm \dtt$.}
\label{fig:demo_PS}
\end{figure*}

The power spectrum of a time series is defined as 
\begin{equation}
    P(\omega) = \langle \widehat{f}(\omega)  \widehat{f}(\omega)^* \rangle \;,
\end{equation}
where $ \widehat{f}(\omega)$ is the Fourier transform of the time series $f(t)$.
In the sections \ref{sec:validation} and \ref{sec:td_withnoise}, we show a number of examples where the intrinsic quasar light curves are simulated using damped random walk and hence they can be described somewhat by the red power spectra ($P(\omega) \propto |\omega|^{-\gamma}$ where $\gamma>0$). In Figure \ref{fig:demo_PS} we illustrate that the method performs equally well for two other types of the power spectrum --  flat (white noise) and blue ($P(\omega) \propto |\omega|^\gamma$, $\gamma>0$) in the left and right panels respectively\footnote{The time series generated from a flat or blue power spectrum (as shown in the top panels of Figure \ref{fig:demo_PS}) may not describe the light curves of quasars. Also note that here we consider high quality data with negligible observational noise and one day cadence for simplicity.}. We correctly recover the time delays by following the pair of prominent secondary minima at $\dtc=\pm \dtt$ (the vertical dashed lines) in the fluctuation curves, $\Sigma(\dtc)$, shown in the bottom panels. This exercise demonstrates that the method can detect the lenses independent of the form of the power spectrum (red/flat/blue) and we do not need to make any assumption in this regard.

\bibliographystyle{aasjournal}
\bibliography{lensed_quasars}

\begin{thebibliography}{}
\expandafter\ifx\csname natexlab\endcsname\relax\def\natexlab#1{#1}\fi
\providecommand{\url}[1]{\href{#1}{#1}}
\providecommand{\dodoi}[1]{doi:~\href{http://doi.org/#1}{\nolinkurl{#1}}}
\providecommand{\doeprint}[1]{\href{http://ascl.net/#1}{\nolinkurl{http://ascl.net/#1}}}
\providecommand{\doarXiv}[1]{\href{https://arxiv.org/abs/#1}{\nolinkurl{https://arxiv.org/abs/#1}}}

\bibitem[{Aghamousa \& Shafieloo(2015)}]{Aghamousa:2014uya}
Aghamousa, A., \& Shafieloo, A. 2015, Astrophys. J., 804, 39,
  \dodoi{10.1088/0004-637X/804/1/39}

\bibitem[{Bag {et~al.}(2021)Bag, Kim, Linder, \& Shafieloo}]{Bag:2020pbg}
Bag, S., Kim, A.~G., Linder, E.~V., \& Shafieloo, A. 2021, Astrophys. J., 910,
  65, \dodoi{10.3847/1538-4357/abe238}

\bibitem[{{Bellm} {et~al.}(2019){Bellm}, {Kulkarni}, {Graham}, {Dekany},
  {Smith}, {Riddle}, {Masci}, {Helou}, {Prince}, {Adams}, {Barbarino},
  {Barlow}, {Bauer}, {Beck}, {Belicki}, {Biswas}, {Blagorodnova}, {Bodewits},
  {Bolin}, {Brinnel}, {Brooke}, {Bue}, {Bulla}, {Burruss}, {Cenko}, {Chang},
  {Connolly}, {Coughlin}, {Cromer}, {Cunningham}, {De}, {Delacroix}, {Desai},
  {Duev}, {Eadie}, {Farnham}, {Feeney}, {Feindt}, {Flynn}, {Franckowiak},
  {Frederick}, {Fremling}, {Gal-Yam}, {Gezari}, {Giomi}, {Goldstein},
  {Golkhou}, {Goobar}, {Groom}, {Hacopians}, {Hale}, {Henning}, {Ho}, {Hover},
  {Howell}, {Hung}, {Huppenkothen}, {Imel}, {Ip}, {Ivezi{\'c}}, {Jackson},
  {Jones}, {Juric}, {Kasliwal}, {Kaspi}, {Kaye}, {Kelley}, {Kowalski},
  {Kramer}, {Kupfer}, {Landry}, {Laher}, {Lee}, {Lin}, {Lin}, {Lunnan},
  {Giomi}, {Mahabal}, {Mao}, {Miller}, {Monkewitz}, {Murphy}, {Ngeow},
  {Nordin}, {Nugent}, {Ofek}, {Patterson}, {Penprase}, {Porter}, {Rauch},
  {Rebbapragada}, {Reiley}, {Rigault}, {Rodriguez}, {van Roestel}, {Rusholme},
  {van Santen}, {Schulze}, {Shupe}, {Singer}, {Soumagnac}, {Stein}, {Surace},
  {Sollerman}, {Szkody}, {Taddia}, {Terek}, {Van Sistine}, {van Velzen},
  {Vestrand}, {Walters}, {Ward}, {Ye}, {Yu}, {Yan}, \& {Zolkower}}]{ztf}
{Bellm}, E.~C., {Kulkarni}, S.~R., {Graham}, M.~J., {et~al.} 2019, \pasp, 131,
  018002, \dodoi{10.1088/1538-3873/aaecbe}

\bibitem[{{Biggio} {et~al.}(2021){Biggio}, {Domi}, {Tosi}, {Vernardos},
  {Ricci}, {Paganin}, \& {Bracco}}]{2021arXiv211001012B}
{Biggio}, L., {Domi}, A., {Tosi}, S., {et~al.} 2021, arXiv e-prints,
  arXiv:2110.01012.
\newblock \doarXiv{2110.01012}

\bibitem[{{Birrer} \& {Treu}(2021)}]{Birrer}
{Birrer}, S., \& {Treu}, T. 2021, \aap, 649, A61,
  \dodoi{10.1051/0004-6361/202039179}

\bibitem[{{Birrer} {et~al.}(2020){Birrer}, {Shajib}, {Galan}, {Millon}, {Treu},
  {Agnello}, {Auger}, {Chen}, {Christensen}, {Collett}, {Courbin}, {Fassnacht},
  {Koopmans}, {Marshall}, {Park}, {Rusu}, {Sluse}, {Spiniello}, {Suyu},
  {Wagner-Carena}, {Wong}, {Barnab{\`e}}, {Bolton}, {Czoske}, {Ding},
  {Frieman}, \& {Van de Vyvere}}]{2020A&A...643A.165B}
{Birrer}, S., {Shajib}, A.~J., {Galan}, A., {et~al.} 2020, \aap, 643, A165,
  \dodoi{10.1051/0004-6361/202038861}

\bibitem[{{Bonvin} {et~al.}(2017){Bonvin}, {Courbin}, {Suyu}, {Marshall},
  {Rusu}, {Sluse}, {Tewes}, {Wong}, {Collett}, {Fassnacht}, {Treu}, {Auger},
  {Hilbert}, {Koopmans}, {Meylan}, {Rumbaugh}, {Sonnenfeld}, \&
  {Spiniello}}]{Bonvin2017}
{Bonvin}, V., {Courbin}, F., {Suyu}, S.~H., {et~al.} 2017, \mnras, 465, 4914,
  \dodoi{10.1093/mnras/stw3006}

\bibitem[{{Browne} {et~al.}(2003){Browne}, {Wilkinson}, {Jackson}, {Myers},
  {Fassnacht}, {Koopmans}, {Marlow}, {Norbury}, {Rusin}, {Sykes}, {Biggs},
  {Blandford}, {de Bruyn}, {Chae}, {Helbig}, {King}, {McKean}, {Pearson},
  {Phillips}, {Readhead}, {Xanthopoulos}, \& {York}}]{Browne2003}
{Browne}, I.~W.~A., {Wilkinson}, P.~N., {Jackson}, N.~J.~F., {et~al.} 2003,
  \mnras, 341, 13, \dodoi{10.1046/j.1365-8711.2003.06257.x}

\bibitem[{{Chambers} {et~al.}(2016){Chambers}, {Magnier}, {Metcalfe},
  {Flewelling}, {Huber}, {Waters}, {Denneau}, {Draper}, {Farrow}, {Finkbeiner},
  {Holmberg}, {Koppenhoefer}, {Price}, {Rest}, {Saglia}, {Schlafly}, {Smartt},
  {Sweeney}, {Wainscoat}, {Burgett}, {Chastel}, {Grav}, {Heasley}, {Hodapp},
  {Jedicke}, {Kaiser}, {Kudritzki}, {Luppino}, {Lupton}, {Monet}, {Morgan},
  {Onaka}, {Shiao}, {Stubbs}, {Tonry}, {White}, {Ba{\~n}ados}, {Bell},
  {Bender}, {Bernard}, {Boegner}, {Boffi}, {Botticella}, {Calamida},
  {Casertano}, {Chen}, {Chen}, {Cole}, {Deacon}, {Frenk}, {Fitzsimmons},
  {Gezari}, {Gibbs}, {Goessl}, {Goggia}, {Gourgue}, {Goldman}, {Grant},
  {Grebel}, {Hambly}, {Hasinger}, {Heavens}, {Heckman}, {Henderson}, {Henning},
  {Holman}, {Hopp}, {Ip}, {Isani}, {Jackson}, {Keyes}, {Koekemoer}, {Kotak},
  {Le}, {Liska}, {Long}, {Lucey}, {Liu}, {Martin}, {Masci}, {McLean}, {Mindel},
  {Misra}, {Morganson}, {Murphy}, {Obaika}, {Narayan}, {Nieto-Santisteban},
  {Norberg}, {Peacock}, {Pier}, {Postman}, {Primak}, {Rae}, {Rai}, {Riess},
  {Riffeser}, {Rix}, {R{\"o}ser}, {Russel}, {Rutz}, {Schilbach}, {Schultz},
  {Scolnic}, {Strolger}, {Szalay}, {Seitz}, {Small}, {Smith}, {Soderblom},
  {Taylor}, {Thomson}, {Taylor}, {Thakar}, {Thiel}, {Thilker}, {Unger},
  {Urata}, {Valenti}, {Wagner}, {Walder}, {Walter}, {Watters}, {Werner},
  {Wood-Vasey}, \& {Wyse}}]{Pan-STARRS1}
{Chambers}, K.~C., {Magnier}, E.~A., {Metcalfe}, N., {et~al.} 2016, arXiv
  e-prints, arXiv:1612.05560.
\newblock \doarXiv{1612.05560}

\bibitem[{{Dalal} \& {Kochanek}(2002)}]{Dalal2002}
{Dalal}, N., \& {Kochanek}, C.~S. 2002, \apj, 572, 25, \dodoi{10.1086/340303}

\bibitem[{{Denissenya} {et~al.}(2022){Denissenya}, {Bag}, {Kim}, {Linder}, \&
  {Shafieloo}}]{Denissenya:2021cpz}
{Denissenya}, M., {Bag}, S., {Kim}, A.~G., {Linder}, E.~V., \& {Shafieloo}, A.
  2022, \mnras, 511, 1210, \dodoi{10.1093/mnras/stac143}

\bibitem[{Dobler {et~al.}(2015)Dobler, Fassnacht, Treu, Marshall, Liao,
  Hojjati, Linder, \& Rumbaugh}]{tdc1}
Dobler, G., Fassnacht, C., Treu, T., {et~al.} 2015, Astrophys. J., 799, 168,
  \dodoi{10.1088/0004-637X/799/2/168}

\bibitem[{{Geiger} \& {Schneider}(1996)}]{Geiger1996}
{Geiger}, B., \& {Schneider}, P. 1996, \mnras, 282, 530,
  \dodoi{10.1093/mnras/282.2.530}

\bibitem[{{Goobar} {et~al.}(2017){Goobar}, {Amanullah}, {Kulkarni}, {Nugent},
  {Johansson}, {Steidel}, {Law}, {M{\"o}rtsell}, {Quimby}, {Blagorodnova},
  {Brandeker}, {Cao}, {Cooray}, {Ferretti}, {Fremling}, {Hangard}, {Kasliwal},
  {Kupfer}, {Lunnan}, {Masci}, {Miller}, {Nayyeri}, {Neill}, {Ofek},
  {Papadogiannakis}, {Petrushevska}, {Ravi}, {Sollerman}, {Sullivan}, {Taddia},
  {Walters}, {Wilson}, {Yan}, \& {Yaron}}]{Goobar:2016uuf}
{Goobar}, A., {Amanullah}, R., {Kulkarni}, S.~R., {et~al.} 2017, Science, 356,
  291, \dodoi{10.1126/science.aal2729}

\bibitem[{{Hirv} {et~al.}(2007){Hirv}, {Eenm{\"a}e}, {Liimets},
  {Liivam{\"a}gi}, \& {Pelt}}]{2007A&A...464..471H}
{Hirv}, A., {Eenm{\"a}e}, T., {Liimets}, T., {Liivam{\"a}gi}, L.~J., \& {Pelt},
  J. 2007, \aap, 464, 471, \dodoi{10.1051/0004-6361:20065537}

\bibitem[{{Hirv} {et~al.}(2011){Hirv}, {Olspert}, \& {Pelt}}]{Hirv2011}
{Hirv}, A., {Olspert}, N., \& {Pelt}, J. 2011, Baltic Astronomy, 20, 125,
  \dodoi{10.1515/astro-2017-0273}

\bibitem[{{Hojjati} {et~al.}(2013){Hojjati}, {Kim}, \& {Linder}}]{Hojjati2013}
{Hojjati}, A., {Kim}, A.~G., \& {Linder}, E.~V. 2013, \prd, 87, 123512,
  \dodoi{10.1103/PhysRevD.87.123512}

\bibitem[{{Huchra} {et~al.}(1985){Huchra}, {Gorenstein}, {Kent}, {Shapiro},
  {Smith}, {Horine}, \& {Perley}}]{Huchra1985}
{Huchra}, J., {Gorenstein}, M., {Kent}, S., {et~al.} 1985, \aj, 90, 691,
  \dodoi{10.1086/113777}

\bibitem[{Jim{\'{e}}nez-Vicente \& Mediavilla(2019)}]{Jim_nez_Vicente_2019}
Jim{\'{e}}nez-Vicente, J., \& Mediavilla, E. 2019, The Astrophysical Journal,
  885, 75, \dodoi{10.3847/1538-4357/ab46b8}

\bibitem[{{Kelly} {et~al.}(2009){Kelly}, {Bechtold}, \&
  {Siemiginowska}}]{Kelly2009}
{Kelly}, B.~C., {Bechtold}, J., \& {Siemiginowska}, A. 2009, \apj, 698, 895,
  \dodoi{10.1088/0004-637X/698/1/895}

\bibitem[{{Kelly} {et~al.}(2015){Kelly}, {Rodney}, {Treu}, {Foley}, {Brammer},
  {Schmidt}, {Zitrin}, {Sonnenfeld}, {Strolger}, {Graur}, {Filippenko}, {Jha},
  {Riess}, {Bradac}, {Weiner}, {Scolnic}, {Malkan}, {von der Linden}, {Trenti},
  {Hjorth}, {Gavazzi}, {Fontana}, {Merten}, {McCully}, {Jones}, {Postman},
  {Dressler}, {Patel}, {Cenko}, {Graham}, \& {Tucker}}]{Kelly:2014mwa}
{Kelly}, P.~L., {Rodney}, S.~A., {Treu}, T., {et~al.} 2015, Science, 347, 1123,
  \dodoi{10.1126/science.aaa3350}

\bibitem[{{Lemon} {et~al.}(2020){Lemon}, {Auger}, {McMahon}, {Anguita},
  {Apostolovski}, {Chen}, {Fassnacht}, {Melo}, {Motta}, {Shajib}, {Treu},
  {Agnello}, {Buckley-Geer}, {Schechter}, {Birrer}, {Collett}, {Courbin},
  {Rusu}, {Abbott}, {Allam}, {Annis}, {Avila}, {Bertin}, {Brooks}, {Burke},
  {Carnero Rosell}, {Carrasco Kind}, {Carretero}, {Costanzi}, {da Costa}, {De
  Vicente}, {Desai}, {Eifler}, {Flaugher}, {Frieman}, {Garc{\'\i}a-Bellido},
  {Gaztanaga}, {Gerdes}, {Gruen}, {Gruendl}, {Gschwend}, {Gutierrez},
  {Honscheid}, {James}, {Kim}, {Krause}, {Kuehn}, {Kuropatkin}, {Lahav},
  {Lima}, {Lin}, {Maia}, {March}, {Marshall}, {Menanteau}, {Miquel}, {Palmese},
  {Paz-Chinch{\'o}n}, {Plazas}, {Roodman}, {Sanchez}, {Schubnell}, {Serrano},
  {Smith}, {Soares-Santos}, {Suchyta}, {Tarle}, \& {Walker}}]{Lemon2020}
{Lemon}, C., {Auger}, M.~W., {McMahon}, R., {et~al.} 2020, \mnras, 494, 3491,
  \dodoi{10.1093/mnras/staa652}

\bibitem[{{Li} {et~al.}(2018){Li}, {Gao}, {Ding}, {Wang}, \&
  {Zhang}}]{2018NatCo...9.3833L}
{Li}, Z.-X., {Gao}, H., {Ding}, X.-H., {Wang}, G.-J., \& {Zhang}, B. 2018,
  Nature Communications, 9, 3833, \dodoi{10.1038/s41467-018-06303-0}

\bibitem[{{Liao} {et~al.}(2017){Liao}, {Fan}, {Ding}, {Biesiada}, \&
  {Zhu}}]{2017NatCo...8.1148L}
{Liao}, K., {Fan}, X.-L., {Ding}, X., {Biesiada}, M., \& {Zhu}, Z.-H. 2017,
  Nature Communications, 8, 1148, \dodoi{10.1038/s41467-017-01152-9}

\bibitem[{{Liao} {et~al.}(2015){Liao}, {Treu}, {Marshall}, {Fassnacht},
  {Rumbaugh}, {Dobler}, {Aghamousa}, {Bonvin}, {Courbin}, {Hojjati}, {Jackson},
  {Kashyap}, {Rathna Kumar}, {Linder}, {Mandel}, {Meng}, {Meylan}, {Moustakas},
  {Prabhu}, {Romero-Wolf}, {Shafieloo}, {Siemiginowska}, {Stalin}, {Tak},
  {Tewes}, \& {van Dyk}}]{tdc2}
{Liao}, K., {Treu}, T., {Marshall}, P., {et~al.} 2015, \apj, 800, 11,
  \dodoi{10.1088/0004-637X/800/1/11}

\bibitem[{{LSST Science Collaboration} {et~al.}(2009){LSST Science
  Collaboration}, {Abell}, {Allison}, {Anderson}, {Andrew}, {Angel}, {Armus},
  {Arnett}, {Asztalos}, {Axelrod}, {Bailey}, {Ballantyne}, {Bankert},
  {Barkhouse}, {Barr}, {Barrientos}, {Barth}, {Bartlett}, {Becker}, {Becla},
  {Beers}, {Bernstein}, {Biswas}, {Blanton}, {Bloom}, {Bochanski}, {Boeshaar},
  {Borne}, {Bradac}, {Brandt}, {Bridge}, {Brown}, {Brunner}, {Bullock},
  {Burgasser}, {Burge}, {Burke}, {Cargile}, {Chandrasekharan}, {Chartas},
  {Chesley}, {Chu}, {Cinabro}, {Claire}, {Claver}, {Clowe}, {Connolly}, {Cook},
  {Cooke}, {Cooray}, {Covey}, {Culliton}, {de Jong}, {de Vries}, {Debattista},
  {Delgado}, {Dell'Antonio}, {Dhital}, {Di Stefano}, {Dickinson}, {Dilday},
  {Djorgovski}, {Dobler}, {Donalek}, {Dubois-Felsmann}, {Durech},
  {Eliasdottir}, {Eracleous}, {Eyer}, {Falco}, {Fan}, {Fassnacht}, {Ferguson},
  {Fernandez}, {Fields}, {Finkbeiner}, {Figueroa}, {Fox}, {Francke}, {Frank},
  {Frieman}, {Fromenteau}, {Furqan}, {Galaz}, {Gal-Yam}, {Garnavich},
  {Gawiser}, {Geary}, {Gee}, {Gibson}, {Gilmore}, {Grace}, {Green}, {Gressler},
  {Grillmair}, {Habib}, {Haggerty}, {Hamuy}, {Harris}, {Hawley}, {Heavens},
  {Hebb}, {Henry}, {Hileman}, {Hilton}, {Hoadley}, {Holberg}, {Holman},
  {Howell}, {Infante}, {Ivezic}, {Jacoby}, {Jain}, {R}, {Jedicke}, {Jee},
  {Garrett Jernigan}, {Jha}, {Johnston}, {Jones}, {Juric}, {Kaasalainen},
  {Styliani}, {Kafka}, {Kahn}, {Kaib}, {Kalirai}, {Kantor}, {Kasliwal},
  {Keeton}, {Kessler}, {Knezevic}, {Kowalski}, {Krabbendam}, {Krughoff},
  {Kulkarni}, {Kuhlman}, {Lacy}, {Lepine}, {Liang}, {Lien}, {Lira}, {Long},
  {Lorenz}, {Lotz}, {Lupton}, {Lutz}, {Macri}, {Mahabal}, {Mandelbaum},
  {Marshall}, {May}, {McGehee}, {Meadows}, {Meert}, {Milani}, {Miller},
  {Miller}, {Mills}, {Minniti}, {Monet}, {Mukadam}, {Nakar}, {Neill}, {Newman},
  {Nikolaev}, {Nordby}, {O'Connor}, {Oguri}, {Oliver}, {Olivier}, {Olsen},
  {Olsen}, {Olszewski}, {Oluseyi}, {Padilla}, {Parker}, {Pepper}, {Peterson},
  {Petry}, {Pinto}, {Pizagno}, {Popescu}, {Prsa}, {Radcka}, {Raddick},
  {Rasmussen}, {Rau}, {Rho}, {Rhoads}, {Richards}, {Ridgway}, {Robertson},
  {Roskar}, {Saha}, {Sarajedini}, {Scannapieco}, {Schalk}, {Schindler},
  {Schmidt}, {Schmidt}, {Schneider}, {Schumacher}, {Scranton}, {Sebag},
  {Seppala}, {Shemmer}, {Simon}, {Sivertz}, {Smith}, {Allyn Smith}, {Smith},
  {Spitz}, {Stanford}, {Stassun}, {Strader}, {Strauss}, {Stubbs}, {Sweeney},
  {Szalay}, {Szkody}, {Takada}, {Thorman}, {Trilling}, {Trimble}, {Tyson}, {Van
  Berg}, {Vanden Berk}, {VanderPlas}, {Verde}, {Vrsnak}, {Walkowicz},
  {Wandelt}, {Wang}, {Wang}, {Warner}, {Wechsler}, {West}, {Wiecha},
  {Williams}, {Willman}, {Wittman}, {Wolff}, {Wood-Vasey}, {Wozniak}, {Young},
  {Zentner}, \& {Zhan}}]{lsst1}
{LSST Science Collaboration}, {Abell}, P.~A., {Allison}, J., {et~al.} 2009,
  arXiv e-prints, arXiv:0912.0201.
\newblock \doarXiv{0912.0201}

\bibitem[{{LSST Science Collaboration} {et~al.}(2017){LSST Science
  Collaboration}, {Marshall}, {Anguita}, {Bianco}, {Bellm}, {Brandt},
  {Clarkson}, {Connolly}, {Gawiser}, {Ivezic}, {Jones}, {Lochner}, {Lund},
  {Mahabal}, {Nidever}, {Olsen}, {Ridgway}, {Rhodes}, {Shemmer}, {Trilling},
  {Vivas}, {Walkowicz}, {Willman}, {Yoachim}, {Anderson}, {Antilogus}, {Angus},
  {Arcavi}, {Awan}, {Biswas}, {Bell}, {Bennett}, {Britt}, {Buzasi},
  {Casetti-Dinescu}, {Chomiuk}, {Claver}, {Cook}, {Davenport}, {Debattista},
  {Digel}, {Doctor}, {Firth}, {Foley}, {Fong}, {Galbany}, {Giampapa}, {Gizis},
  {Graham}, {Grillmair}, {Gris}, {Haiman}, {Hartigan}, {Hawley}, {Hlozek},
  {Jha}, {Johns-Krull}, {Kanbur}, {Kalogera}, {Kashyap}, {Kasliwal}, {Kessler},
  {Kim}, {Kurczynski}, {Lahav}, {Liu}, {Malz}, {Margutti}, {Matheson},
  {McEwen}, {McGehee}, {Meibom}, {Meyers}, {Monet}, {Neilsen}, {Newman},
  {O'Dowd}, {Peiris}, {Penny}, {Peters}, {Poleski}, {Ponder}, {Richards},
  {Rho}, {Rubin}, {Schmidt}, {Schuhmann}, {Shporer}, {Slater}, {Smith},
  {Soares-Santos}, {Stassun}, {Strader}, {Strauss}, {Street}, {Stubbs},
  {Sullivan}, {Szkody}, {Trimble}, {Tyson}, {de Val-Borro}, {Valenti},
  {Wagoner}, {Wood-Vasey}, \& {Zauderer}}]{lsst2}
{LSST Science Collaboration}, {Marshall}, P., {Anguita}, T., {et~al.} 2017,
  arXiv e-prints, arXiv:1708.04058.
\newblock \doarXiv{1708.04058}

\bibitem[{{Magain} {et~al.}(1998){Magain}, {Courbin}, \&
  {Sohy}}]{1998ApJ...494..472M}
{Magain}, P., {Courbin}, F., \& {Sohy}, S. 1998, \apj, 494, 472,
  \dodoi{10.1086/305187}

\bibitem[{Mao \& Schneider(1998)}]{Mao:1997ek}
Mao, S.-d., \& Schneider, P. 1998, Mon. Not. Roy. Astron. Soc., 295, 587,
  \dodoi{10.1046/j.1365-8711.1998.01319.x}

\bibitem[{Metcalf \& Madau(2001)}]{Metcalf:2001ap}
Metcalf, R.~B., \& Madau, P. 2001, Astrophys. J., 563, 9,
  \dodoi{10.1086/323695}

\bibitem[{{Millon} {et~al.}(2020){Millon}, {Courbin}, {Bonvin}, {Paic},
  {Meylan}, {Tewes}, {Sluse}, {Magain}, {Chan}, {Galan}, {Joseph}, {Lemon},
  {Tihhonova}, {Anderson}, {Marmier}, {Chazelas}, {Lendl}, {Triaud}, \&
  {Wyttenbach}}]{Millon:2020xab}
{Millon}, M., {Courbin}, F., {Bonvin}, V., {et~al.} 2020, \aap, 640, A105,
  \dodoi{10.1051/0004-6361/202037740}

\bibitem[{Mortlock {et~al.}(1999)Mortlock, Webster, \& Francis}]{Mortlock1999}
Mortlock, D.~J., Webster, R.~L., \& Francis, P.~J. 1999, Monthly Notices of the
  Royal Astronomical Society, 309, 836,
  \dodoi{10.1046/j.1365-8711.1999.02872.x}

\bibitem[{Oguri(2007)}]{Oguri_2007}
Oguri, M. 2007, The Astrophysical Journal, 660, 1, \dodoi{10.1086/513093}

\bibitem[{{Oguri}(2019)}]{2019RPPh...82l6901O}
{Oguri}, M. 2019, Reports on Progress in Physics, 82, 126901,
  \dodoi{10.1088/1361-6633/ab4fc5}

\bibitem[{{Oguri} \& {Marshall}(2010)}]{2010MNRAS.405.2579O}
{Oguri}, M., \& {Marshall}, P.~J. 2010, \mnras, 405, 2579,
  \dodoi{10.1111/j.1365-2966.2010.16639.x}

\bibitem[{Oguri {et~al.}(2014)Oguri, Rusu, \& Falco}]{Oguri2014}
Oguri, M., Rusu, C.~E., \& Falco, E.~E. 2014, Monthly Notices of the Royal
  Astronomical Society, 439, 2494, \dodoi{10.1093/mnras/stu106}

\bibitem[{{Pelt} {et~al.}(1996){Pelt}, {Kayser}, {Refsdal}, \&
  {Schramm}}]{Pelt1996}
{Pelt}, J., {Kayser}, R., {Refsdal}, S., \& {Schramm}, T. 1996, \aap, 305, 97.
\newblock \doarXiv{astro-ph/9501036}

\bibitem[{Peng {et~al.}(1999)Peng, Impey, Falco, Kochanek, Lehar, McLeod, Rix,
  Keeton, \& Munoz}]{Peng_1999}
Peng, C.~Y., Impey, C.~D., Falco, E.~E., {et~al.} 1999, The Astrophysical
  Journal, 524, 572, \dodoi{10.1086/307860}

\bibitem[{{Pindor}(2005)}]{Pindor2005}
{Pindor}, B. 2005, \apj, 626, 649, \dodoi{10.1086/430048}

\bibitem[{{Planck Collaboration} {et~al.}(2020){Planck Collaboration},
  {Aghanim}, {Akrami}, {Ashdown}, {Aumont}, {Baccigalupi}, {Ballardini},
  {Banday}, {Barreiro}, {Bartolo}, {Basak}, {Battye}, {Benabed}, {Bernard},
  {Bersanelli}, {Bielewicz}, {Bock}, {Bond}, {Borrill}, {Bouchet}, {Boulanger},
  {Bucher}, {Burigana}, {Butler}, {Calabrese}, {Cardoso}, {Carron},
  {Challinor}, {Chiang}, {Chluba}, {Colombo}, {Combet}, {Contreras}, {Crill},
  {Cuttaia}, {de Bernardis}, {de Zotti}, {Delabrouille}, {Delouis}, {Di
  Valentino}, {Diego}, {Dor{\'e}}, {Douspis}, {Ducout}, {Dupac}, {Dusini},
  {Efstathiou}, {Elsner}, {En{\ss}lin}, {Eriksen}, {Fantaye}, {Farhang},
  {Fergusson}, {Fernandez-Cobos}, {Finelli}, {Forastieri}, {Frailis},
  {Fraisse}, {Franceschi}, {Frolov}, {Galeotta}, {Galli}, {Ganga},
  {G{\'e}nova-Santos}, {Gerbino}, {Ghosh}, {Gonz{\'a}lez-Nuevo}, {G{\'o}rski},
  {Gratton}, {Gruppuso}, {Gudmundsson}, {Hamann}, {Handley}, {Hansen},
  {Herranz}, {Hildebrandt}, {Hivon}, {Huang}, {Jaffe}, {Jones}, {Karakci},
  {Keih{\"a}nen}, {Keskitalo}, {Kiiveri}, {Kim}, {Kisner}, {Knox},
  {Krachmalnicoff}, {Kunz}, {Kurki-Suonio}, {Lagache}, {Lamarre}, {Lasenby},
  {Lattanzi}, {Lawrence}, {Le Jeune}, {Lemos}, {Lesgourgues}, {Levrier},
  {Lewis}, {Liguori}, {Lilje}, {Lilley}, {Lindholm}, {L{\'o}pez-Caniego},
  {Lubin}, {Ma}, {Mac{\'\i}as-P{\'e}rez}, {Maggio}, {Maino}, {Mandolesi},
  {Mangilli}, {Marcos-Caballero}, {Maris}, {Martin}, {Martinelli},
  {Mart{\'\i}nez-Gonz{\'a}lez}, {Matarrese}, {Mauri}, {McEwen}, {Meinhold},
  {Melchiorri}, {Mennella}, {Migliaccio}, {Millea}, {Mitra},
  {Miville-Desch{\^e}nes}, {Molinari}, {Montier}, {Morgante}, {Moss}, {Natoli},
  {N{\o}rgaard-Nielsen}, {Pagano}, {Paoletti}, {Partridge}, {Patanchon},
  {Peiris}, {Perrotta}, {Pettorino}, {Piacentini}, {Polastri}, {Polenta},
  {Puget}, {Rachen}, {Reinecke}, {Remazeilles}, {Renzi}, {Rocha}, {Rosset},
  {Roudier}, {Rubi{\~n}o-Mart{\'\i}n}, {Ruiz-Granados}, {Salvati}, {Sandri},
  {Savelainen}, {Scott}, {Shellard}, {Sirignano}, {Sirri}, {Spencer},
  {Sunyaev}, {Suur-Uski}, {Tauber}, {Tavagnacco}, {Tenti}, {Toffolatti},
  {Tomasi}, {Trombetti}, {Valenziano}, {Valiviita}, {Van Tent}, {Vibert},
  {Vielva}, {Villa}, {Vittorio}, {Wandelt}, {Wehus}, {White}, {White},
  {Zacchei}, \& {Zonca}}]{Planck:2018vyg}
{Planck Collaboration}, {Aghanim}, N., {Akrami}, Y., {et~al.} 2020, \aap, 641,
  A6, \dodoi{10.1051/0004-6361/201833910}

\bibitem[{Pooley {et~al.}(2009)Pooley, Rappaport, Blackburne, Schechter,
  Schwab, \& Wambsganss}]{Pooley_2009}
Pooley, D., Rappaport, S., Blackburne, J., {et~al.} 2009, The Astrophysical
  Journal, 697, 1892, \dodoi{10.1088/0004-637x/697/2/1892}

\bibitem[{{Press} {et~al.}(1992){Press}, {Rybicki}, \& {Hewitt}}]{Press1992}
{Press}, W.~H., {Rybicki}, G.~B., \& {Hewitt}, J.~N. 1992, \apj, 385, 416,
  \dodoi{10.1086/170952}

\bibitem[{Refsdal(1964)}]{Refsdal1964_2}
Refsdal, S. 1964, Monthly Notices of the Royal Astronomical Society, 128, 307,
  \dodoi{10.1093/mnras/128.4.307}

\bibitem[{Refsdal \& Bondi(1964)}]{Refsdal1964_1}
Refsdal, S., \& Bondi, H. 1964, Monthly Notices of the Royal Astronomical
  Society, 128, 295, \dodoi{10.1093/mnras/128.4.295}

\bibitem[{Riess {et~al.}(2019)Riess, Casertano, Yuan, Macri, \&
  Scolnic}]{Riess:2019cxk}
Riess, A.~G., Casertano, S., Yuan, W., Macri, L.~M., \& Scolnic, D. 2019,
  Astrophys. J., 876, 85, \dodoi{10.3847/1538-4357/ab1422}

\bibitem[{{Rodney} {et~al.}(2021){Rodney}, {Brammer}, {Pierel}, {Richard},
  {Toft}, {O'Connor}, {Akhshik}, \& {Whitaker}}]{2021NatAs.tmp..164R}
{Rodney}, S.~A., {Brammer}, G.~B., {Pierel}, J. D.~R., {et~al.} 2021, Nature
  Astronomy, 5, 1118, \dodoi{10.1038/s41550-021-01450-9}

\bibitem[{{Saha} {et~al.}(2006){Saha}, {Coles}, {Macci{\`o}}, \&
  {Williams}}]{Saha}
{Saha}, P., {Coles}, J., {Macci{\`o}}, A.~V., \& {Williams}, L. L.~R. 2006,
  \apjl, 650, L17, \dodoi{10.1086/507583}

\bibitem[{Shafieloo(2007)}]{Shafieloo:2007cs}
Shafieloo, A. 2007, Mon. Not. Roy. Astron. Soc., 380, 1573,
  \dodoi{10.1111/j.1365-2966.2007.12175.x}

\bibitem[{Shafieloo {et~al.}(2006)Shafieloo, Alam, Sahni, \&
  Starobinsky}]{Shafieloo:2005nd}
Shafieloo, A., Alam, U., Sahni, V., \& Starobinsky, A.~A. 2006, Mon. Not. Roy.
  Astron. Soc., 366, 1081, \dodoi{10.1111/j.1365-2966.2005.09911.x}

\bibitem[{Shafieloo \& Clarkson(2010)}]{Shafieloo:2009hi}
Shafieloo, A., \& Clarkson, C. 2010, Phys. Rev. D, 81, 083537,
  \dodoi{10.1103/PhysRevD.81.083537}

\bibitem[{{Shu} {et~al.}(2021){Shu}, {Belokurov}, \& {Evans}}]{Shu2020}
{Shu}, Y., {Belokurov}, V., \& {Evans}, N.~W. 2021, \mnras, 502, 2912,
  \dodoi{10.1093/mnras/stab241}

\bibitem[{{Springer} \& {Ofek}(2021{\natexlab{a}})}]{Springer:2021yhe}
{Springer}, O.~M., \& {Ofek}, E.~O. 2021{\natexlab{a}}, \mnras, 506, 864,
  \dodoi{10.1093/mnras/stab1600}

\bibitem[{{Springer} \& {Ofek}(2021{\natexlab{b}})}]{Springer:2021jhm}
---. 2021{\natexlab{b}}, \mnras, 508, 3166, \dodoi{10.1093/mnras/stab2432}

\bibitem[{{Suyu} {et~al.}(2020){Suyu}, {Huber}, {Ca{\~n}ameras}, {Kromer},
  {Schuldt}, {Taubenberger}, {Y{\i}ld{\i}r{\i}m}, {Bonvin}, {Chan}, {Courbin},
  {N{\"o}bauer}, {Sim}, \& {Sluse}}]{2020A&A...644A.162S}
{Suyu}, S.~H., {Huber}, S., {Ca{\~n}ameras}, R., {et~al.} 2020, \aap, 644,
  A162, \dodoi{10.1051/0004-6361/202037757}

\bibitem[{{Tewes} {et~al.}(2013){Tewes}, {Courbin}, \& {Meylan}}]{Tewes2013}
{Tewes}, M., {Courbin}, F., \& {Meylan}, G. 2013, \aap, 553, A120,
  \dodoi{10.1051/0004-6361/201220123}

\bibitem[{{Treu}(2010)}]{Treu2010}
{Treu}, T. 2010, \araa, 48, 87, \dodoi{10.1146/annurev-astro-081309-130924}

\bibitem[{{Treu} \& {Marshall}(2016)}]{2016A&ARv..24...11T}
{Treu}, T., \& {Marshall}, P.~J. 2016, \aapr, 24, 11,
  \dodoi{10.1007/s00159-016-0096-8}

\bibitem[{{Treu} {et~al.}(2018){Treu}, {Agnello}, {Baumer}, {Birrer},
  {Buckley-Geer}, {Courbin}, {Kim}, {Lin}, {Marshall}, {Nord}, {Schechter},
  {Sivakumar}, {Abramson}, {Anguita}, {Apostolovski}, {Auger}, {Chan}, {Chen},
  {Collett}, {Fassnacht}, {Hsueh}, {Lemon}, {McMahon}, {Motta}, {Ostrovski},
  {Rojas}, {Rusu}, {Williams}, {Frieman}, {Meylan}, {Suyu}, {Abbott},
  {Abdalla}, {Allam}, {Annis}, {Avila}, {Banerji}, {Brooks}, {Carnero Rosell},
  {Carrasco Kind}, {Carretero}, {Castander}, {D'Andrea}, {da Costa}, {De
  Vicente}, {Doel}, {Eifler}, {Flaugher}, {Fosalba}, {Garc{\'\i}a-Bellido},
  {Goldstein}, {Gruen}, {Gruendl}, {Gutierrez}, {Hartley}, {Hollowood},
  {Honscheid}, {James}, {Kuehn}, {Kuropatkin}, {Lima}, {Maia}, {Martini},
  {Menanteau}, {Miquel}, {Plazas}, {Romer}, {Sanchez}, {Scarpine}, {Schindler},
  {Schubnell}, {Sevilla-Noarbe}, {Smith}, {Smith}, {Soares-Santos}, {Sobreira},
  {Suchyta}, {Swanson}, {Tarle}, {Thomas}, {Tucker}, \& {Walker}}]{Treu2018}
{Treu}, T., {Agnello}, A., {Baumer}, M.~A., {et~al.} 2018, \mnras, 481, 1041,
  \dodoi{10.1093/mnras/sty2329}

\bibitem[{Verde {et~al.}(2019)Verde, Treu, \& Riess}]{Verde:2019ivm}
Verde, L., Treu, T., \& Riess, A.~G. 2019, Nature Astron., 3, 891,
  \dodoi{10.1038/s41550-019-0902-0}

\bibitem[{{Wong} {et~al.}(2020){Wong}, {Suyu}, {Chen}, {Rusu}, {Millon},
  {Sluse}, {Bonvin}, {Fassnacht}, {Taubenberger}, {Auger}, {Birrer}, {Chan},
  {Courbin}, {Hilbert}, {Tihhonova}, {Treu}, {Agnello}, {Ding}, {Jee},
  {Komatsu}, {Shajib}, {Sonnenfeld}, {Blandford}, {Koopmans}, {Marshall}, \&
  {Meylan}}]{Wong:2019kwg}
{Wong}, K.~C., {Suyu}, S.~H., {Chen}, G. C.~F., {et~al.} 2020, \mnras, 498,
  1420, \dodoi{10.1093/mnras/stz3094}

\end{thebibliography}
\end{document}